\documentclass[a4paper,12pt]{article}
\usepackage[utf8]{inputenc}
\usepackage[top=1in, bottom=1in, left=1in, right=1in]{geometry}
\usepackage{color}
\usepackage{amsmath}
\usepackage{amssymb}
\usepackage{amsthm}
\usepackage{graphics}
\usepackage{graphicx}
\usepackage{epstopdf}
\usepackage{epsfig}
\usepackage{comment}
\usepackage{multirow} 
\usepackage{tikz}
\usepackage{natbib}
\usepackage{stix}
\usepackage{array}
\usepackage[doublespacing]{setspace}
\usepackage{xr}
\usepackage[normalem]{ulem}
\usepackage{hyperref}
\usepackage{booktabs}        
\usepackage{multirow}        
\usepackage{multicol}        
\usepackage{pdflscape}  

\usepackage{titlesec}
\titleformat*{\section}{\large\bfseries}
\titleformat*{\subsection}{\large\bfseries}
\titleformat*{\subsubsection}{\normalsize\bfseries}

\title{\Large Using a Two-Parameter Sensitivity Analysis Framework to Efficiently
Combine Randomized and Non-randomized Studies}

\author{\normalsize Ruoqi Yu$^1$\footnote{Equal contribution} \footnote{Address for correspondence: Ruoqi Yu, University of Illinois Urbana-Champaign, 605 E. Springfield Ave., Champaign, IL 61820, USA. Email: ruoqi.yu.ry@gmail.com}, Bikram Karmakar$^2$$^*$, Jessica Vandeleest$^3$, Eleanor Bimla Schwarz$^4$\\ \normalsize $^1$Department of Statistics, University of Illinois Urbana-Champaign, Champaign, IL 61820, USA\\\normalsize $^2$Department of Statistics, University of Wisconsin-Madison, Madison, WI 53706, USA\\\normalsize $^3$California National Primate Research Center, University of California Davis, Davis, CA 95616, USA\\\normalsize $^4$Department of Medicine, University of California San Francisco, San Francisco, CA 94143, USA}

\date{}

\newtheorem{theorem}{Theorem}
\newtheorem{proposition}{Proposition}

\newtheorem{assumption}{Assumption}

\newcommand{\X}{\mathcal{X}} 

\makeatletter
\newcommand{\dwidetilde}[1]{{%
  \mathpalette\double@widetilde{#1}%
}}
\newcommand{\double@widetilde}[2]{%
  \sbox\z@{$\m@th#1\widetilde{#2}$}%
  \ht\z@=.9\ht\z@
  \widetilde{\box\z@}%
}

\titlespacing*{\section}{0pt}{*1}{*0.5}

\renewenvironment{abstract}
 {\small
  \begin{center}
  \bfseries \abstractname\vspace{-.5em}\vspace{0pt}
  \end{center}
  \list{}{
    \setlength{\leftmargin}{.05cm}%
    \setlength{\rightmargin}{\leftmargin}%
  }%
  \item\relax}
 {\endlist}
 
\makeatother
\begin{document}

\begin{onehalfspacing}
   \maketitle     
\end{onehalfspacing}

\vspace{-20pt}
\begin{abstract}
Causal inference is vital for informed decision-making across fields such as biomedical research and social sciences. Randomized controlled trials (RCTs) are considered the gold standard for internal validity of inferences, whereas observational studies (OSs) often provide the opportunity for greater external validity. However, both data sources have inherent limitations preventing their use for broadly valid statistical inferences: RCTs may lack generalizability due to their selective eligibility criterion, and OSs are vulnerable to unobserved confounding. This paper proposes an innovative approach to integrate RCT and OS that borrows the other study's strengths to remedy each study's limitations. The method uses a novel triplet matching algorithm to align RCT and OS samples and a new two-parameter sensitivity analysis framework to quantify internal and external validity biases. This combined approach yields causal estimates that are more robust to hidden biases than OSs alone and provides reliable inferences about the treatment effect in the general population. We apply this method to investigate the effects of lactation on maternal health using a small RCT and a long-term observational health records dataset from the California National Primate Research Center. This application demonstrates the practical utility of our approach in generating scientifically sound and actionable causal estimates.
\end{abstract}

\noindent%
{\it Keywords:}  Causal inference, generalizability bias, matching, sensitivity analysis, unmeasured confounding.
\vfill


\section{Introduction}
\subsection{Causal inference and two data sources}
 
In the context of causal inference, internal validity and external validity are two critical concepts that help ensure the reliability and generalizability of research findings \citep{cook_campbell_1979}. Internal validity refers to the extent to which a study accurately identifies the true causal relationships within the study itself, controlling for the influence of other factors such as confounding variables and measurement errors \citep{brewer2000research}. Researchers strive to establish strong internal validity to ensure that their findings are trustworthy and credible. In contrast, external validity, also known as generalizability, refers to the applicability of findings to broader populations or contexts \citep{degtiar2023review}. External validity is necessary to determine whether those findings have broader applicability of the research findings. Striking the right balance between internal and external validity is essential for producing scientifically sound results that are both relevant and actionable.

These two concepts of validity of causal inference are closely tied to the two primary statistical methods for causality: randomized controlled trials (RCTs) and non-physically-randomized observational studies (OSs). RCTs, often regarded as the gold standard for causal inference, excel in internal validity due to the random assignment of treatments, which reduces the impact of confounding. However, their strictly controlled eligibility criteria can compromise 
external validity, making it difficult to generalize the findings to broader populations 
\citep{rothwell2005external}. Moreover, due to their high cost and logistical complexities, 
RCTs often have smaller sample sizes, which can undermine the power of the statistical analysis. On the other hand, a wealth of observational data has become increasingly accessible to scholars through national surveys, administrative claims databases, and electronic health records. OSs often excel in 
external validity, since they typically boast expansive sample sizes and better reflect the diversity of the population. Nevertheless, the existence of potential unobserved confounding variables due to a non-random and unknown assignment of treatments can threaten 
internal validity, making researchers hesitate to ascribe causal interpretations to their conclusions \citep{rosenbaum2002observational}.

Given these challenges, the central question arises: how can researchers combine the strengths of RCTs and OSs to achieve more robust causal estimates? Specifically, how can we estimate the treatment effect on a target population by aggregating the internal validity of RCTs with the external validity of OSs?

Recent literature has considered ways of combining RCT and OS to improve the internal or external validity of the inference. One line of research uses OSs to gain insights into the target population's characteristics, allowing researchers to adjust RCT inferences accordingly to increase external validity \citep{cole2010generalizing,stuart2011use,tipton2013improving,pearl2022external,hartman2015sample,dahabreh2019generalizing}. Another line of research focuses on enhancing the efficiency of RCT estimates by incorporating observational data \citep{gagnon2023precise,wu2022integrative}.
Researchers have also delved into the bias-variance trade-off between RCTs and OSs \citep{chen2021minimax,yang2023elastic}.
However, despite this progress, existing methods often make simplifying assumptions, e.g., assuming 
{the positivity between the RCT and OS populations}.
In reality, RCTs are often conducted on a selective population, either for convenience or higher statistical power. Thus, the support of 
participants' characteristics in an RCT 
may only be a nonrepresentative subset of the support of the target population's characteristics. Although \cite{Zivich2024} tackled this 
nonpositivity issue with one covariate, the situation can be more complicated in practice. Furthermore, the existing approaches fall short of addressing both the internal and external validity
biases present in the two data sources.

To address these limitations, we propose a novel method that combines RCT and OS data while acknowledging the inherent limitations of each study design. Specifically, for OSs, we introduce a sensitivity analysis approach for unmeasured confounding under Rosenbaum's sensitivity analysis framework for a general blocked design, focusing on testing the weak null hypothesis for a population average treatment effect. This analysis quantifies the extent to which the inference is robust to hidden biases from possible unmeasured confounding.
For RCTs, we present a new sensitivity analysis model to account for generalizability bias (i.e., external validity bias), which arises when the RCT sample is not representative of the target population 
due to limited support. This analysis provides confidence intervals that account for a specified level of generalizability bias. Finally, we develop a method that combines the two sensitivity analyses, addressing both internal and external validity biases.  
The combining method creates a calibrated confidence interval that is valid under specified levels of generalizability bias and hidden bias from unmeasured confounding.
We develop a triplet matching algorithm that aligns samples from the RCT and OS, facilitating our new two-parameter sensitivity analysis framework. While the combined inference does not remove the internal and external validity biases when both are present, we show that it is more robust to these biases than either of the individual inferences. Furthermore, the combined two-parameter sensitivity analysis confidence intervals tend to be shorter than either of the two single-parameter sensitivity analysis confidence intervals. Thus, our results demonstrate that it is always preferable to use the combined inference rather than the data sources separately in practice.

\subsection{Lactation and maternal health: Primate data from both sources}
Lactation, whether a mother breastfeeds her newborns, is a decision that mothers need to make for every baby they deliver. The US Centers for Disease Control and Prevention (CDC), the American Academy of Pediatrics (AAP), the American College of Obstetrics and Gynecology (ACOG), and the World Health Organization (WHO) all recommend exclusively breastfeeding for the first 6 months of their infant’s life and continuing breastfeeding for at least 2 years.
However, current CDC data show that only 84.1\% of US mothers initiate breastfeeding, just 59.8\% breastfeed for 6 months, and only 27.2\% exclusively breastfeed for 6 months. 

In humans, OSs suggest that pregnancy without lactation (e.g., formula feeding) is associated with adverse health outcomes for mothers, including maternal weight retention and increasing obesity over time \citep{harder2005duration,von1999breast}. However, due to the possibility of unmeasured confounding that must be acknowledged with any OS, it remains speculative that lactation plays a significant role in determining maternal health in later life. On the other hand, RCTs related to the care and feeding of human infants are limited by ethical considerations. For example, \cite{oken2013effects} conducted a cluster RCT that promoted longer breastfeeding duration among women who had already chosen to breastfeed, but designing an RCT that directly manipulates whether women begin breastfeeding is neither feasible nor ethical. 
Therefore, data from animal studies can play a critical role in understanding how lactation
may affect maternal health across the lifespan.

Specifically, to explore the causal impact of first-time non-lactation on maternal weight with a nonhuman primate model, a small RCT with 18 monkeys stratified in 6 matched sets was conducted at the California National Primate Research Center (CNPRC). Each matched set included one treated unit (no lactation) with parity ranging from second to fifth offspring, aged 6-8 years, that had lactated in previous pregnancies (the treated females had to have reared all but the most recent infant). Each treated unit had two matched controls, matching on parity, age, weight ($+/-1$ kg), and lactation history (control animals had to have reared all infants she had birthed).
While the RCT was restricted to specific age and parity ranges, our focus is on the general population beyond the subjects satisfying the selection criteria, aiming to use the primate data to inform human studies. 
In addition, the RCT with such a small sample size may not be able to detect subtle effects that a large study can detect due to the lack of power. All procedures were approved by the University of California, Davis IACUC.

In addition, the CNRPC maintains a long-term database of health records for all animals. 
Records include information gathered from birth to death, including weights (taken at approximately 6-month intervals), animal locations, and reproductive histories. 
Of particular interest for this project are records involving the outdoor breeding colony, which consists of 24 half-acre enclosures containing social groups of 80--120 animals of multiple age/sex classes. Enclosures had either grass or gravel substrate with multiple enrichment objects.   
Animals were provided ad libitum access to food and water and additional produce enrichment 1-2 times per week.  Reproductive-age females (age 3--18 years) in this colony usually get pregnant yearly, and there are approximately 600 infants born each year, although, as with humans, not all pregnancies go to term. 
We use data for conception and female weight from 2009--2019, which involves 2116 mother monkeys. We focus on the sample with information on the lactation status and non-missing weight measurements at both 3 and 6 months postpartum. The treated group includes the first non-lactation conception record for monkeys that are non-lactating. The control group includes the conception records with always lactated history. If there are multiple conception records for a monkey, we keep the one with maximum parity since the treated monkeys have typically delivered more babies than the control monkeys. In sum, we have 591 primates in the observational data, each with one conception record. With the OS, we can adjust the confounding effects from the observed covariates, i.e., age, parity, and baseline weight before pregnancy. However, the potential unobserved confounders can still bias the estimated causal effects.

To understand and address the limitations of two data sources, we apply the newly proposed matching design and two-parameter sensitivity analysis method to combine the complementary strengths, aiming to quantify the internal and external validity biases and provide more robust causal conclusions than working with a single data source.

\section{Notation and Framework}
\label{sec_notation}

Let $\Omega$ denote the probability space of units where unit $k$ has a vector of pre-treatment covariates $X_k$ and potential outcomes $Y_k(1)$ and $Y_k(0)$ according to whether it is exposed to the treatment or not \citep{rubin1974}. Let $S_k$ be an indicator so that $S_k=1$ denotes that the unit $k$ is selected for the RCT. 
We assume that the selection in the RCT is based only on the covariate values. 
This assumption is formally stated as follows:
\begin{assumption}
    The selection indicator $S_k$ is independent of the potential outcomes given the covariates; following the notation of \cite{dawid1979}, we require $S_k \perp Y_k(1), Y_k(0) \mid X_k$.
\end{assumption}
\noindent This assumption usually holds in practice as researchers enroll units in an RCT by considering their collected information. 
In our primate data, the selection into RCT only depends on the monkeys' age, parity, weight, and lactation history, which are all recorded by CNRPC.

We further allow that the RCT has a smaller support for the covariate values. Define $e(x) = \Pr( S_k=1 \mid X_k=x)$ for the selection probability into the RCT for a unit with characteristic $x$.    We make the following assumption.
\begin{assumption}
    $0\leq e(x)<1$ for all $x$.
\end{assumption}
\noindent Thus, each covariate value $x$ in the whole support has a positive probability of being represented in the OS. On the other hand, $e(x)$ may be 0 for some covariate values $x$, leading to no opportunity of units with that $x$ in the RCT, perhaps because the RCT's eligibility criteria do not allow it. In particular, the covariate space with $e(x)>0$ denotes the  common support
$\mathcal{X}$ between the RCT and the OS. 
The {common support} is a subset of the support of the covariate for the OS. This is likely in practice and is often the reason for preferring a large OS to report a generalizable result when there is concern about treatment effect 
heterogeneity between the common support and the whole support. For instance, the monkeys in our primate RCT have a parity of 2-5 offspring, an age of 6-8 years old, a weight of 5-10 kg, and have lactated in previous pregnancies. Nevertheless, our main focus is the general population beyond the selection criteria in order to gain insights into human beings from the primate data. 
In rare cases where the RCT support does not fully overlap with the OS support, we will only utilize the RCT units whose covariates belong to the covariate support of the OS units, denoting the intersection of the covariate supports as $\mathcal{X}$. 

To keep track of the technical details across the two studies, we use the indexing $k$ for the general population unit, $l$ for the OS units, and $m$ for the RCT units.

The conditional probability distribution of $(X_k, Y_k(1), Y_k(0))$ over $\Omega$ given $S_k=0$ equals the population distribution of the corresponding variables in the OS. Specifically, there are $n_o$ observational study units $l=1,\ldots,n_o$ which are independently and identically distributed (i.i.d.) and drawn from a population distribution such that the law of $(X^o_l, Y^o_l(1), Y^o_l(0))$ is the same as the conditional law of $(X_k, Y_k(1), Y_k(0))$ given $S_k=0$. The unit $l$ also has a treatment indicator $Z^o_l$ i.i.d. across $l$. 
Under the stable unit treatment value assumption (SUTVA) for the OS, requiring no interference between subjects and no hidden treatment versions, we can write the observed outcome $Y^o_l = Z^o_lY^o_l(1) + (1-Z^o_l)Y^o_l(0)$.

Parallelly, the conditional distribution of $(X_k, Y_k(1), Y_k(0))$ given $S_k=1$ equals the population distribution of the corresponding quantities in the RCT. Suppose there are $n_r$ units in the RCT. The information $(X^r_m, Y^r_m(1), Y^r_m(0))$, for $m=1,\ldots,n_r$, is drawn i.i.d. from a distribution with the law that matches the conditional law of $(X_k, Y_k(1), Y_k(0))$ given $S_k=1$. Additionally, the treatment indicators $Z^r_1, \ldots, Z^r_{n_r}$ generate the observed outcomes  $Y^r_m = Z^r_mY^r_m(1) + (1-Z^r_m)Y^r_m(0)$. Unlike in the OS, we do not assume $Z^r_m$s are independent. This allows general randomization designs, e.g., completely randomized designs and block designs.

The above framework with the common probability space $\Omega$ binds the two studies but allows for the complexities of the two studies by not including the treatment indicators for either the OS or the RCT in $\Omega$.
Due to the physical randomization, the RCT satisfies 
$\{{   Y^r_m(1), Y^r_m(0)} : m=1,\ldots,n_r\}\perp \{Z^r_m: m=1,\ldots,n_r\}\text{ given }\{X^r_m: m=1,\ldots,n_r\},$
and we know its randomization process, i.e., the probability distribution of $\{Z^r_m: m=1,\ldots,n_r\}$ given $\{X^r_m: m=1,\ldots,n_r\}$. On the other hand, the framework allows unmeasured confounders, say, $u_l^o$s, in the OS. Thus, the treatment assignment $Z^o_l$ may depend on $\{Y^o_l(1), Y^o_l(0)\}$ even after conditioning on $X^o_l$, violating the no unmeasured confounders assumption. We define the propensity score for the OS as 
$\pi(x) := \Pr (Z^o_l=1\mid X^o_l=x)$. 

We are interested in estimating the average treatment effect on the observational treated population (ATOT)
\vspace{-1em}
\begin{equation}\label{eq_ATOT}
    \beta^\star = E\{ Y^o_l(1) - Y^o_l(0)\mid Z^o_l=1\}.
    \vspace{-0.5em}
\end{equation}
The goal is to quantify the internal and external validity biases in estimating the ATOT $\beta^\star$ using the single data sources and provide a robust estimate of $\beta^\star$ by efficiently combining the RCT and OS.

In Section~\ref{sec_matching}, we propose a triplet matching design that supports the new inference framework in Section~\ref{sec_inference}. The performance of the proposed methods is evaluated using numerical experiments in Section~\ref{sec_simulation}. A thorough analysis of the primate data is presented in Section~\ref{sec_analysis}.

\section{Design: A New Matching Design to Integrate RCT and OS}\label{sec_matching}
\subsection{A triplet matching algorithm for data integration}

The primary idea of matching in OS is to construct matched sets consisting of treated and control units that are similar in terms of observed covariates, thereby mimicking a stratified randomized experiment. These matched sets can take on various forms, such as one treated unit paired with one control, one treated unit paired with a fixed number of controls, or even one treated unit paired with a variable number of controls \citep{rosenbaum1989optimal,smith1997matching,hansen2004full,lu2004optimal,stuart2008matching,zubizarreta2012using,pimentel2015variable,yu2020matching}. 
As a design-based method, matching offers transparent and interpretable results, which enhances the objectivity of the causal inferences \citep{rubin2008objective}. For a more comprehensive overview of matching methods, refer to \cite{stuart2010matching} and \cite{rosenbaum2020modern}.

While matching between OS treated and OS control units is routine, a critical part of our matching design is matching OS treated and OS control units along with the RCT units. We propose a matching algorithm that matches across these three groups. 

For the OS, we allow each treated unit to be matched to a variable number of control units determined by the investigators. 
For instance, investigators can choose to form matched pairs with similar treated and control units from the OS. 
For consistent estimation of our ATOT from these matched OS units, we need a sufficient number of control units with similar covariate values for each treated unit; \cite{savje2022inconsistency} proved the matching estimator's inconsistency when this is not true.

The design decisions of how RCT units should be matched are driven by our target parameter ATOT. The ATOT can be separated into two parts: the 
ATOT in the {common support $\mathcal{X}$ and in the external support from the OS $\mathcal{X}^c$ (i.e., the difference between the OS support and the RCT support)}. The RCT units are non-informative regarding the latter. Moreover, while 
they inform us of the former, the standard estimator may not be consistent since the covariate distributions differ between the RCT units and the OS treated units in $\mathcal{X}$. In the following, we propose a solution to this design problem.

Let $D_l$ be the indicator for whether the corresponding covariate value $X_l$ lies within $\mathcal{X}$ for OS unit $l$, $l=1,\ldots,n_o$. 
To leverage the information from the RCT, we define the generalization score of each RCT unit to the OS treated group as $\nu(x)=\pi_o(x)(1-{e}(x))/{e}(x)$, where ${\pi}_o(x)$ is the propensity score for the OS within the {common support} and ${e}(x)$ is the selection probability for the RCT. Similar to the concept of the "entire number" introduced by \cite{yoon2009new}, the generalization score represents the average number of OS treated units in the common support
that are available to be matched to an RCT unit with covariate value $x$. 
Since the generalization score is typically unknown, we need to estimate it in practice and use this estimate $\hat{\nu}(X)$ to calcultate the number of treated individuals from the OS for matching with each RCT unit. 
As a result, to ensure that the matched observational data and RCT are similar in the 
{common support} and reflect the covariate distribution in the {common support} of the observational treated population, we perform variable ratio matching {\citep{pimentel2015variable}} based on the generalization score. 
Let $n_{o1}^+$ and $n_{o1}^-$ denote the number of OS treated units in the {common support} and in the {external support}, respectively. We create $C_m=\lceil n_{o1}^+\hat\nu(X_m)/\sum_{m=1}^{{  n_r}}(\hat\nu(X_m))\rceil$ copies for each RCT unit $m$. The weighted RCT sample has a similar covariate distribution to the OS treated group in the common support $\mathcal{X}$, so we can treat them as two samples drawn from the same distribution. Our goal in the triplet matching process is to make a matched OS control group in the common support have a distribution that mirrors these two samples.

For the technical convenience of matching, we create imaginary units so that the weighted RCT group and the OS treated group have the same sample size. Specifically, since $\sum_{m=1}^{n_r}C_m\geq n_{o1}^+$, we create $\sum_{m=1}^{n_r} C_m-n_{o1}^+$ "imaginary" units in the OS treated group for the common support $\mathcal{X}$. 
For the external support $\mathcal{X}^c$, we create $n_{o1}^-$ imaginary RCT units, the same number as there are OS treated units in $\mathcal{X}^c$ ($n_{o1}^-$), so that the OS treated and controls matched to the same imaginary RCT unit form a matched set. Then, we apply a modification of the three-way approximate matching algorithm in \cite{karmakar2019using} to implement this matching process. In the matching process, we set the distance between the imaginary units and any other units to be a large penalty so that matched sets using only non-imaginary units are close.
We discard the matched sets of imaginary OS treated units in the inference stage in the common support as well as the imaginary RCT units in the external support.

Although the proposed matching design focuses on a binary treatment, it can be developed in parallel to analyze data with multiple treatment groups. The implementation of the matching algorithm from \cite{karmakar2019using} already allows for any $k$-many treatment groups.

We now introduce the notation for our matched data. Let $i$ index our $I$ matched sets. Suppose matched set $i$ contains $J_i$ OS units and zero or one RCT unit. Let $ij$, for $j=1,\ldots, J_i$, denote the OS units in the matched set $i$. The matched structure looks different according to whether the units belong to the {common support} $\X$ or not. However, each matched set $i$ contains exactly one OS treated unit such that $\sum_{j=1}^{J_i} Z^o_{ij}=1$. Each OS control unit is included in at most one matched set. Each RCT unit is included in zero, one, or more matched sets. There are $C_m$ matched sets that includes RCT unit $m$, $m=1,\ldots,n_r$. 

\subsection{Illustration of the matching procedure}
To demonstrate the proposed triplet matching algorithm, consider the toy example in Figure~\ref{fig_matchingillustration}. Colors are used to denote different covariate values. The original data is displayed in the first panel. The common support $\X$ includes the red and orange units, but the OS population also includes blue and purple units, which are not represented in the RCT population. In the first step of matching, we choose $J_i=2$ for the OS and calculate the generalization score $\nu(X_m)$ for each RCT unit and the corresponding number of copies $C_m$ required for duplication to create the weighted RCT group. A red RCT unit has ${e}(x) = 2/9$ and ${   \pi_o(x)}=3/7$; hence, it should be matched to $(3/7)(1-2/9)/(2/9) = 1.5$ treated units in the OS, i.e., $\nu(red)=1.5$.
An orange RCT unit has ${e}(x) = 1/7$ and ${   \pi_o(x)}=1/3$; hence, it should be matched to $(1/3)(1-1/7)/(1/7) = 2$ treated units in the OS, i.e., $v(orange)=2$. Correspondingly, we calculate the number of copies as $C=\lceil5\times1.5/(1.5+1.5+2)\rceil=2$ for the two red RCT units and $C=\lceil5\times2/(1.5+1.5+2)\rceil=2$ for the orange RCT unit. Then in step 2, we add imaginary units labeled in gray to make the OS treated group and the weighted RCT group the same size. Finally, in step 3, we perform a 1-1-1 matching. The constructed matched sets are labeled with black dotted lines. In practice, investigators can also change the matching ratio for the OS based on their OS data structure. In the inference stage, we exclude the fourth matched set with the imaginary OS treated unit. Therefore, the inclusion of imaginary units is only for technical convenience and has no practical effects on the inferences.

\begin{figure}[htbp]
	\centering
	\begin{tikzpicture}[scale=0.65]
\scriptsize
		\node[inner sep=0] at (1.5,2) {Step 0: Original Data};
		\node[inner sep=0] at (0,1) {OS Treated};
		\node[inner sep=0] at (1.5,1) {RCT};
		\node[inner sep=0] at (3,1) {OS Control};
		
		\filldraw[red!50!white] (0,0) circle (0.1);
		\filldraw[red!50!white] (0,-0.5) circle (0.1);
		\filldraw[red!50!white] (0,-1) circle (0.1);
		\filldraw[orange!50!white] (0,-1.5) circle (0.1);
		\filldraw[orange!50!white] (0,-2) circle (0.1);
		
		\filldraw[red!50!white] (1.5,0) circle (0.1);
		\filldraw[red!50!white] (1.5,-0.5) circle (0.1);
		\filldraw[orange!50!white] (1.5,-1) circle (0.1);
		
		\filldraw[red!50!white] (3,0) circle (0.1);
		\filldraw[red!50!white] (3,-0.5) circle (0.1);
		\filldraw[red!50!white] (3,-1) circle (0.1);
		\filldraw[red!50!white] (3,-1.5) circle (0.1);
		\filldraw[orange!50!white] (3,-2) circle (0.1);
		\filldraw[orange!50!white] (3,-2.5) circle (0.1);
		\filldraw[orange!50!white] (3,-3) circle (0.1);
		\filldraw[orange!50!white] (3,-3.5) circle (0.1);
		
		\draw[line width=1, dashed] (-1,-4) -- (4,-4);
		
		\filldraw[cyan!50!white] (0,-4.5) circle (0.1);
		\filldraw[cyan!50!white] (0,-5) circle (0.1);
		\filldraw[purple!50!white] (0,-5.5) circle (0.1);
		\filldraw[purple!50!white] (0,-6) circle (0.1);
		\filldraw[purple!50!white] (0,-6.5) circle (0.1);
		\filldraw[purple!50!white] (0,-7) circle (0.1);
		
		\filldraw[cyan!50!white] (3,-4.5) circle (0.1);
		\filldraw[cyan!50!white] (3,-5) circle (0.1);
		\filldraw[cyan!50!white] (3,-5.5) circle (0.1);
		\filldraw[cyan!50!white] (3,-6) circle (0.1);
		\filldraw[cyan!50!white] (3,-6.5) circle (0.1);
		\filldraw[cyan!50!white] (3,-7) circle (0.1);
		\filldraw[purple!50!white] (3,-7.5) circle (0.1);
		\filldraw[purple!50!white] (3,-8) circle (0.1);
		\filldraw[purple!50!white] (3,-8.5) circle (0.1);
		\filldraw[purple!50!white] (3,-9) circle (0.1);
		\filldraw[purple!50!white] (3,-9.5) circle (0.1);
		
		\node[inner sep=0] at (7.5,2) {Step 1: Duplicate units};
		\node[inner sep=0] at (6,1) {OS Treated};
		\node[inner sep=0] at (7.5,1) {RCT};
		\node[inner sep=0] at (9,1) {OS Control};
		
		\filldraw[red!50!white] (6,0) circle (0.1);
		\filldraw[red!50!white] (6,-0.5) circle (0.1);
		\filldraw[red!50!white] (6,-1) circle (0.1);
		\filldraw[orange!50!white] (6,-1.5) circle (0.1);
		\filldraw[orange!50!white] (6,-2) circle (0.1);
		
		\filldraw[red!50!white] (7.5,0) circle (0.1);
		\filldraw[red!50!white] (7.5,-0.5) circle (0.1);
		\filldraw[red!50!white] (7.5,-1) circle (0.1);
        \filldraw[red!50!white] (7.5,-1.5) circle (0.1);
		\filldraw[orange!50!white] (7.5,-2) circle (0.1);
		\filldraw[orange!50!white] (7.5,-2.5) circle (0.1);
		\draw[black, thick] (7.2,0.2) rectangle (7.8,-0.7);
        \draw[black, thick] (7.2,-0.8) rectangle (7.8,-1.7);
        \draw[black, thick] (7.2,-1.8) rectangle (7.8,-2.7);
        
		\filldraw[red!50!white] (9,0) circle (0.1);
		\filldraw[red!50!white] (9,-0.5) circle (0.1);
		\filldraw[red!50!white] (9,-1) circle (0.1);
		\filldraw[red!50!white] (9,-1.5) circle (0.1);
		\filldraw[orange!50!white] (9,-2) circle (0.1);
		\filldraw[orange!50!white] (9,-2.5) circle (0.1);
		\filldraw[orange!50!white] (9,-3) circle (0.1);
		\filldraw[orange!50!white] (9,-3.5) circle (0.1);
		
		\draw[line width=1, dashed] (5,-4) -- (10,-4);
		
		\filldraw[cyan!50!white] (6,-4.5) circle (0.1);
		\filldraw[cyan!50!white] (6,-5) circle (0.1);
		\filldraw[purple!50!white] (6,-5.5) circle (0.1);
		\filldraw[purple!50!white] (6,-6) circle (0.1);
		\filldraw[purple!50!white] (6,-6.5) circle (0.1);
		\filldraw[purple!50!white] (6,-7) circle (0.1);

		\filldraw[cyan!50!white] (9,-4.5) circle (0.1);
		\filldraw[cyan!50!white] (9,-5) circle (0.1);
		\filldraw[cyan!50!white] (9,-5.5) circle (0.1);
		\filldraw[cyan!50!white] (9,-6) circle (0.1);
		\filldraw[cyan!50!white] (9,-6.5) circle (0.1);
		\filldraw[cyan!50!white] (9,-7) circle (0.1);
		\filldraw[purple!50!white] (9,-7.5) circle (0.1);
		\filldraw[purple!50!white] (9,-8) circle (0.1);
		\filldraw[purple!50!white] (9,-8.5) circle (0.1);
		\filldraw[purple!50!white] (9,-9) circle (0.1);
		\filldraw[purple!50!white] (9,-9.5) circle (0.1);
		
				\node[inner sep=0] at (13.5,2) {Step 2: Add imaginary units};
		\node[inner sep=0] at (12,1) {OS Treated};
		\node[inner sep=0] at (13.5,1) {RCT};
		\node[inner sep=0] at (15,1) {OS Control};
		
		\filldraw[red!50!white] (12,0) circle (0.1);
		\filldraw[red!50!white] (12,-0.5) circle (0.1);
		\filldraw[red!50!white] (12,-1) circle (0.1);
		\filldraw[black!50!white] (12,-1.5) circle (0.1);
		\filldraw[orange!50!white] (12,-2) circle (0.1);
  \filldraw[orange!50!white] (12,-2.5) circle (0.1);
		
		\filldraw[red!50!white] (13.5,0) circle (0.1);
		\filldraw[red!50!white] (13.5,-0.5) circle (0.1);
		\filldraw[red!50!white] (13.5,-1) circle (0.1);
		\filldraw[red!50!white] (13.5,-1.5) circle (0.1);
		\filldraw[orange!50!white] (13.5,-2) circle (0.1);
  \filldraw[orange!50!white] (13.5,-2.5) circle (0.1);
		
		\filldraw[red!50!white] (15,0) circle (0.1);
		\filldraw[red!50!white] (15,-0.5) circle (0.1);
		\filldraw[red!50!white] (15,-1) circle (0.1);
		\filldraw[red!50!white] (15,-1.5) circle (0.1);
		\filldraw[orange!50!white] (15,-2) circle (0.1);
		\filldraw[orange!50!white] (15,-2.5) circle (0.1);
		\filldraw[orange!50!white] (15,-3) circle (0.1);
		\filldraw[orange!50!white] (15,-3.5) circle (0.1);
		
		\draw[line width=1, dashed] (11,-4) -- (16,-4);
		
		\filldraw[cyan!50!white] (12,-4.5) circle (0.1);
		\filldraw[cyan!50!white] (12,-5) circle (0.1);
		\filldraw[purple!50!white] (12,-5.5) circle (0.1);
		\filldraw[purple!50!white] (12,-6) circle (0.1);
		\filldraw[purple!50!white] (12,-6.5) circle (0.1);
		\filldraw[purple!50!white] (12,-7) circle (0.1);
		
		\filldraw[black!50!white] (13.5,-4.5) circle (0.1);
		\filldraw[black!50!white] (13.5,-5) circle (0.1);
		\filldraw[black!50!white] (13.5,-5.5) circle (0.1);
		\filldraw[black!50!white] (13.5,-6) circle (0.1);
		\filldraw[black!50!white] (13.5,-6.5) circle (0.1);
		\filldraw[black!50!white] (13.5,-7) circle (0.1);
		
		\filldraw[cyan!50!white] (15,-4.5) circle (0.1);
		\filldraw[cyan!50!white] (15,-5) circle (0.1);
		\filldraw[cyan!50!white] (15,-5.5) circle (0.1);
		\filldraw[cyan!50!white] (15,-6) circle (0.1);
		\filldraw[cyan!50!white] (15,-6.5) circle (0.1);
		\filldraw[cyan!50!white] (15,-7) circle (0.1);
		\filldraw[purple!50!white] (15,-7.5) circle (0.1);
		\filldraw[purple!50!white] (15,-8) circle (0.1);
		\filldraw[purple!50!white] (15,-8.5) circle (0.1);
		\filldraw[purple!50!white] (15,-9) circle (0.1);
		\filldraw[purple!50!white] (15,-9.5) circle (0.1);

        				\node[inner sep=0] at (19.5,2) {Step 3: Perform matching};
		\node[inner sep=0] at (18,1) {OS Treated};
		\node[inner sep=0] at (19.5,1) {RCT};
		\node[inner sep=0] at (21,1) {OS Control};
		
		\filldraw[red!50!white] (18,0) circle (0.1);
		\filldraw[red!50!white] (18,-0.5) circle (0.1);
		\filldraw[red!50!white] (18,-1) circle (0.1);
		\filldraw[black!50!white] (18,-1.5) circle (0.1);

        \draw[line width=1, dotted] (18.1,0) -- (19.5,0);
        \draw[line width=1, dotted] (19.6,0) -- (21,0);
        \draw[line width=1, dotted] (18.1,-0.5) -- (19.5,-0.5);
        \draw[line width=1, dotted] (19.6,-0.5) -- (21,-0.5);
        \draw[line width=1, dotted] (18.1,-1) -- (19.5,-1);
        \draw[line width=1, dotted] (19.6,-1) -- (21,-1);
        \draw[line width=1, dotted] (18.1,-1.5) -- (19.5,-1.5);
        \draw[line width=1, dotted] (19.6,-1.5) -- (21,-1.5);
        
		\filldraw[orange!50!white] (18,-2) circle (0.1);

  \filldraw[orange!50!white] (18,-2.5) circle (0.1);

        \draw[line width=1, dotted] (18.1,-2) -- (19.5,-2);
        \draw[line width=1, dotted] (19.6,-2) -- (21,-2);
        \draw[line width=1, dotted] (18.1,-2.5) -- (19.5,-2.5);
        \draw[line width=1, dotted] (19.6,-2.5) -- (21,-2.5);
        
		\filldraw[red!50!white] (19.5,0) circle (0.1);
		\filldraw[red!50!white] (19.5,-0.5) circle (0.1);
		\filldraw[red!50!white] (19.5,-1) circle (0.1);
		\filldraw[red!50!white] (19.5,-1.5) circle (0.1);
		\filldraw[orange!50!white] (19.5,-2) circle (0.1);
  \filldraw[orange!50!white] (19.5,-2.5) circle (0.1);
		
		\filldraw[red!50!white] (21,0) circle (0.1);
		\filldraw[red!50!white] (21,-0.5) circle (0.1);
		\filldraw[red!50!white] (21,-1) circle (0.1);
		\filldraw[red!50!white] (21,-1.5) circle (0.1);
		\filldraw[orange!50!white] (21,-2) circle (0.1);
		\filldraw[orange!50!white] (21,-2.5) circle (0.1);
		\filldraw[orange!50!white] (21,-3) circle (0.1);
		\filldraw[orange!50!white] (21,-3.5) circle (0.1);
		
		\draw[line width=1, dashed] (17,-4) -- (22,-4);
		
		\filldraw[cyan!50!white] (18,-4.5) circle (0.1);
		\filldraw[cyan!50!white] (18,-5) circle (0.1);

        \draw[line width=1, dotted] (18.1,-5) -- (19.5,-5);
        \draw[line width=1, dotted] (19.6,-5) -- (21,-5);
        \draw[line width=1, dotted] (18.1,-4.5) -- (19.5,-4.5);
        \draw[line width=1, dotted] (19.6,-4.5) -- (21,-4.5);
        
		\filldraw[purple!50!white] (18,-5.5) circle (0.1);
		\filldraw[purple!50!white] (18,-6) circle (0.1);
		\filldraw[purple!50!white] (18,-6.5) circle (0.1);
		\filldraw[purple!50!white] (18,-7) circle (0.1);

        \draw[line width=1, dotted] (18.1,-5.5) -- (19.5,-5.5);
        \draw[line width=1, dotted] (19.6,-5.5) -- (21,-7.5);
        \draw[line width=1, dotted] (18.1,-6) -- (19.5,-6);
        \draw[line width=1, dotted] (19.6,-6) -- (21,-8);
        \draw[line width=1, dotted] (18.1,-6.5) -- (19.5,-6.5);
        \draw[line width=1, dotted] (19.6,-6.5) -- (21,-8.5);
        \draw[line width=1, dotted] (18.1,-7) -- (19.5,-7);
        \draw[line width=1, dotted] (19.6,-7) -- (21,-9);
        
		\filldraw[black!50!white] (19.5,-4.5) circle (0.1);
		\filldraw[black!50!white] (19.5,-5) circle (0.1);
		\filldraw[black!50!white] (19.5,-5.5) circle (0.1);
		\filldraw[black!50!white] (19.5,-6) circle (0.1);
		\filldraw[black!50!white] (19.5,-6.5) circle (0.1);
		\filldraw[black!50!white] (19.5,-7) circle (0.1);
		
		\filldraw[cyan!50!white] (21,-4.5) circle (0.1);
		\filldraw[cyan!50!white] (21,-5) circle (0.1);
		\filldraw[cyan!50!white] (21,-5.5) circle (0.1);
		\filldraw[cyan!50!white] (21,-6) circle (0.1);
		\filldraw[cyan!50!white] (21,-6.5) circle (0.1);
		\filldraw[cyan!50!white] (21,-7) circle (0.1);
		\filldraw[purple!50!white] (21,-7.5) circle (0.1);
		\filldraw[purple!50!white] (21,-8) circle (0.1);
		\filldraw[purple!50!white] (21,-8.5) circle (0.1);
		\filldraw[purple!50!white] (21,-9) circle (0.1);
		\filldraw[purple!50!white] (21,-9.5) circle (0.1);
	\end{tikzpicture}
 \caption{A toy example illustrating how our triplet matching structure is set up. Colors represent distinct covariate values, and the dashed horizontal line separates the common support and external support between the OS and RCT. Units in the same rectangle are duplicated copies of the same unit. Units in gray represent imaginary units. Units belonging to the same matched sets are connected with dotted lines. }
  \label{fig_matchingillustration}
\end{figure}

\section{Inference: A Novel Two-parameter Sensitivity Analysis Model}\label{sec_inference}
\subsection{Inference from the OS: sensitivity analysis for unmeasured confounders}

\subsubsection{A brief introduction to sensitivity analysis for OS}

We first focus on inferences with the OS alone, using the notation based on the line of work by Rosenbaum and others \citep{rosenbaum1987sensitivity,rosenbaum2002observational, hsu2013calibrating, visconti2018handling}. 
Let $\mathcal{F}=\{ ( Y^o_{ij}(1), Y^o_{ij}(0), X^o_{ij}, u^o_{ij}): j=1,\ldots, J_i; i=1, \ldots, I\}$ denote the collection of all potential outcomes and covariates for the matched data and $\mathcal{Z} = \{ Z_{ij}^o\in\{0,1\}: i=1, \ldots, I, j=1, \ldots, J_i \text{ so that } \sum_{j=1}^{J_i} Z_{ij}^o = 1 \text{ for all } i\}$ denote all possible 1-to-$J_i$ designs.
The matched data defines an OS block design where matching ensures necessary adjustment for the observed covariates so that $X_{ij}^o=X_{ij'}^o$ for all $i$ and $j\neq j'$. Write the conditional probability of the $j$th individual in the $i$th matched set as 
\vspace{-1em}
$$\eta_{ij}:=\Pr(Z^o_{ij}=1\mid \mathcal{F}, \mathcal{Z}), \text{  with } \sum_{j=1}^{J_i} \eta_{ij} = 1. \vspace{-0.5em}$$

If there are no unmeasured confounders, then $\eta_{ij}=1/J_i$ for all $i$, and it specifies a probability distribution over $\mathcal{Z}$. We can use this probability distribution to perform randomization-based inference for any sharp null hypothesis of no treatment effect where all the potential outcomes can be calculated under the null.  
A primary benefit of randomization-based inference is that we do not require model specifications for the outcome variable, which may be incorrect. 

However, we are interested in inference regarding $\beta^\star$, and a point null hypothesis regarding $\beta^\star$ is not a sharp null hypothesis -- several different sets of values of the potential outcomes can have the same $\beta^\star$. Below we propose a randomization-based inference for the Neyman null hypothesis 
\vspace{-1em}
$$H_0: \beta^\star=\beta^\star_0, \quad\textrm{ for some fixed value }\beta_0^\star. \vspace{-0.5em}$$

When there are unmeasured confounders, i.e., $u^o_{ij}\neq u^o_{ij'}$ for some $j\neq j'$, the probabilities $\eta_{ij}\neq 1/J_i$. Further, because of the unmeasured confounders, the probabilities are unknown. A sensitivity analysis for unmeasured confounders relaxes the assumption of no unmeasured confounders to different degrees and provides inference regarding a hypothesis or an estimand that is valid under this relaxation. 
A significant amount of work exists on design-based sensitivity analysis methods for difference test statistics and study designs for a sharp null hypothesis; see, e.g., \cite{rosenbaum1987sensitivity,rosenbaum2010design,rosenbaum2015bahadur} and references therein. 
For Neyman's null hypothesis, relatively less is known regarding sensitivity analysis methods for different designs 
\citep{fogarty2017randomization, fogarty2020studentized, zhao2019sensitivity}. In line with these works, we propose a sensitivity analysis method for the Neyman null and, hence, a confidence interval for our ATOT in our blocked design. 
The inference method we propose below is a new contribution and may be of separate interest to researchers who need to conduct a sensitivity analysis regarding the ATOT in a general blocked observational study design.

We will consider the Neyman null hypothesis $H_0: \beta^\star=\beta^\star_0$. 
We follow Rosenbaum's sensitivity analysis model that says that for a sensitivity parameter $\Gamma\geq 1$ 
\vspace{-1em}
\begin{equation}\label{eq_gamma_model}
\frac{1}{\Gamma} \leq \frac{\eta_{ij}}{\eta_{ij^\prime}} \leq \Gamma, \text{  with } \sum_{j=1}^{J_i} \eta_{ij} = 1,\vspace{-0.5em}
\end{equation}
for all set $i$ and all $j$th and $j^\prime$th unit in that set. 

To understand the role of $\Gamma$, note that, when $\Gamma=1$, the odd is 1, i.e., $\eta_{ij}=\eta_{ij'}=1/J_i$, and there is no unmeasured confounding. When $\Gamma>1$, the ratio may be different from 1, indicating an effect of unmeasured confounding. For example, when $\Gamma=1.1$, the ratio is in $[1/1.1, 1.1]=[.91, 1.1]$. In other words, even after adjusting for the observed covariates by matching, because of an imbalance of unmeasured confounders, an individual may be $10\%$ more likely or $9\%$ less likely to receive the treatment compared to another unit in its matched set. The larger $\Gamma$ is, the more we allow the effect of unmeasured confounding. An observed difference of the outcome between the treated and control group in the OS may be statistically significant if $\Gamma=1$, i.e., assuming no unmeasured confounders, but may become insignificant for $\Gamma=1.1$ if we find that the observed effect can be created under the null using a treatment assignment that prefers to assign treatment to units with larger potential outcomes with just a 10\% higher probability. Such a finding 
is concerning since the observed significant effect disappears under a small unmeasured confounding, 
while a small amount of unmeasured confounding is hard to dismiss in most OSs. In contrast, the effect of heavy smoking on lung cancer only became insignificant when $\Gamma>6.5$ \citep[][Table 4.1]{rosenbaum2002observational}.
Rosenbaum's sensitivity model \eqref{eq_gamma_model} may be equivalently written in a semiparametric model for the probability $\Pr(Z_{ij}^o=1\mid X_{ij}^o, u_{ij}^o)$, where $\Gamma$ appears as a coefficient of $u_{ij}^o$; see 
Supplement~S1.1 for details. In a sensitivity analysis, we shall ask if a hypothesized value $\beta^\star_0$ for the ATOT is plausible for a given level of unmeasured confounding. Then the set of plausible $\beta^\star_0$ values at a given significance level will create a confidence set for ATOT, allowing for $\Gamma$ level of unmeasured confounding. The larger $\Gamma$ is, the wider the set of values that becomes plausible for ATOT with a wider range of allowed effects of unmeasured confounders, and the confidence set becomes bigger.

\subsubsection{Sensitivity analysis for ATOT in a general block design}

Fix a value of $\Gamma$. Let $\boldsymbol{\eta}_i = (\eta_{i1},\ldots,\eta_{iJ_i})$. Then, for testing $H_0: \beta^\star = \beta_0^\star$, there is a specific choice $\widetilde{\boldsymbol{\eta}}_i^{(\beta_0^\star)}$ satisfying \eqref{eq_gamma_model} that is important to us. 
These are called {\it separable approximations} of the {\it most extreme} probabilities in the sense that they make the null hypothesis most difficult to reject under a $\Gamma$ level of unmeasured confounding \citep{gastwirth2000asymptotic}. 
These separable approximations are called {separable} because the calculation of $\widetilde{\boldsymbol{\eta}}_i^{(\beta_0^\star)}$ only requires information on matched set $i$; thus, separable across different strata. Further, they are approximations because the desired {extreme} case happens only in large samples as the number of blocks goes to infinity. However, 
the approximation error is reasonably small in finite samples \citep{rosenbaum2018}. A third fact about these $\widetilde{\boldsymbol{\eta}}_i^{(\beta_0^\star)}$ that is crucial for the validity of our method is that this approximate choice of extreme probabilities is in fact exact for $i\leq I_0$ for some $I_0$. We describe the computation of $\widetilde{\boldsymbol{\eta}}_i^{(\beta_0^\star)}$ in 
Supplement~S1.2.

In the following, we describe our testing procedure for testing the Neyman null $H_0$. 

For stratum $i$, let 
\vspace{-1em}
$$\hat\tau_i =\sum_j Z^o_{ij}(Y^o_{ij}-Z^o_{ij}\beta_0^\star) - (J_i-1)^{-1} \sum_j (1-Z^o_{ij})(Y^o_{ij}-Z^o_{ij}\beta_0^\star),\vspace{-1em}$$
be the difference of the averages of the outcomes offset by $Z^o_{ij}\beta_0^\star$ between the treated and control units in set $i$. In case of a constant additive treatment effect, $(Y^o_{ij}-Z^o_{ij}\beta_0^\star)$ are called the adjusted outcomes. However, we do not assume a constant additive treatment effect.

Subtract from this difference term an estimate of its extreme value under the specified bias and define 
\vspace{-1em}
$$\widetilde\tau_i^{(\beta_0^\star)} = \hat\tau_i - \Big\{\sum_j \widetilde\eta_{ij}^{(\beta_0^\star)}(Y^o_{ij}-Z^o_{ij}\beta_0^\star) - (J_i-1)^{-1} \sum_j (1-\widetilde\eta_{ij}^{(\beta_0^\star)})(Y^o_{ij}-Z^o_{ij}\beta_0^\star) \Big\}.\vspace{-1em}$$
We use $\sum_i \widetilde\tau_i^{(\beta_0^\star)}/I$, the average of $\hat\tau_i$ centered with respect to its extreme average value across the strata, as our test statistic for testing $H_0$ versus $H_1: \beta^\star > \beta_0^\star$.  
The distribution of this statistic is not known exactly since the distribution of the treatment assignment that depends upon the unmeasured confounders $u_{ij}^o$ is also unknown. Rather, we show, as part of the proof of Theorem \ref{thm:1} below, that, when the number of strata is large, the distribution of  the test statistic is approximately stochastically dominated by a centered normal distribution with variance 
equal to the sample variance of the $I$ many $\widetilde\tau_i^{(\beta_0^\star)}$ values over the sample size $I$. 
Thus, an asymptotically valid upper-sided $(1-\alpha)100\%$ confidence interval can be constructed by inverting the test that rejects $H_0: \beta^\star = \beta_0^\star$ in favor of 
$H_1: \beta^\star > \beta_0^\star$ when
\vspace{-1em}
\begin{equation}\label{eq_ciGamma}
 \frac{1}{I}\sum_{i=1}^I \widetilde\tau_i^{(\beta_0^\star)} > z_{1-\alpha} se(\sum_{i=1}^I\widetilde\tau_i^{(\beta_0^\star)}/I), \vspace{-0.5em}
\end{equation}
where $se(\sum_{i=1}^I\widetilde\tau_i^{(\beta_0^\star)}/I)=\sqrt{\frac{1}{I(I-1)}\sum_i \{\widetilde\tau_i^{(\beta_0^\star)}\}^2 - \frac{1}{I^2(I-1)}\{\sum_i \widetilde\tau_i^{(\beta_0^\star)}\}^2}$. 
The test mimics a standardized test based on {$(1-\alpha)$th standard }normal quantiles, $z_{1-\alpha}$. { 
For example, if $\alpha=0.05$, we reject the hypothesized $\beta^\star_0$ as a plausible value for  ATOT if the ratio of our test statistic to its standard error is greater than $1.96$.}
For $\Gamma>1$, the test is generally asymptotically conservative for a treatment assignment distribution satisfying \eqref{eq_gamma_model}.

\begin{theorem}\label{thm:1}
Suppose \eqref{eq_gamma_model} (or equivalently, the semiparametric model 
\textnormal{(S1.1)}) holds for a given $\Gamma\geq 1$. Under 
Assumption S1 stated in the supplementary materials, for $\alpha<.5$, \eqref{eq_ciGamma} gives an asymptotically valid $(1-\alpha)100\%$ upper-sided confidence interval for ATOT $\beta^\star$.
\end{theorem}

Similarly, we construct a lower-sided confidence interval by 
inverting the hypothesis testing for $H_0: \beta^\star = \beta^\star_0$ vs $H_1: \beta^\star < \beta^\star_0$ that rejects $H_0$ in favor of $H_1$ if 
\vspace{-1em}
\begin{equation}\label{eq_ciGammaL}
\frac{1}{I}\sum_i \dwidetilde\tau_i^{(\beta_0^\star)}  < z_{\alpha} se(\sum_{i=1}^I\dwidetilde\tau_i^{(\beta_0^\star)}/I), \vspace{-1em}
\end{equation}
where $\dwidetilde\tau_i^{(\beta_0^\star)} = \hat\tau_i + \Big\{\sum_j \dwidetilde\eta_{ij}^{(\beta_0^\star)}(Y^o_{ij}-Z^o_{ij}\beta_0^\star) - (J_i-1)^{-1} \sum_j (1-\dwidetilde\eta_{ij}^{(\beta_0^\star)})(Y^o_{ij}-Z^o_{ij}\beta_0^\star) \Big\}$ and $se(\sum_{i=1}^I\dwidetilde\tau_i^{(\beta_0^\star)}/I)=\sqrt{\frac{1}{I(I-1)}\sum_i \{\dwidetilde\tau_i^{(\beta_0^\star)}\}^2 - \frac{1}{I^2(I-1)}\{\sum_i \dwidetilde\tau_i^{(\beta_0^\star)}\}^2}$. {   The primary difference between equations \eqref{eq_ciGamma} and \eqref{eq_ciGammaL} is that in \eqref{eq_ciGammaL} we ``center" $\hat{\tau}_i$ by adding to it an estimate of its smallest average value under the sensitivity analysis model. Thus we use a different set of extreme probabilities, $\dwidetilde\eta_{ij}^{(\beta_0^\star)}$, defined in 
Supplement~S1.2, which are analogous to $\widetilde\eta_{ij}^{(\beta_0^\star)}$ but for testing against the less than
alternative $H_1:\beta^\star<\beta^\star_0$.}

We use numerical methods to find the confidence interval by inverting the test; details are discussed in 
Supplement~S1.3. 
Putting them together $[\beta_L^o,\beta_U^o]$ is an approximate $(1-2\alpha)100\%$ confidence interval under specified bias $\Gamma$.

\subsection{Inference from the RCT}  \label{subsec:rctinf}

\subsubsection{Large sample inference}
In this section, we discuss the inference from the RCT part of the design. 
Let 
\vspace{-1em}
$$\tilde\theta_{m} = \Pr(Z^r_m=1 \mid X_1^r, \ldots, X_{n_r}^r) \vspace{-1em}$$
denote the known treatment assignment probability for RCT unit $m$. We start with design-based inference for the RCT. Note, though, that the standard design-based inference for the RCT is not necessarily consistent with our target estimand $\beta^\star$ when there is effect heterogeneity and the support of the RCT is smaller than that of the OS. 

Recall that the matched design matches OS treated units to a certain number of RCT units on the {common support}. The RCT unit $m$ is copied $C_m$ times in our design, for $m=1,\ldots,n_r$. Then our estimator for the ATOT on the common support is 
\vspace{-1em}
\begin{equation}\label{eq_atot_rct}
\widehat{\beta^r_\X} := \frac{1}{\sum_m C_m}\sum_m C_m\left\{ \frac{Z^r_mY^r_m}{\tilde\theta_{m}} - \frac{(1-Z^r_m)Y^r_m}{1-\tilde\theta_{m}} \right\}. \vspace{-1em}
\end{equation}
The estimator may also be written $\frac{1}{\sum_m C_m}\sum_{m: Z_m^r=1} C_m \frac{Y^r_m}{\tilde\theta_{m}} - \frac{1}{\sum_m C_m}\sum_{m: Z_m^r=0} C_m \frac{Y^r_m}{1-\tilde\theta_{m}}$. Thus, it is the difference of the weighted averages, with weights being the number of copies of the units, of the $Y_m^r/\tilde{\theta}_m$ for the treated units and $Y_m^r/(1-\tilde{\theta}_m)$ for the control units.

The randomization of the RCT will ensure that this estimator is consistent for the 
ATOT on the  {common support} $\mathcal{X}$, i.e., for $E\{Y_k^o(1)-Y_k^o(0) \mid X_k^o \in \X, Z^o_k=1\}$. 
Theorem \ref{thm_rct_consistency} below establishes the consistency of the estimator for a general randomization design. More specifically, for completely randomized and stratified designs, the estimator is approximately normally distributed in large samples. This is proved through Theorem \ref{thm_rct_clt} below.

\begin{theorem}\label{thm_rct_consistency}
Under 
Assumption S2 stated in the supplementary materials,
as $n_r\rightarrow\infty$,  under appropriate moment conditions on the distribution of the potential outcomes, $\widehat{\beta^r_\X}$ converges in probability to $E\{Y_k^o(1)-Y_k^o(0) \mid X_k^o \in \X, Z^o_k=1\}$.
\end{theorem}

\begin{theorem}\label{thm_rct_clt}
Under 
Assumption S3 (or Assumption S4) stated in the supplementary materials, for a completely randomized design (or a stratified design), as $n_r\rightarrow\infty$,  $\sqrt{n_r}[\widehat{\beta^r_\X}-E\{Y_k^o(1)-Y_k^o(0) \mid X_k^o\in \X, Z^o_k=1\}]$ converges to a centered normal random variable.
\end{theorem}

The required assumptions are mostly regularity conditions on the potential outcomes' distributions and the weights $C_m$. As the estimators are connected to an estimand from the OS and the RCT does not see the units selected into the OS, we also assume the following.
\begin{assumption}
    The OS units' treatment effects are independent of unmeasured confounders in the  {common support} $\mathcal{X}$, i.e., $(Y_l(1) - Y_l(0)) \perp Z_l^o\mid X_l, S_l=0$.
\end{assumption} 
This assumption is needed for the estimates from the OS on its treated individuals to carry transferable information to the RCT. It does not require that there be no unmeasured confounding. The assumption holds broadly if $X_l$ captures all treatment effect heterogeneity.  Its plausibility relies on domain knowledge. The assumption that all effect modifiers are observed is intrinsic to much of the related literature \citep{yang2023elastic}.

\subsubsection{Inference from small RCTs}
The above theorems are large sample results, and Theorem \ref{thm_rct_clt} may be used to construct large sample confidence intervals by estimating the variance of the asymptotic normal distribution. For finite samples, however, we need to rely on randomization-based inference for the RCT. The tradeoff is that the randomization inference assumes a constant treatment effect. For randomization inference with less restrictive assumptions on the treatment effect, see \cite{su2024treatment} and \cite{caughey2023randomisation}. 

To construct confidence intervals, let $\tilde{Z}_{1,s}^r, \ldots, \tilde{Z}_{n_r,s}^r$, for $s=1,\ldots,S$ be $S$ Monte Carlo samples from the randomization distribution $\Pr(Z_{1}^r, \ldots, Z_{n_r}^r \mid  X_1^r, \ldots, X_{n_r}^r)$. Consider the constant additive treatment effect $Y^r_m(1) = Y^r_m(0)+\beta_0$ with the hypothesized ATOT in the  {common support} as $\beta_0$. Let $\widetilde{Y^r_m} = Y^r_m-Z^r_m \beta_0$ be the adjusted outcomes. Calculate the $S$ values
$$t(s, \beta_0):= \frac{1}{\sum_m C_m}\sum_m C_m\left\{ \frac{\tilde{Z}^r_mY^r_m}{\tilde\theta_{m}} - \frac{(1-\tilde{Z}^r_m)Y^r_m}{1-\tilde\theta_{m}} \right\}.$$
Reject the hypothesized treatment effect $\beta_0$ as plausible with type-I probability $\alpha$ if $$\frac{1}{\sum_m C_m}\sum_m C_m\left\{ \frac{{Z}^r_m\tilde{Y}^r_m}{\tilde\theta_{m}} - \frac{(1-{Z}^r_m)\tilde{Y}^r_m}{1-\tilde\theta_{m}} \right\},$$
is outside of the $\alpha/2$-th quantile and $(1-\alpha/2)$-th quantile of the $S$-many $t(s, \beta_0)$ values. The $(1-\alpha)\times 100\%$ level confidence interval is constructed by pooling all the plausible $\beta_0$ values. A point estimate is found by the Hodges-Lehman estimator \citep{lehmann2006nonparametrics}.

\subsubsection{Sensitivity analysis for generalizability bias}

The above method provides inference for the average treatment effect on the treated units in the RCT. However, in the  {external support}, the treatment effect can be different. Hence, the RCT may give an inconsistent estimate of $\beta^\star$. We consider a sensitivity analysis model for the potential generalizability bias outside the  {common support}.  Consider sensitivity parameter $\Delta\geq 0$ such that 
\begin{equation}\label{eq_rct_sens_model}
\Big|E\{Y^o_l(1) - Y^o_l(0)\mid Z^o_l=1, X^o_l\in\X \}- E\{Y^o_l(1) - Y^o_l(0)\mid Z^o_l=1\} \Big| \leq\Delta.
\end{equation}
Thus, $\Delta$ bounds the difference in the target estimand ATOT and the ATOT on the common support, which $\widehat{\beta^r_\X}$ consistently estimates.  Notice that $\Delta=0$ indicates no bias due to 
 external support, while $\Delta>0$ measures 
generalizability bias. 
Note that \eqref{eq_rct_sens_model} is equivalent to bounding the effect heterogeneity between the  {common support and external support} as
$$\Big|E\{Y^o_l(1) - Y^o_l(0)\mid Z^o_l=1, X^o_l\in\X \}- E\{Y^o_l(1) - Y^o_l(0)\mid Z^o_l=1, X^o_l\in\X^c\} \Big| \leq \frac{\Delta}{\Pr(X^o_l\in\X^c\mid Z_l^o=1)},$$
when $\Pr(X^o_l\in\X^c\mid Z_l^o=1)>0$. We denote this rescaled bound by $\widetilde\Delta$. By the Bayes formula, the denominator is $\Pr(X^o_l\in\X^c\mid Z_l^o=1)=
\Pr(X^o_l\in\X^c)\Pr(Z_l^o=1\mid X^o_l\in\X^c)/\Pr(Z_l^o=1)$. So that, $\Pr(X^o_l\in\X^c\mid Z_l^o=1)=0$ only when { the external support is empty}, i.e., $\Pr(X^o_l\in\X^c)=0$. Next, if there is significant overlap between the common support and the support of the OS covariates, the denominator is small. Consequently, a small $\Delta$ value will capture the same effect heterogeneity when there is significant overlap between those supports, as a large $\Delta$ value when there is limited overlap between those supports. For example, when $\Pr(X^o_l\in\X)=.75$, $\Delta=0.1$ gives a ratio $\Delta/\Pr(X^o_l\in\X)$ is $0.4$ while, when $\Pr(X^o_l\in\X)=.25$, $\Delta=0.3$ gives the ratio is again $0.4$.
In addition, the sensitivity parameter $\Delta$ also depends on the scale of the outcome, e.g., if the outcomes are divided by $10$, the $\Delta$ value should also be divided by $10$. This is unlike the sensitivity parameter $\Gamma$, which is scale-free. Thus, it might be more appropriate to determine the scale of the sensitivity analysis at the scale of the standard deviation of the outcome. One can use the parametrization $\Delta^\prime=\Delta/\sqrt{S_t^2 + S_c^2}$ with $S_t^2$ and $S_c^2$ being the sample variances of the OS treated and control units in our matched sample, respectively.

For a given $\Delta>0$, instead of a single point estimate, we can provide two extreme point estimates $\widehat{\beta^r_\X}-\Delta$ and $\widehat{\beta^r_\X}+\Delta$. Theorem \ref{thm_rct_consistency} ensures that under \eqref{eq_rct_sens_model}, the ATOT $\beta^\star$ will be inside the two asymptotic limits of the two extreme point estimates. 
The corresponding confidence interval will be wider than the design-based confidence interval by subtracting $\Delta$ from the lower limit and adding $\Delta$ to the upper limit. In practice, one can choose an increasing sequence of values of $\Delta$ and report the corresponding confidence intervals under those bounds on the generalizability bias. It may be informative, for example, to report the level of generalizability bias at which the confidence interval includes zero, indicating a statistically insignificant ATOT.

\subsection{Combining inferences from the OS and RCT: A two-parameter sensitivity analysis framework}

The OS and the RCT have complementary strengths. The OS is representative of a bigger population and has a larger sample size, while the RCT is the gold standard because of the random assignment of the treatment. At the same time, the OS is susceptible to unmeasured confounding. Our proposed sensitivity analysis to unmeasured confounding allows us to judge the effect of unmeasured confounding on our inference. The RCT's strength can help improve the sensitivity analysis of an OS in the absence of generalizability bias. On the other hand, the RCT is susceptible to generalizability bias because the treatment effect may be different in regions outside of the covariate support of the RCT. Our proposed sensitivity analysis for generalizability bias allows us to infer the effect given a bound on the generalizability bias. Because of the larger sample size, the OS can help improve the sensitivity analysis of an RCT in the absence of unmeasured confounding. Below, we consider situations where we allow both bias due to unmeasured confounding and generalizability bias in simultaneous sensitivity analysis. 

To describe how we combine the two studies, fix $\Gamma$ and $\Delta$ values in our two sensitivity analysis models \eqref{eq_gamma_model} and \eqref{eq_rct_sens_model} respectively, throughout this section. The combining method is based on sensitivity analysis $p$-values, while the resultant goal is still to create a confidence interval for $\beta^\star$ which we get by inverting the combined $p$-values. For other methods using multiple sensitivity parameters to quantify separate biases and combine them in other contexts, see \cite{Karmakar1929} and \cite{Karmakar2148}.

The sensitivity analysis $p$-value for testing $H_0: \beta^\star = \beta_0^\star$ vs $H_1: \beta^\star > \beta_0^\star$ from the OS calculates 
\begin{equation}\label{eq_os_pval}
    p^o_{\beta_0^\star} = \sup_{\tilde\beta^\star \geq \beta_0^\star}  1 - \Phi^{-1}\left( \frac{ \frac{1}{I}\sum_i \widetilde\tau_i^{(\beta_0^\star)}}{se(\sum_{i=1}^I\widetilde\tau_i^{(\beta_0^\star)}/I)} \right). \vspace{-.5em}
\end{equation}
The supremum is used for technical reasons to ensure that the $p$-values are monotone in $\beta_0^\star$. It is only necessary for the proof of Theorem \ref{thm_combinedefficiency} and not required for the validity of the combined confidence interval as established in Theorem \ref{thm_combinedinference}.

Let $p^r_{\beta_0^\star}$ denote the sensitivity analysis $p$-value for testing  $H_0: \beta^\star = \beta_0^\star$ vs $H_1: \beta^\star > \beta_0^\star$ from the RCT. Calculate this $p$-value by first defining the test statistic
\vspace{-0.5em}
$$T_{\beta^\star_0}(Z^r_1, \ldots, Z^r_{n_r}) = \frac{1}{\sum_m C_m}\sum_m C_m\left\{ \frac{Z^r_m(Y^r_m - \beta_0^\star + \Delta)}{\tilde\theta_{m}} - \frac{(1-Z^r_m)Y^r_m}{1-\tilde\theta_{m}} \right\}. \vspace{-0.5em}$$
The statistic can be understood as a difference of weighted averages of some adjusted outcomes between the treated and control units, since $T_{\beta^\star_0}(Z^r_1, \ldots, Z^r_{n_r})$ is equal to $\frac{1}{\sum_m C_m}\sum_{m: Z^r_m=1}  C_m \frac{Y^r_m - \beta_0^\star + \Delta}{\tilde\theta_{m}} - \frac{1}{\sum_m C_m}\sum_{m: Z^r_m=0} C_m \frac{Y^r_m}{1-\tilde\theta_{m}}$. The adjusted outcome is $Y^r_m - Z_m^r(\beta_0^\star - \Delta)$, which is $(Y^r_m - \beta_0^\star + \Delta)$ for a treated unit and $Y^r_m$ for a control unit. 
The inference process uses randomization inference and requires a constant additive treatment effect for the RCT units. Start by drawing $S$ Monte Carlo samples $\tilde{Z}_{1,s}^r, \ldots, \tilde{Z}_{n_r,s}^r$, for $s=1,\ldots,S$ from the randomization distribution $\Pr(Z_{1}^r, \ldots, Z_{n_r}^r \mid  X_1^r, \ldots, X_{n_r}^r)$. For each draw, calculate the test statistic under the resampled treatment assignment $T_{\tilde\beta^\star_0}(\tilde{Z}_{1,s}^r, \ldots, \tilde{Z}_{n_r,s}^r)$. Thereby, calculate the sensitivity analysis $p$-value by calculating the average number of these statistics that are greater than the observed statistic: 
\begin{equation}\label{eq_rct_pval}
    p^r_{\beta_0^\star} :=  \sup_{\tilde\beta^\star \geq \beta_0^\star} \frac1{S+1}\Big[1+\sum_s \mathbb{I}\big\{ T_{\tilde\beta^\star_0}(\tilde{Z}_{1,s}^r, \ldots, \tilde{Z}_{n_r,s}^r) > T_{\tilde\beta^\star_0}(Z^r_1, \ldots, Z^r_{n_r})\big\}\Big],
\end{equation}
where $\mathbb{I}\{\cdot\}$ is the indicator function. The supremum is used for technical reasons to ensure that the $p$-values are monotone in $\beta_0^\star$. We add one to the numerator and denominator to avoid a zero $p$-value, which may occur if the $p$-value is too small. Alternatively, for large RCTs, we can calculate the $p$-value using the large sample result in Theorem \ref{thm_rct_clt}.

We combine the two sensitivity analyses using the test statistic that is the product of the two sensitivity analysis $p$-values. 
Specifically, we calculate the combined level $(1-\alpha)$ confidence interval as $(-\infty, \widehat\beta_{U,combined,\alpha}^\star]$, where
\begin{equation}\label{eq_combined_ci}
\widehat\beta_{U,combined,\alpha}^\star = \sup\{ \beta_0^\star : p^o_{\beta_0^\star} \times p^r_{\beta_0^\star} \geq \kappa_\alpha \}. \vspace{-1em}
\end{equation}
Here, $\kappa_\alpha = \exp(-\chi^2_{4;1-\alpha}/2)$; $\chi^2_{4;1-\alpha}$ is the $(1-\alpha)$th quantile of the $\chi^2$ distribution with $4$ degrees of freedom. The subscript $U$ emphasizes the upper confidence limit. Details of the critical level calculation are discussed in 
Supplement~S2.
This corresponds to a confidence interval created from Fisher's $p$-value that combines the two $p$-values. However, the sensitivity analysis $p$-values are not uniformly distributed.  The following Theorem establishes the validity of the above confidence interval, which is conservative when $\Gamma>1$ or $\Delta>0$.

\begin{theorem}\label{thm_combinedinference}
Under the sensitivity analysis models, if  $p^o_{\beta_0^\star}$ and $p^r_{\beta_0^\star}$ are valid sensitivity analysis $p$-values for the RCT and OS respectively, then 
the resulting interval $(-\infty, \widehat\beta_{U,combined,\alpha}^\star]$ is an asymptotically valid level $(1-\alpha)100\%$ confidence interval for $\beta^\star$.
\end{theorem}

Next, we show that the combined confidence interval is better -- in a sense that will be made concrete below -- than the individual confidence intervals for the same confidence level. Let $(-\infty, \widehat\beta^\star_{U, OS, \alpha}]$ and $(-\infty, \widehat\beta^\star_{U, RCT, \alpha}]$ denote $(1-\alpha)100\%$ confidence intervals 
for using single data sources. In particular,
$\widehat\beta^\star_{U, OS, \alpha}=\sup\Big\{\beta_0^\star: p^o_{\beta_0^\star}\geq \alpha\Big\}$ and $\widehat\beta^\star_{U, RCT, \alpha}=\sup\Big\{\beta_0^\star: p^r_{\beta_0^\star}\geq \alpha\Big\}.$ 
The theoretical result considers an asymptotic situation where the OS and RCT both increase in size, perhaps at different rates.

Let $s$ be a common index for a paired sequence of studies: $OS_s$ and $RCT_s$. We have in our mind that as $s\rightarrow \infty$, the sizes of $OS_s$ and $RCT_s$ both go to infinity. Let $p^{os_s}_{\beta_0^\star}$ and $p^{rct_s}_{\beta_0^\star}$ be the $p$-values corresponding to the two studies. Let $\alpha_s\rightarrow 0$ be a sequence that gives an increasing sequence of $(1-\alpha_s)\times 100$\% confidence levels. We make the following set of assumptions, which are, in general, mild. 

\begin{assumption}\label{assm_combinedefficiency}
    \normalfont
    
    \begin{enumerate}
        \item[4.1] The two sequences of $p$-values $p^{os_s}_{\beta_0^\star}$ and $p^{rct_s}_{\beta_0^\star}$ are monotone in $\beta_0^\star$.
        \item[4.2] $p^{os_s}_{\beta_0^\star}$ and $p^{rct_s}_{\beta_0^\star}$ are continuous in $\beta_0^\star$.
        \item[4.3] $\lim_{s\rightarrow \infty} [\widehat\beta^\star_{U, OS_s, \alpha_s}-\widehat\beta^\star_{U, RCT_s, \alpha_s} ] = 0$.
        Thus, $p^{os_s}_{\beta_0^\star}\rightarrow 0$ and $p^{rct_s}_{\beta_0^\star}\rightarrow 0$ for any $\beta_0^\star>\lim_{s\rightarrow \infty}\widehat\beta^\star_{U, OS_s, \alpha_s}$.
    \end{enumerate}
\end{assumption}

Assumption 4.1 is enforced by the supremums in defining the $p^{os_s}_{\beta_0^\star}$ and $p^{rct_s}_{\beta_0^\star}$ in \eqref{eq_os_pval} and \eqref{eq_rct_pval} respectively. Assumption 4.2 is made for convenience and may be removed at the cost of more cumbersome proof of Theorem \ref{thm_combinedefficiency}. 
Assumption 4.3 says that the sensitivity parameters $\Gamma$ and $\Delta$ in the two sensitivity models are comparable in the sense that the corresponding confidence intervals converge to the same interval. This is the case where one wishes to judge if there is a gain by pooling the strengths of the two inferences. Alternatively, if the situation is such that the upper limit for the $OS_s$ is smaller than that of the upper limit for the $RCT_s$ in large enough samples, then the combined interval will converge to the confidence interval for the $OS_s$. Similarly, if the upper limit for the $RCT_s$ is smaller than that of the upper limit for the $OS_s$ in large enough samples, then the combined interval will converge to the confidence interval for the $RCT_s$. 

\begin{theorem}
    \label{thm_combinedefficiency}
    Under Assumption \ref{assm_combinedefficiency}, when the sensitivity analysis models \eqref{eq_gamma_model} and \eqref{eq_rct_sens_model} hold with sensitivity parameters $\Gamma$ and $\Delta$, respectively, we have $\widehat\beta_{U,combined_s,\alpha_s}^\star < \min\Big\{ \widehat\beta^\star_{U, OS_s, \alpha_s},\widehat\beta^\star_{U, RCT_s, \alpha_s} \Big\}$ for large enough $s$.
\end{theorem}
The theorem says that the combined confidence interval will be strictly contained in the individual confidence intervals constructed from $OS_s$ and $RCT_s$, i.e., the combined inference is asymptotically more efficient than either study considered separately. This is an asymptotic result with the sample sizes increasing to infinity while the confidence levels also increase to 100\%. The setting is along the line of work on design sensitivity and the Bahadur efficiency of comparing tests where we increase the sample size to infinity and decrease the type I error rates to zero 
\citep{karmakar2019integrating,rosenbaum2015bahadur}. However, we see shorter confidence intervals by using the combined method compared to the intervals based on the individual analyses for finite samples as well.

The upper-sided confidence interval by combining the OS and RCT is calculated similarly. First, we compute the $p$-values $\tilde{p}^o_{\beta^\star}$ and $\tilde{p}^r_{\beta^\star}$ for the OS and RCT separately for testing $H_0: \beta^\star = \beta_0^\star$ vs $H_1: \beta^\star < \beta_0^\star$.\footnote{Since we have worked out the calculations of the $p$-values for the greater than alternatives in \eqref{eq_os_pval} and \eqref{eq_rct_pval}, an easy way to calculate these $p$-values for the less than alternative is by first transforming the outcomes $Y^o_{ij}$ and $Y^r_m$ to $-Y^o_{ij}$ and $-Y^r_m$. Then, calculating $p$-values for testing $H_0: \beta^\star = -\beta_0^\star$ vs $H_1: \beta^\star > -\beta_0^\star$.}
Subsequently, $[\widehat\beta^\star_{L, combined, \alpha}, \infty)$ gives the combined upper-sided confidence interval where $\hat{\beta^\star}_{L, combined, \alpha} = \inf\{ \beta_0^\star : \tilde{p}^o_{\beta_0^\star} \times \tilde{p}^r_{\beta_0^\star} \geq \kappa_\alpha \}$. Finally, the $(1-\alpha)\times 100$\% two-sided confidence interval is $[\widehat\beta^\star_{L, combined, \alpha/2}, \widehat\beta^\star_{U, combined, \alpha/2}]$. For two-sided confidence intervals, the previous theorem says that the length of the combined interval will be smaller than the lengths of the confidence intervals for the OS and RCT when sample sizes are large, and the confidence level is close to 100\%. In Section \ref{sec_simulation}, we compare the average lengths of these different confidence intervals in finite samples with the typical 95\% confidence level.

\section{Simulation Study}\label{sec_simulation}

We consider the following data-generating process in the population that allows us to generate data with unmeasured confounding bias in the OS and generalizability bias in the RCT with true values of the bias levels $\Gamma^\star$ and $\Delta^\star$ respectively. There are five observed covariates $(X_1,\dots, X_5)$ independently distributed and each following the standard normal distribution, and one unobserved covariate $U$ independent from the observed covariates and following the standard normal distribution. The potential outcome under control is $Y(0)=10+4X_1-2X_2+3X_5+U+\epsilon$, where $\epsilon\sim N(0,1)$ and under treatment is $Y(1)=Y(0)+\widetilde\Delta^\star I(\textrm{Unit belongs to the  {common support} $\X$})$, where $\widetilde\Delta^\star=\Delta^\star/\Pr(X^o\in\X^c\mid Z^o=1)$ as discussed in \S\ref{subsec:rctinf}. In the common support $\X$, the probability of selecting into the RCT is $expit(-1.5+0.1X_1+0.1X_2-0.3X_4)$. Once selected into either the RCT or OS, the probability of being assigned to treatment is $1/2$ in the RCT and $expit(-2-0.3X_1+0.1X_3-0.2X_5+\log(\Gamma^\star)U)$ in the OS. Our simulation study creates several data-generating models by varying the total sample size $N$, the bias-controlling parameters $\Gamma^\star$ and $\Delta^\star$, and the common support in various contexts. 

\subsection{Validity of theoretical results}\label{subsec:sim_theory}
In the first set of simulated experiments, we investigate the impact of four factors on the validity of our theoretical results in \S~\ref{sec_inference}. The first factor is the total sample size $N$ of the collected data (including both RCT and OS). We consider a sample size similar to our primate data in the real data analysis, $N=500$, and a larger sample size $N=1000$. The second factor is the formation of the common support. We consider three inclusion criteria for the  {common support}: extensive, moderate, or limited, such that the inclusion criterion is the entire domain ($I(X_1\geq-\infty)$), majority domain ($I(X_1\geq-1)$), or a half domain ($I(X_1\geq0)$), respectively. The third factor is the external validity parameter, with $\Delta^\star=0,0.2,0.5$ for no bias, small bias, and large bias, respectively. The last factor is the internal validity parameter, with $\Gamma^\star=1,1.2,1.5$ for no bias, small bias, and large bias, respectively. 
Thus, in total, there are $2\times 3\times 3\times 3 = 54$ data-generating models.

For each simulated data from one such model, we first construct matched samples, consisting of matched sets with one OS treated unit, one OS control unit, and a variable number of RCT units depending on the generalization score as described in \S~\ref{sec_matching}. To evaluate the match quality, we use the maximum absolute standardized mean differences in the  {common support and external support}. From the results in 
Table S1 in the supplementary materials, we can observe that our matching procedure greatly reduces the large standardized mean differences in all cases. The match quality improves as the sample size increases.

In the inference stage, we construct 95\% confidence intervals in three ways: using the RCT and OS individually, and combining both of them. To further adjust for the remaining imbalances in the matched data, the inferences first calculate the residuals of regressing $Y$ on $X_1,\dots, X_5$ and the matched set indicator and then use these residuals instead of the original outcomes in all the formulas described in our inference methods.

To study the validity of our theoretical results, we start by calculating these intervals by setting our sensitivity parameters for the inferences as the true values of sensitivity parameters, i.e., $(\Delta,\Gamma)=(\Delta^\star,\Gamma^\star)$. 
The empirical coverage rates and the average lengths of the 95\% confidence intervals are summarized in Table~\ref{tb:simcitrue}. 

Several of our theoretical understandings of the proposed method are validated by these results. First, we can observe that with correctly specified sensitivity parameters, all three types of confidence intervals achieve a coverage probability of around 95\%. Second, the combined confidence intervals are shorter than or have a similar length to the OS confidence interval, which are much shorter than the RCT confidence intervals due to the small sample size of the RCT. Third, as the biases increase in the data-generating model (i.e., the sensitivity parameter values increase), all three confidence intervals become longer. Finally, as expected, all confidence intervals become shorter as the sample size increases. 

\begin{table}[ht]
\centering
\caption{Confidence interval with true parameters $(\Delta,\Gamma)=(\Delta^\star,\Gamma^\star)$: Simulated coverage rates and average lengths of 95\% confidence intervals by using RCT, OS, and combined analysis. Calculated based on 1000 simulated datasets from each data-generating model in each row and each of the three types of common support between the RCT and OS's covariate spaces.}\label{tb:simcitrue}
\resizebox{\textwidth}{!}{
\begin{tabular}{llccccccccccc}
\hline\hline
&&\multicolumn{11}{c}{Confidence Interval Coverage Rate: $N=500$}\\\hline
&&\multicolumn{3}{c}{All common support}&&\multicolumn{3}{c}{Majority common support}&&\multicolumn{3}{c}{Limited common support}\\\cline{3-5}\cline{7-9}\cline{11-13}
&&RCT & OS & Combined &&RCT & OS & Combined &&RCT & OS & Combined \\\hline
$\Delta^\star=0$&$\Gamma^\star=1$&0.95 & 0.96 & 0.95 && 0.95 & 0.95 & 0.95 && 0.95 & 0.94 & 0.95 \\ 
$\Delta^\star=0$&$\Gamma^\star=1.2$& 0.95 & 0.97 & 0.97 && 0.96 & 0.97 & 0.97 && 0.95 & 0.97 & 0.97 \\ 
$\Delta^\star=0$&$\Gamma^\star=1.5$& 0.95 & 0.95 & 0.95 && 0.95 & 0.95 & 0.95 && 0.95 & 0.96 & 0.97 \\ 
$\Delta^\star=0.2$&$\Gamma^\star=1$&0.99 & 0.95 & 0.98 && 0.98 & 0.94 & 0.97 && 0.98 & 0.94 & 0.97 \\ 
$\Delta^\star=0.2$&$\Gamma^\star=1.2$& 0.99 & 0.98 & 0.99 && 0.98 & 0.98 & 0.98 && 0.98 & 0.97 & 0.98 \\ 
$\Delta^\star=0.2$&$\Gamma^\star=1.5$&0.98 & 0.97 & 0.98 && 0.98 & 0.97 & 0.97 && 0.98 & 0.96 & 0.98 \\ 
$\Delta^\star=0.5$&$\Gamma^\star=1$&1.00 & 0.95 & 0.99 && 0.99 & 0.91 & 0.97 && 0.99 & 0.94 & 0.99 \\ 
$\Delta^\star=0.5$&$\Gamma^\star=1.2$& 1.00 & 0.98 & 0.99 && 0.99 & 0.98 & 0.99 && 0.99 & 0.97 & 0.99 \\ 
$\Delta^\star=0.5$&$\Gamma^\star=1.5$& 1.00 & 0.98 & 0.99 && 1.00 & 0.98 & 0.99 && 0.99 & 0.97 & 0.98 \\ 
   \hline
\hline&&\multicolumn{11}{c}{Confidence Interval Coverage Rate: $N=1000$}\\\hline
&&\multicolumn{3}{c}{All common support}&&\multicolumn{3}{c}{Majority common support}&&\multicolumn{3}{c}{Limited common support}\\\cline{3-5}\cline{7-9}\cline{11-13}
&&RCT & OS & Combined &&RCT & OS & Combined &&RCT & OS & Combined \\\hline
$\Delta^\star=0$&$\Gamma^\star=1$& 0.94 & 0.96 & 0.95 && 0.95 & 0.95 & 0.95 && 0.94 & 0.95 & 0.95 \\ 
$\Delta^\star=0$&$\Gamma^\star=1.2$& 0.95 & 0.97 & 0.96 && 0.95 & 0.96 & 0.96 && 0.95 & 0.95 & 0.96 \\ 
$\Delta^\star=0$&$\Gamma^\star=1.5$& 0.95 & 0.96 & 0.95 && 0.95 & 0.95 & 0.95 && 0.95 & 0.94 & 0.95 \\ 
$\Delta^\star=0.2$&$\Gamma^\star=1$&0.99 & 0.96 & 0.99 && 0.98 & 0.95 & 0.97 && 0.97 & 0.95 & 0.97 \\ 
$\Delta^\star=0.2$&$\Gamma^\star=1.2$& 0.99 & 0.98 & 0.99 && 0.99 & 0.97 & 0.97 && 0.98 & 0.96 & 0.97 \\ 
$\Delta^\star=0.2$&$\Gamma^\star=1.5$&0.99 & 0.96 & 0.98 && 0.98 & 0.96 & 0.97 && 0.97 & 0.95 & 0.96 \\ 
$\Delta^\star=0.5$&$\Gamma^\star=1$&1.00 & 0.95 & 0.99 && 0.99 & 0.93 & 0.98 && 0.98 & 0.95 & 0.98 \\ 
$\Delta^\star=0.5$&$\Gamma^\star=1.2$& 1.00 & 0.98 & 1.00 && 0.99 & 0.98 & 0.99 && 0.99 & 0.97 & 0.98 \\ 
$\Delta^\star=0.5$&$\Gamma^\star=1.5$& 1.00 & 0.97 & 0.99 && 0.99 & 0.98 & 0.99 && 0.98 & 0.95 & 0.96 \\ 
\hline\hline
&&\multicolumn{11}{c}{Confidence Interval Length: $N=500$}\\\hline
&&\multicolumn{3}{c}{All common support}&&\multicolumn{3}{c}{Majority common support}&&\multicolumn{3}{c}{Limited common support}\\\cline{3-5}\cline{7-9}\cline{11-13}
&&RCT & OS & Combined &&RCT & OS & Combined &&RCT & OS & Combined \\\hline
$\Delta^\star=0$&$\Gamma^\star=1$&1.66 & 1.04 & 0.92 && 1.84 & 1.02 & 0.93 & &2.50 & 0.99 & 0.97 \\ 
$\Delta^\star=0$&$\Gamma^\star=1.2$& 1.66 & 1.30 & 1.08 && 1.84 & 1.28 & 1.10 && 2.49 & 1.24 & 1.16 \\ 
$\Delta^\star=0$&$\Gamma^\star=1.5$& 1.64 & 1.61 & 1.23 && 1.81 & 1.59 & 1.27 && 2.44 & 1.56 & 1.36 \\ 
$\Delta^\star=0.2$&$\Gamma^\star=1$ & 2.07 & 1.04 & 1.04 && 2.28 & 1.04 & 1.05 && 2.91 & 0.99 & 1.05 \\ 
$\Delta^\star=0.2$&$\Gamma^\star=1.2$& 2.06 & 1.30 & 1.22 && 2.28 & 1.30 & 1.25 && 2.90 & 1.25 & 1.26 \\ 
$\Delta^\star=0.2$&$\Gamma^\star=1.5$& 2.04 & 1.61 & 1.39 && 2.26 & 1.63 & 1.47 && 2.85 & 1.56 & 1.49 \\ 
$\Delta^\star=0.5$&$\Gamma^\star=1$& 2.68 & 1.04 & 1.18 && 3.08 & 1.13 & 1.23 && 3.56 & 1.00 & 1.14 \\
$\Delta^\star=0.5$&$\Gamma^\star=1.2$&2.67 & 1.31 & 1.38 & &3.07 & 1.42 & 1.48 && 3.55 & 1.27 & 1.38 \\ 
$\Delta^\star=0.5$&$\Gamma^\star=1.5$& 2.66 & 1.62 & 1.59 && 3.05 & 1.78 & 1.78 && 3.50 & 1.59 & 1.64 \\    \hline
&&\multicolumn{11}{c}{Confidence Interval Length: $N=1000$}\\\hline
&&\multicolumn{3}{c}{All common support}&&\multicolumn{3}{c}{Majority common support}&&\multicolumn{3}{c}{Limited common support}\\\cline{3-5}\cline{7-9}\cline{11-13}
&&RCT & OS & Combined &&RCT & OS & Combined &&RCT & OS & Combined \\\hline
$\Delta^\star=0$&$\Gamma^\star=1$&1.18 & 0.76 & 0.66 && 1.31 & 0.74 & 0.67 && 1.80 & 0.71 & 0.71 \\ 
$\Delta^\star=0$&$\Gamma^\star=1.2$& 1.18 & 1.03 & 0.83 && 1.31 & 1.01 & 0.85 && 1.78 & 0.99 & 0.91 \\ 
$\Delta^\star=0$&$\Gamma^\star=1.5$& 1.16 & 1.36 & 0.96 && 1.29 & 1.34 & 1.00 && 1.76 & 1.31 & 1.11 \\ 
$\Delta^\star=0.2$&$\Gamma^\star=1$ &1.58 & 0.76 & 0.78 && 1.74 & 0.75 & 0.77 && 2.20 & 0.72 & 0.78 \\
$\Delta^\star=0.2$&$\Gamma^\star=1.2$& 1.58 & 1.03 & 0.97 && 1.74 & 1.03 & 1.00 && 2.19 & 0.99 & 1.00 \\ 
$\Delta^\star=0.2$&$\Gamma^\star=1.5$&1.56 & 1.36 & 1.14 && 1.72 & 1.37 & 1.22 && 2.17 & 1.32 & 1.25 \\ 
$\Delta^\star=0.5$&$\Gamma^\star=1$& 2.19 & 0.76 & 0.91 && 2.47 & 0.81 & 0.90 && 2.84 & 0.73 & 0.84 \\ 
$\Delta^\star=0.5$&$\Gamma^\star=1.2$& 2.19 & 1.04 & 1.13 && 2.47 & 1.12 & 1.19 && 2.83 & 1.01 & 1.10 \\ 
$\Delta^\star=0.5$&$\Gamma^\star=1.5$& 2.17 & 1.37 & 1.34 && 2.45 & 1.48 & 1.50 && 2.80 & 1.34 & 1.39 \\ 
\hline\hline
\end{tabular}}
\end{table}

\subsection{Sensitivity parameter choices}
The previous set of results assumed the true values of the sensitivity parameters that generated the datasets. Since the actual degree of bias is never known in practice, here, in a second set of simulations, we focus on one of the settings considered in the previous subsection, with moderate biases, $(\Delta^\star,\Gamma^\star)=(0.2,1.2)$, majority  {common support} and a sample size of $N=1000$. We compare the three confidence intervals specifying the sensitivity parameters as $\Delta=0,0.2,0.4$ or $0.6$ and $\Gamma=1,1.2,1.5$ or $1.8$ for the inference. We evaluate the inference quality using the coverage rate and average length of the 95\% confidence intervals over 1000 repetitions.

\begin{figure}[!ht]
\centering
\includegraphics[width=\linewidth]{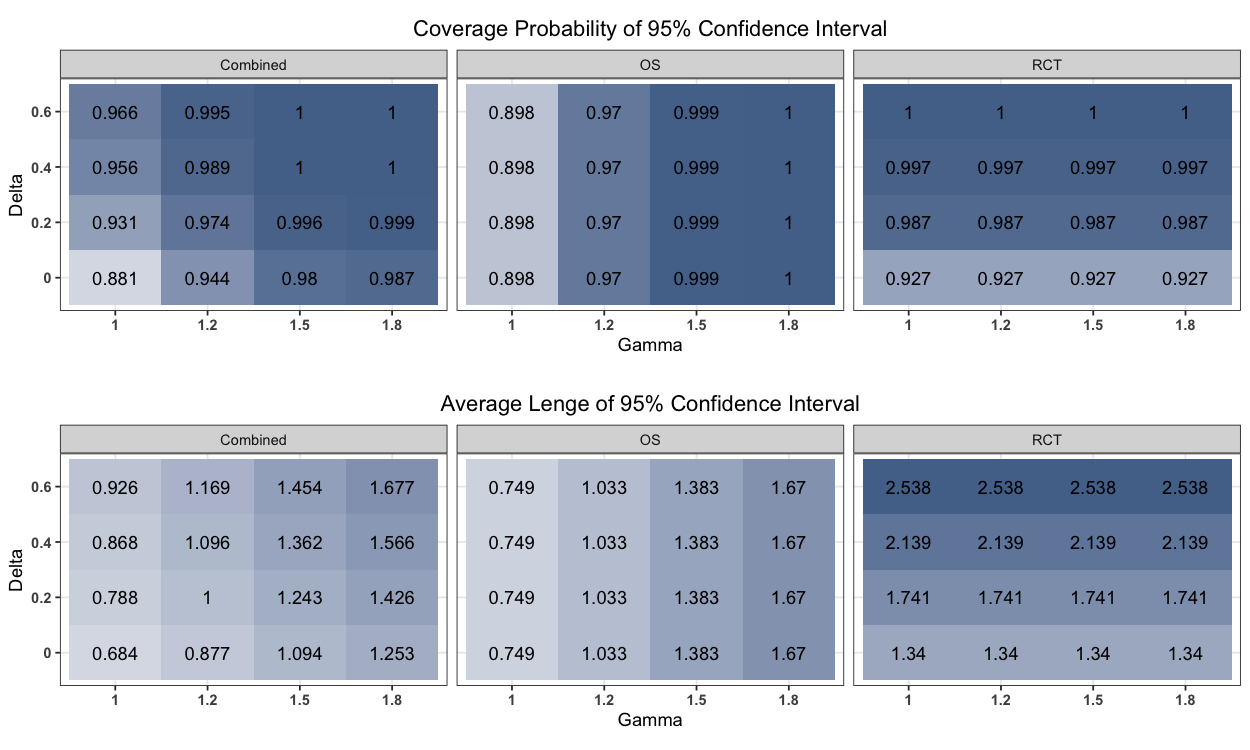}
\caption{Coverage rates and average lengths for confidence intervals with various combinations of sensitivity parameters, estimated based on 1000 repetitions. The data are generated with $N=1000,\Delta^\star=0.2,\Gamma^\star=1.2$, and a majority  {common support} between RCT and OS's covariate spaces.}\label{fig:ci}
\end{figure}

The results are summarized in Figure~\ref{fig:ci}. We can observe that with any fixed $\Delta$, the RCT confidence intervals have the same coverage probabilities and average length when $\Gamma$, which is specific to the OS, varies, but the OS confidence intervals have higher coverage probabilities and average length as $\Gamma$ increases. Similarly, with any fixed $\Gamma$, the OS confidence intervals have the same coverage probabilities and average length when $\Delta$, which is specific to the RCT, varies, but the RCT confidence intervals have higher coverage probabilities and average length as $\Delta$ increases. When either sensitivity parameter increases, the combined intervals improve the coverage probabilities with a wider confidence interval. 

It is of interest to compare the coverage rates and average lengths of the combined confidence intervals to those of the confidence intervals from a single data source. When both sensitivity parameters are larger than or equal to the true values, $\Delta \geq \Delta^\star=0.2$ and $\Gamma\geq \Gamma^\star=1.2$, all three intervals have coverage rates above 95\%. However, the combined interval has robust performances unless both $\Gamma<\Gamma^\star=1.2$ and $\Delta<\Delta^\star=0.2$, while the OS intervals consistently undercover if $\Gamma<\Gamma^\star=1.2$, irrespective of the $\Delta$ values, and the RCT intervals consistently undercover if $\Delta<\Delta^\star=0.2$, irrespective of the $\Gamma$ values. At the same time, the average length of the combined interval tends to be comparable or even shorter than the individual intervals when both sensitivity parameters are larger than or equal to the true values. Thus, it is generally safer to use the combined interval than either of the two data sources alone, and it is preferable to use the combined interval than the worst-performing of the single data sources.

\subsection{Power of sensitivity analysis}
\begin{figure}[!ht]
    \centering
    \includegraphics[width=.95\linewidth]{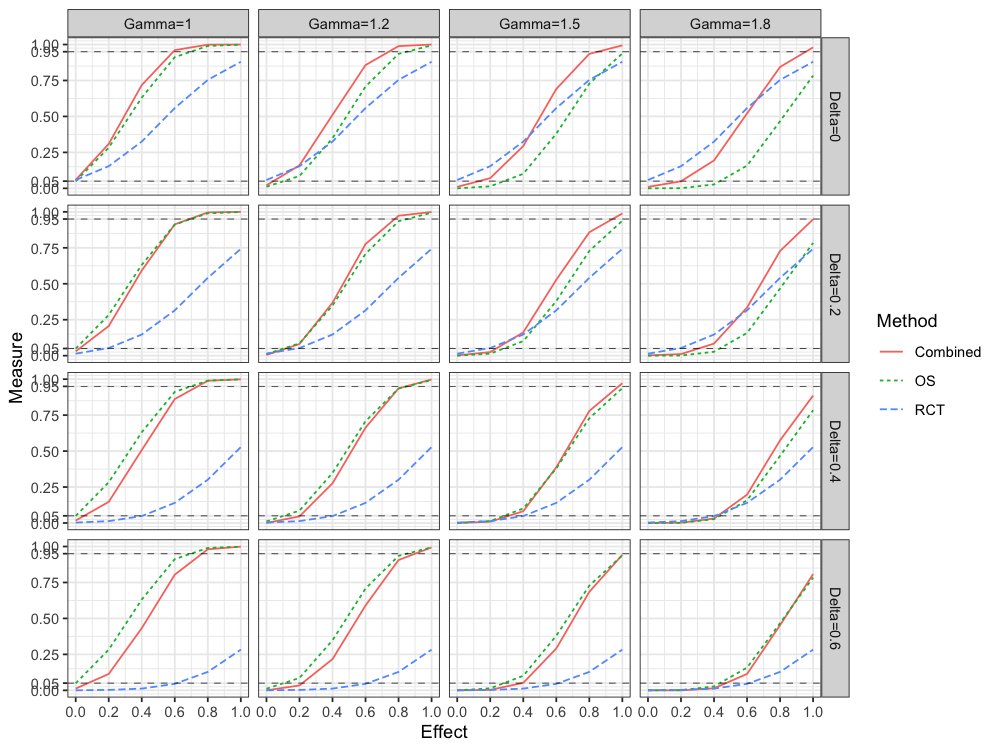}
    \caption{Simulated power curves for the RCT, OS, and the combined method when there is no bias: $\Delta^\star=0,\Gamma^\star=1$. Total sample size $N=1000$ and a majority common support between RCT and OS's covariate spaces.}
    \label{fig:power}
\end{figure}

Aiming to evaluate the statistical power of the three inferential methods, we consider the same model as introduced at the beginning of the section with no bias ($\Delta^\star=0$ and $\Gamma^\star=1$) and a majority common support. We vary the treatment effect from $0, 0.2, 0.4, 0.6, 0.8$, and $1$, so that the potential outcome under treatment is $Y(1)=Y(0)+\tau$ for $\tau = 0, 0.2, 0.4, 0.6, 0.8$ or $1$. We study the power of the proposed method with various choices of sensitivity parameters for the analysis, $\Delta=0,0.2,0.4,0.6$ and $\Gamma=1,1.2,1.5,1.8$. The results in Figure~\ref{fig:power} show that the combined method can control the Type I error well in all cases. As $\Gamma$ increases, the power of using OS alone drops; as $\Delta$ increases, the power of using RCT alone drops; but the combined method keeps robust performance. 

\section{Analysis of CNPRC Primate Dataset} \label{sec_analysis}
We now return to the CNPRC primate dataset to investigate the effects of lactation on postpartum obesity. Recall that the RCT includes 18 monkeys stratified into 6 matched sets. Each matched set has one treated unit (no lactation) and two control units, matched on several factors: parity, age, weight ($+/-1$ kg), and lactation history. In the OS, there are 231 treated monkeys and 360 control monkeys, and covariate data on age, parity, and baseline weight prior to pregnancy. 

To leverage the strengths of both the RCT and the OS, we first apply the proposed matching method that accounts for generalization scores. Specifically, we constructed matched sets consisting of one OS treated unit, one OS control unit, and zero or one copy of an RCT unit. Table~\ref{tb:covbal} shows the covariate balances in terms of absolute standardized mean differences before and after matching, with a noticeable reduction after matching, indicating improved covariate balance across the groups.

\begin{figure}
    \centering
    \includegraphics[width=0.8\linewidth]{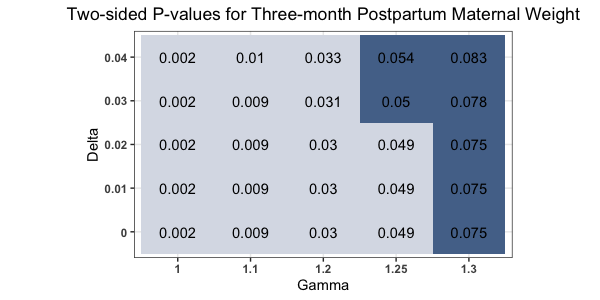}
    \caption{Two-sided $p$-values for testing that there is no effect of lactation on three-month postpartum maternal weight. Values greater than 0.05 are labeled in navy, indicating insignificant evidence of an effect at those levels of biases.}
    \label{fig:pval}
\end{figure}

We estimate the ATOT using the residuals after covariate adjustment to further adjust the residual imbalances. 
The results are summarized in Table~\ref{tb:real}. Due to the limited sample size of the RCT, the RCT confidence intervals are the widest, and the combined confidence interval is a bit longer than the OS confidence interval. Results from the combined analysis suggest that lactation has a modest positive effect on three months postpartum maternal weight.
Specifically, we are 95\% confident that lactation increases maternal weight by between 0.09 kg and 0.44 kg. These results appear to be robust, even when accounting for a small generalization bias in the RCT ($\Delta = 0.02$) and a moderate hidden bias due to unmeasured confounders in the OS ($\Gamma = 1.25$).  See Figure~\ref{fig:pval} for how the sensitivity parameters are determined. However, lactation has no significant effect on six months postpartum maternal weight. For comparison, the RCT data alone did not yield statistically significant results, likely due to the limited sample size. 
While the OS data alone led to the same conclusion as the combined analysis, the results were more sensitive to hidden bias, with $\Gamma = 1.23$, for three months postpartum maternal weight.

In summary, our analysis of the CNPRC primate data supports the conclusion that lactation leads to a modest increase in maternal weight three months postpartum, but no significant effect is observed at six months. The initial weight gain may be attributed to various physiological factors associated with lactation, such as hormonal changes and caloric retention. However, as lactation progresses, increased maternal energy expenditure, along with other factors such as dietary adjustments, physical activity, and metabolic adaptations, could offset the initial weight gain. This may explain the absence of significant weight differences at six months postpartum. These findings, however, contrast with some prior human research, which suggests that breastfeeding is associated with a reduction in postpartum weight retention at six months or longer \citep{baker2008breastfeeding,hebeisen2024prospective,loy2024breastfeeding}. While we do not find any significant weight reduction, one possible explanation for the suggested weight gain is that 
breastfeeding could reduce visceral adiposity \citep{mcclure2011breastfeeding,mcclure2012maternal}. To further investigate this hypothesis, future studies, including large-scale RCTs, are needed to verify these conjectures.

\begin{table}[ht]
	\centering
	\caption{Covariate balance before and after matching: Covariate means and standardized mean differences. }\label{tb:covbal}
	\resizebox{\textwidth}{!}{
	\begin{tabular}{lccccccc}
		\hline\hline
		\multicolumn{8}{c}{ {Common Support}: Covariate Mean}\\\hline
		&\multicolumn{3}{c}{Before Matching} &&\multicolumn{3}{c}{After Matching} \\\cline{2-4}\cline{6-8}
        &OS  &   & OS   && OS  &  & OS \\
		&Treated & RCT  &  Control  &&  Treated & RCT  &  Control  \\\hline
		Age &6.91 &6.92&6.77&&6.91 &6.39 &6.74\\
		Parity& 2.62 &3.83&2.59&&2.62 &3.04 &2.58\\
		Pre-pregnancy weight& 7.38& 7.98 &7.60&&7.38 &7.26& 7.49\\\hline
		\multicolumn{8}{c}{ {Common Support}: Absolute Standardized Mean Differences}\\\hline
		&\multicolumn{3}{c}{Before Matching} &&\multicolumn{3}{c}{After Matching} \\\cline{2-4}\cline{6-8}
		  & OS \ \ \ \  & OS \ \ \ \  & OS Treated    && OS \ \ \ \ & OS \ \ \ \  & OS Treated \\
		& \multicolumn{1}{l}{Treated -- RCT}  &  \multicolumn{1}{l}{Control -- RCT}  &  -- OS Control   && \multicolumn{1}{l}{Treated -- RCT} &  \multicolumn{1}{l}{Control -- RCT} & -- OS Control  \\\hline
		Age &0.00 &0.05 &0.05&&0.19&0.12&0.06\\
		Parity&0.60&0.62 &0.01&&0.21&0.23&0.02\\
		Pre-pregnancy weight & 0.35& 0.22 &0.13&&0.07&0.13&0.06\\\hline\hline
		\multicolumn{8}{c}{ {External Support}: Covariate Mean}\\\hline
		&\multicolumn{3}{c}{Before Matching} &&\multicolumn{3}{c}{After Matching} \\\cline{2-4}\cline{6-8}
		&OS  &   & OS   && OS  &  & OS \\
		&Treated & RCT  &  Control  &&  Treated & RCT  &  Control   \\\hline
		Age &7.24&-& 6.11&&7.24&-& 7.16\\
		Parity& 1.99 &-&1.65&&1.99 &-&2.20\\
		Pre-pregnancy weight & 7.37 &-&7.15&&7.37 &-&7.47\\\hline
		\multicolumn{8}{c}{ {External Support}: Absolute Standardized Mean Differences}\\\hline
		&\multicolumn{3}{c}{Before Matching} &&\multicolumn{3}{c}{After Matching} \\\cline{2-4}\cline{6-8}
        & OS \ \ \ \  & OS \ \ \ \  & OS Treated    && OS \ \ \ \ & OS \ \ \ \  & OS Treated \\
		& \multicolumn{1}{l}{Treated -- RCT}  &  \multicolumn{1}{l}{Control -- RCT}  &  -- OS Control   && \multicolumn{1}{l}{Treated -- RCT} &  \multicolumn{1}{l}{Control -- RCT} & -- OS Control  \\\hline
		Age &-&-&0.40&&-&-&0.03\\
		Parity&-&-&0.17&&-&-&0.10\\
		Pre-pregnancy weight &-&-& 0.13&&-&-&0.06\\\hline\hline
	\end{tabular}}
\end{table}

\begin{table}[ht]
	\centering
	\caption{95\% confidence intervals for the effect of lactation}\label{tb:real}
	\begin{tabular}{lll}
			\hline\hline
			\multicolumn{3}{c}{6-month postpartum weight}\\\hline
RCT &$\Delta=0$& $[-0.49,0.54]$\\
OS & $\Gamma=1$& $[-0.24,0.05]$\\
Combined & $\Delta=0,\Gamma=1$ &$[-0.25,0.08]$\\\hline
			\multicolumn{3}{c}{3-month postpartum weight}\\\hline
			RCT &$\Delta=0$&$[-1.40,0.25]$\\
			OS & $\Gamma=1$& $[-0.40,-0.10]$\\
			OS & $\Gamma=1.23$& $[-0.49,-0.00]$\\
			Combined & $\Delta=0,\Gamma=1$ & $[-0.44,-0.09]$\\
			Combined & $\Delta=0.02,\Gamma=1.25$ & $[-0.54,-0.00]$\\\hline\hline
		\end{tabular}
\end{table}

\section*{Acknowledgments}
This work was supported by the National Science Foundation grant number DMS-2015250 and the CNPRC base grant number P51OD011107.

\section*{Data Availability}
The CNPRC primate dataset cannot be shared due to confidentiality restrictions. Code for simulating a comparable dataset, 
implementing the proposed method, and replicating all numerical experiments is available at the Github repository\\ \href{https://github.com/ruoqiyu/CombinedCausalInference/} {https://github.com/ruoqiyu/CombinedCausalInference/}. 

\clearpage

\begin{center}
	{
		\Large Supplementary Materials
	}
\end{center}
	
	\setcounter{section}{0}
	\setcounter{equation}{0}
	\setcounter{figure}{0}
	\setcounter{table}{0}
	\setcounter{proposition}{0}
	\setcounter{assumption}{0}
	\def\theequation{S\arabic{section}.\arabic{equation}}
	\def\thesection{S\arabic{section}}
	\def\thefigure{S\arabic{figure}}
	\def\thetable{S\arabic{table}}
	\def\theproposition{S\arabic{proposition}}
	\def\theassumption{S\arabic{assumption}}

	\section{Details of the sensitivity analysis of the observational study}
	\subsection{Equivalent form of Rosenbaum's sensitivity model} \label{supp:equi}
	Rosenbaum's sensitivity model specification 
	(2)
	may be equivalently written in the following semiparametric model. Following \cite{rosenbaum2020modern}, define principal unobserved covariate $\Pr(Z^o_i = 1 \mid Y_i^o(1), Y^o_i(0), X^o_i)=:v^o_i\in[0,1]$, so that treatment assignment is always ignorable (or unconfounded) given $(X^o_i,v^o_i)$. Then,  Proposition 12 of \cite{rosenbaum2002observational} shows that for some function $\varphi_x(v^o_i):=u^o_i\in[0,1]$,
	\begin{equation}\label{eq_sens_exp_model}
		\Pr(Z_i^o=1\mid Y_i^o(1), Y^o_i(0), X^o_i=x, v^o_i) = \frac{\exp\{ \kappa(x) + \log(\Gamma) u^o_i\}}{1+ \exp\{ \kappa(x) + \log(\Gamma) u^o_i\}},
	\end{equation}
	where $\kappa$ is some unknown function $\kappa$ that depends of the potential outcomes. This model clarifies the role of $\Gamma$, appearing in the coefficient of $u_i^o$, as encoding the effect of the unmeasured confounder. 
	
	\subsection{Calculation of the separable approximation of the extreme $p$-values} \label{supp:sep}
	We give the details of the calculation of the separable approximation of the extreme treatment assignment probabilities $\widetilde{\boldsymbol{\eta}}_i^{(\beta_0^\star)}$s described in Section 4.1. Fix $\beta_0^\star$. Define $\tilde{Y}^o_{ij} = {Y}^o_{ij} - \beta_0^\star Z_{ij}^o$ as the adjusted outcomes. Let $\tilde{Y}^o_{i(1)} \leq \cdots \leq \tilde{Y}^o_{i(J_i)}$ be the sorted values of the adjusted outcomes. From here onwards, let $(1),\ldots, (J_i)$ denote the indices that give the ordered adjusted outcomes. Consider the set of $(J_i-1)$ vectors $\mathcal{P}$ of $\boldsymbol{\eta}_i = (\eta_{i(1)}, \ldots, \eta_{i(J_i)})$ where
	$\eta_{i(1)} = \cdots = \eta_{i(m)} = 1/(m + ((J_i - m) * \Gamma)$ and  $\eta_{i(m+1)} = \cdots = \eta_{i(J_i)} = \Gamma/(m + ((J_i - m) * \Gamma)$ for $m=1,\ldots, J_i-1$. For each $\boldsymbol{\eta}_i\in \mathcal{P}$, calculate
	$\mu(\boldsymbol{\eta}_i) = \sum_{j=1}^{J_i} \tilde{Y}^o_{i(j)}\eta_{i(j)}$
	and $\sigma^2(\boldsymbol{\eta}_i) = \sum_{j=1}^{J_i} \{\tilde{Y}^o_{i(j)}\}^2\eta_{i(j)}
	- \mu(\boldsymbol{\eta}_i)^2$.
	
	Search for $\boldsymbol{\eta}_i$s in $\mathcal{P}$ that maximizes $\mu(\boldsymbol{\eta}_i)$. If there are multiple such vectors that maximize these means, choose the one among this set that maximizes $\sigma^2(\boldsymbol{\eta}_i)$. This choice of $\boldsymbol{\eta}_i$ gives our separable approximation probabilities $\widetilde{\boldsymbol{\eta}}_i^{(\beta_0^\star)}$s.
	
	Separable probabilities $\widetilde{\boldsymbol{\eta}}_i^{(\beta_0^\star)}$ are used to test for greater than alternative $H_1:\beta^\star>\beta^\star_0$. Separable probabilities $\dwidetilde{\boldsymbol{\eta}}_i^{(\beta_0^\star)}$ are calculated similarly but to test for the less than alternative. The calculation simply redefines the adjusted outcomes as $\dwidetilde{Y}^o_{ij} = -{Y}^o_{ij} + \beta_0^\star Z_{ij}^o$. Notice that this corresponds to multiplying the outcomes by $-1$ so that the null is $H_0:-\beta^\star=-\beta^\star_0$ and the alternative is again a greater than alternative $H_1:-\beta^\star>-\beta^\star_0$.
	
	\subsection{Confidence interval construction} \label{supp:ci}
	We use numerical methods to find the confidence interval by converting the test for the OS. The process will proceed by first fixing the desired confidence level $\alpha$. Then a root finding method finds the limit of the upper-sided confidence interval $[\beta_L^o, \infty)$, that solves for 
	(3) 
	of the main text, with equality in place of the inequality, when we write $\beta_L^o$ in place of $\beta_0^\star$; the subscript $L$ emphasizes that it is the lower limit of the interval. Similarly, a root finding method solves for limit of the lower-sided confidence interval, $(-\infty, \beta_U^o]$, that solves for 
	(4) 
	in the main text, with equality in place of the inequality, when we write $\beta_U^o$ in place of $\beta_0^\star$; the subscript $U$ emphasizes that it is the upper limit of the interval. Most statistical software, including \texttt{R}, provides a root finding tool, e.g., the \texttt{uniroot} function in \texttt{R}. 
	
	\section{Details of the combined analysis} \label{supp:critical}
	The critical level $\kappa_\alpha$ follows from the fact that negative two times the logarithm of the product of two independent uniform random variables on $(0,1)$ has a $\chi^2$ distribution with $4$ degrees of freedom. Thus, if $p^o_{\beta_0^\star}$ and $p^r_{\beta_0^\star}$ were uniformly distributed under the null, after some calculations, we would get the cutoff $\kappa_\alpha = \exp(-\chi^2_{4;1-\alpha}/2)$ for the test statistic $p^o_{\beta_0^\star}\times p^r_{\beta_0^\star}$.

	\section{Additional simulation study}
	\subsection{Covariate balance for simulation study in 
		\S5.1}
	
	To evaluate the match quality for the simulation study in 
	\S5.1, we use the maximum absolute standardized mean differences in the common support and external support and summarize the results in Table~\ref{tb:sim1balance}. We can observe that our matching procedure greatly reduces the large standardized mean differences in all cases. The match quality improves as the sample size increases.  
	
	\begin{table}[ht!]
		\centering
		\caption{Covariate balance: Average maximum absolute standardized mean differences in the common support and external support, over 1000 simulations.}\label{tb:sim1balance}
		\resizebox{\textwidth}{!}{
			\begin{tabular}{llcccccccccccccc}
				\hline\hline
				&&\multicolumn{14}{c}{Maximum Absolute Standardized Mean Differences: $N=500$}\\\hline
				&&\multicolumn{2}{c}{   {All common support}}&&\multicolumn{5}{c}{   {Majority common support}}&&\multicolumn{5}{c}{   {Limited common support}}\\\cline{3-4}\cline{6-10}\cline{12-16}
				&&\multicolumn{2}{c}{   {Common support}}&& \multicolumn{2}{c}{   {Common support}}&&\multicolumn{2}{c}{   {External support}}&& \multicolumn{2}{c}{   {Common support}}&&\multicolumn{2}{c}{   {External support}}\\\cline{3-4}\cline{6-7}\cline{9-10}\cline{12-13}\cline{15-16}
				&&Before & After & &Before & After &&Before & After &&Before & After &&Before & After\\\hline
				$\Delta^\star=0$&$\Gamma^\star=1$&0.55 & 0.26 &   & 0.51 & 0.27 & &0.53 & 0.30 && 0.58 & 0.34 && 0.34 & 0.15 \\ 
				$\Delta^\star=0$&$\Gamma^\star=1.2$&0.55 & 0.26 &   & 0.51 & 0.26 && 0.53 & 0.31 && 0.58 & 0.34& & 0.34 & 0.15 \\ 
				$\Delta^\star=0$&$\Gamma^\star=1.5$&0.54 & 0.25 &   & 0.50 & 0.25 & &0.51 & 0.30 && 0.56 & 0.33 && 0.34 & 0.15 \\ 
				$\Delta^\star=0.2$&$\Gamma^\star=1$&0.55 & 0.26 &   & 0.51 & 0.27 && 0.53 & 0.30 && 0.58 & 0.34 && 0.34 & 0.15 \\ 
				$\Delta^\star=0.2$&$\Gamma^\star=1.2$&0.55 & 0.26 &  & 0.51 & 0.26 && 0.53 & 0.31 && 0.58 & 0.34 & &0.34 & 0.15 \\ 
				$\Delta^\star=0.2$&$\Gamma^\star=1.5$&0.54 & 0.25 & & 0.50 & 0.25 && 0.51 & 0.30 && 0.56 & 0.33 && 0.34 & 0.15 \\ 
				$\Delta^\star=0.5$&$\Gamma^\star=1$& 0.55 & 0.26 &  & 0.51 & 0.27 && 0.53 & 0.30 & &0.58 & 0.34 && 0.34 & 0.15 \\ 
				$\Delta^\star=0.5$&$\Gamma^\star=1.2$&0.55 & 0.26 &  & 0.51 & 0.26 && 0.53 & 0.31 && 0.58 & 0.34 && 0.34 & 0.15 \\ 
				$\Delta^\star=0.5$&$\Gamma^\star=1.5$&0.54 & 0.25 & & 0.50 & 0.25 && 0.51 & 0.30 && 0.56 & 0.33 && 0.34 & 0.15 \\ 
				\hline
				&&\multicolumn{14}{c}{Maximum Absolute Standardized Mean Differences: $N=1000$}\\\hline
				&&\multicolumn{2}{c}{   {All common support}}&&\multicolumn{5}{c}{   {Majority common support}}&&\multicolumn{5}{c}{   {Limited common support}}\\\cline{3-4}\cline{6-10}\cline{12-16}
				&&\multicolumn{2}{c}{   {Common support}}&& \multicolumn{2}{c}{   {Common support}}&&\multicolumn{2}{c}{   {External support}}&& \multicolumn{2}{c}{   {Common support}}&&\multicolumn{2}{c}{   {External support}}\\\cline{3-4}\cline{6-7}\cline{9-10}\cline{12-13}\cline{15-16}
				&&Before & After &&Before & After &&Before & After &&Before & After &&Before & After\\\hline
				$\Delta^\star=0$&$\Gamma^\star=1$&0.48 & 0.18 &  & 0.44 & 0.18 &&  0.40 & 0.19 &&  0.47 & 0.23 &&  0.28 & 0.10 \\ 
				$\Delta^\star=0$&$\Gamma^\star=1.2$&0.48 & 0.18 &  & 0.44 & 0.18 & & 0.40 & 0.19 & & 0.47 & 0.23 & & 0.28 & 0.10 \\ 
				$\Delta^\star=0$&$\Gamma^\star=1.5$&0.47 & 0.17 &  & 0.43 & 0.17 &&  0.39 & 0.19 & & 0.46 & 0.22 &&  0.27 & 0.10 \\  
				$\Delta^\star=0.2$&$\Gamma^\star=1$&0.48 & 0.18 & & 0.44 & 0.18 &&  0.40 & 0.19 & & 0.47 & 0.23 &&  0.28 & 0.10 \\ 
				$\Delta^\star=0.2$&$\Gamma^\star=1.2$&0.48 & 0.18 &  & 0.44 & 0.18 & & 0.40 & 0.19 &&  0.47 & 0.23 &&  0.28 & 0.10 \\ 
				$\Delta^\star=0.2$&$\Gamma^\star=1.5$&0.47 & 0.17 &  & 0.43 & 0.17 &&  0.39 & 0.19 &&  0.46 & 0.22 &&  0.27 & 0.10 \\ 
				$\Delta^\star=0.5$&$\Gamma^\star=1$&0.48 & 0.18 &  & 0.44 & 0.18 &&  0.40 & 0.19 & & 0.47 & 0.23 &&  0.28 & 0.10 \\ 
				$\Delta^\star=0.5$&$\Gamma^\star=1.2$&0.48 & 0.18 & & 0.44 & 0.18 &&  0.40 & 0.19 &&  0.47 & 0.23 & & 0.28 & 0.10 \\ 
				$\Delta^\star=0.5$&$\Gamma^\star=1.5$& 0.47 & 0.17 & & 0.43 & 0.17 & & 0.39 & 0.19 &&  0.46 & 0.22 & & 0.27 & 0.10 \\  
				\hline\hline
		\end{tabular}}
	\end{table}
	
	\vspace{-10pt}
	
	\newpage

	\subsection{Comparison with other candidate methods}
	
	In this section, we compare our proposed combined method with two integrated inference approaches from the literature: the elastic integrative analysis of \cite{yang2023elastic} and the integrative R-learner of \cite{wu2022integrative}. Specifically, we focus on the majority common support scenario described in 
	\S5.1, and evaluate performance based on mean squared error (MSE) and confidence interval coverage for estimating the ATOT. 
	
	We further add two classical methods that calibrate an RCT using covariate data from the OS to estimate the ATOT. The first method, due to \cite{hartman2015sample}, uses a matching followed by a weighting method, while the second method, due to \cite{stuart2011use}, uses a propensity score-based method. Importantly, both methods aim to estimate the ATOT, which is also our target estimand. However, unlike our method, they do not use the outcome data from the OS.
	
	\subsubsection{Results when the covariate support of RCT is completely in OS support}
	
	The results are presented in Table~\ref{tb:compare}. First, we focus on the left half of the table, deferring discussion of the other half to the following subsection. For the current simulation, we use the same simulation model as in the main text, with a total sample size $1000$, five covariates, and a `majority' common support between the RCT and OS's covariate supports (i.e., the RCT covariate space spans only 50\% of the OS's covariate space). 
	
	The results show that the combined method using the true sensitivity parameters ($\Delta = \Delta^\star$, $\Gamma = \Gamma^\star$) generally outperforms both state-of-the-art benchmarks in terms of both MSE and confidence interval coverage. When the sensitivity parameters are unknown, using default values ($\Delta=0, \Gamma=1$) in the combined method still yields comparable or smaller MSE relative to the elastic integrative analysis and integrative R-learner under these simulation settings. 
	
	The classical methods show notably poor MSE. This is not unexpected for a few reasons. First, these use the relation between the outcome and exposure in the RCT, while using the OS to only calibrate the covariate distribution. Thus, the effective sample size is much smaller. Second, when the RCT covariates do not span the whole covariate support of the OS, as is the case with this simulation setting, these methods may fail to calibrate the RCT covariate distribution correctly. Finally, both the weighting and the propensity score methods using a small RCT sample are usually very noisy.
	
	Both these methods give at least the nominal coverage in our simulation. The weighting method gives a much higher coverage, often close to 100\% coverage. However, there is no known theoretical result that establishes when these methods will provide the desired coverage.
	
	Let's then only focus on the proposed method and the two state-of-the-art methods in the first four columns of the table.
	\color{black}
	In terms of confidence interval coverage, when there is no generalizability bias ($\Delta^\star=0$), all methods perform similarly under no unmeasured confounding ($\Gamma^\star=1$). However, as the level of unmeasured confounding increases ($\Gamma^\star>1$), the competing methods become more sensitive to the misspecification of $\Gamma$, leading to slightly reduced coverage (and falling below the nominal level).  
	This is unexpected as those methods target the average treatment effect in the RCT population, which is the same as the ATOT when $\Delta^\star=0$. Thus, we should expect a consistent estimation and nominal coverage by these methods. In contrast, the proposed combined method maintains relatively stable coverage across these settings. However, incorrectly specifying sensitivity parameters by setting $\Delta=0, \Gamma=1$ leads to undercoverage with the proposed combining method.
	On the other hand, under a generalizability (external validity) bias, i.e., $\Delta^\star=0.2$ or $0.5$, the proposed method provides above the nominal coverage if the hidden bias from unmeasured confounding is correctly specified. The competing methods have significant undercoverage, which deteriorates with larger external validity bias or internal validity bias. This behavior is expected, as those methods are only able to estimate the average treatment effect in the RCT population and this estimand differs from the ATOT when $\Delta^\star\neq 0$.
	Although using the default sensitivity parameters (i.e., assuming $\Gamma=1, \Delta=0$) gives undercoverage for ATOT using the combined method, the combined method generally has similar or higher coverages than the competing methods. 
	
	\subsubsection{Results when the covariate support of RCT is not completely in OS support}
	
	We consider a second simulation model that allows some units of the RCT to have characteristics that are not represented in any OS units. This allows us to evaluate the performance of the methods when there is likely a shift in characteristics from the RCT units to the OS units. We design this simulation setting by allowing 10\% of the RCT units to be `outside' of the OS support. These units have a constant shift in the control potential outcome values.
	
	More specifically, we consider the following data-generating mechanism. There are six observed covariates $(X_1,\dots, X_5, X_6)$ independently distributed, the first five following the standard normal distribution and $X_6$ is Bernoulli(0.10).  When $X_6$ is 1, the unit is placed into the RCT. Thus, this part of RCT, about 10\% of the total sample size, is outside of OS support. On the other hand, when $X_1< -1$, the unit is put into OS.
	If $X_6$ is 0 or $X_1> -1$, the probability of selecting into the RCT is $expit(-1.5+0.1X_1+0.1X_2-0.3X_4)$ for any covariate value. Once selected into either the RCT or OS, the probability of being assigned to treatment is $1/2$ in the RCT and $expit(-2-0.3X_1+0.1X_3-0.2X_5+\log(\Gamma^\star)U)$ in the OS.
	The potential outcome under control $Y(0)=10+4X_1-2X_2+3X_5+2X_6+U+\epsilon$, where $\epsilon\sim N(0,1)$ and under treatment is $Y(1)=Y(0)+\widetilde\Delta^\star I(\textrm{Unit belongs to the  {common support} $\X$})$, where $\widetilde\Delta^\star=\Delta^\star/\Pr(X^o\in\X^c\mid Z^o=1)$. The covariate $X_6$ creates the covariate shift between the RCT and OS so that the RCT units may have covariate values that fall outside of the OS support. $X_6$ also affects the potential outcomes.

	The MSEs and empirical coverage rates are presented in the right half of Table~\ref{tb:compare}. 
	The proposed method with correctly specified sensitivity parameters performs the best, giving smaller MSEs than all other methods along with empirical coverages of at least 95\% for all levels of unmeasured confounding and generalizability bias. The MSE results are not very sensitive to the covariate shift for all methods except the PS method. 
	The classical methods still had notably worse MSEs compared to the proposed method and state-of-the-art methods.
	The weighting method suffers from a very large MSE and an overly conservative coverage rate. 
	The elastic integrative method provides below nominal coverage even when there is no unmeasured confounding or generalizability bias. In the same setup, the other methods provide close to nominal coverage. Thus, the elastic integrative method is clearly sensitive to a covariate shift in the RCT outside the OS support.

	\begin{landscape}
		\begin{table}[htbp]
			\centering
			\caption{Mean squared errors and 95\% confidence interval coverage probabilities for estimating the ATOT comparing the proposed combined method with other competing state-of-the-art and classical methods. Simulation conducted with a total sample size of $1000$. The RCT inside the OS support shares 50\% of the OS's covariate support.}
			\label{tb:compare}
			\begin{tabular}{p{.25in}p{.25in}|r|rrrrrr|r|rrrrrr}
				\toprule
				& \multicolumn{1}{p{.25in}}{} & \multicolumn{1}{r}{} & \multicolumn{6}{c|}{RCT completely inside OS support} &       & \multicolumn{6}{c}{A fraction ($\sim$40\%) of RCT is outside OS support} \\
				\cmidrule{4-16}          & \multicolumn{1}{r}{} & \multicolumn{1}{r}{} & \multicolumn{13}{c}{Mean Squared Error} \\
				\midrule
				& \multicolumn{1}{p{.25in}}{} & \multicolumn{1}{r}{} & \multicolumn{2}{|c}{Proposed combined} &       &       &       &       &       & \multicolumn{2}{|c}{Proposed combined} &       &       &       &  \\
				\cmidrule{4-5}\cmidrule{11-12}    \multicolumn{1}{p{.25in}}{$\delta^\star$} & \multicolumn{1}{p{.25in}|}{$\Gamma^\star$} &       & \multicolumn{1}{p{.5in}}{$\Gamma=\Gamma^\star$} & \multicolumn{1}{p{.5in}}{$\Gamma=1$} & \multicolumn{1}{l}{} & \multicolumn{1}{l}{} & \multicolumn{1}{l}{} & \multicolumn{1}{l|}{} &       & \multicolumn{1}{p{.5in}}{$\Gamma=\Gamma^\star$} & \multicolumn{1}{p{.5in}}{$\Gamma=1$} & \multicolumn{1}{l}{} & \multicolumn{1}{l}{} & \multicolumn{1}{l}{} & \multicolumn{1}{l}{} \\
				\multicolumn{1}{p{.25in}}{} & \multicolumn{1}{p{.25in}|}{} &       & \multicolumn{1}{p{.5in}}{$\delta=\delta^\star$} & \multicolumn{1}{p{.5in}}{$\delta=0$} & \multicolumn{1}{l}{Elastic} & \multicolumn{1}{l}{R-Learner} & \multicolumn{1}{l}{Weighting} & \multicolumn{1}{l|}{PS} &       & \multicolumn{1}{p{.5in}}{$\delta=\delta^\star$} & \multicolumn{1}{p{.5in}}{$\delta=0$} & \multicolumn{1}{l}{Elastic} & \multicolumn{1}{l}{R-Learner} & \multicolumn{1}{l}{Weighting} & \multicolumn{1}{l}{PS} \\
				\cmidrule{1-2}\cmidrule{4-9}\cmidrule{11-16}    0     & 1     &       & 0.004 & 0.004 & 0.063 & 0.014 & 1.842 & 4.368 &       & 0.006 & 0.006 & 0.033 & 0.009 & 1.739 & 2.660 \\
				0     & 1.2   &       & 0.008 & 0.004 & 0.066 & 0.029 & 1.893 & 4.628 &       & 0.011 & 0.007 & 0.034 & 0.026 & 1.802 & 2.841 \\
				0     & 1.5   &       & 0.021 & 0.005 & 0.079 & 0.103 & 2.121 & 5.184 &       & 0.026 & 0.007 & 0.043 & 0.071 & 1.916 & 3.166 \\
				0.2   & 1     &       & 0.023 & 0.015 & 0.087 & 0.022 & 1.672 & 4.387 &       & 0.025 & 0.018 & 0.042 & 0.014 & 1.836 & 2.412 \\
				0.2   & 1.2   &       & 0.043 & 0.014 & 0.094 & 0.048 & 1.774 & 4.710 &       & 0.048 & 0.021 & 0.054 & 0.044 & 1.891 & 2.544 \\
				0.2   & 1.5   &       & 0.073 & 0.014 & 0.111 & 0.131 & 2.084 & 5.367 &       & 0.084 & 0.015 & 0.049 & 0.098 & 1.994 & 2.792 \\
				0.5   & 1     &       & 0.043 & 0.029 & 0.274 & 0.082 & 1.904 & 4.608 &       & 0.044 & 0.025 & 0.134 & 0.038 & 1.421 & 2.690 \\
				0.5   & 1.2   &       & 0.077 & 0.027 & 0.272 & 0.118 & 1.971 & 4.879 &       & 0.080 & 0.033 & 0.130 & 0.097 & 1.458 & 2.824 \\
				0.5   & 1.5   &       & 0.126 & 0.026 & 0.276 & 0.241 & 2.223 & 5.446 &       & 0.141 & 0.023 & 0.144 & 0.190 & 1.533 & 3.074 \\
				\midrule
				& \multicolumn{1}{p{.25in}}{} &       & \multicolumn{13}{c}{Empirical Coverage at 95\% Confidence Level } \\
				\midrule
				& \multicolumn{1}{p{.25in}}{} &       & \multicolumn{2}{c}{Proposed combined} &       &       &       &       &       & \multicolumn{2}{c}{Proposed combined} &       &       &       &  \\
				\cmidrule{4-5}\cmidrule{11-12}    \multicolumn{1}{p{.25in}}{$\delta^\star$} & \multicolumn{1}{p{.25in}|}{$\Gamma^\star$} &       & \multicolumn{1}{p{.5in}}{$\Gamma=\Gamma^\star$} & \multicolumn{1}{p{.5in}}{$\Gamma=1$} & \multicolumn{1}{l}{} & \multicolumn{1}{l}{} & \multicolumn{1}{l}{} & \multicolumn{1}{l|}{} &       & \multicolumn{1}{p{.5in}}{$\Gamma=\Gamma^\star$} & \multicolumn{1}{p{.5in}}{$\Gamma=1$} & \multicolumn{1}{l}{} & \multicolumn{1}{l}{} & \multicolumn{1}{l}{} & \multicolumn{1}{l}{} \\
				\multicolumn{1}{p{.25in}}{} & \multicolumn{1}{p{.25in}|}{} &       & \multicolumn{1}{p{.5in}}{$\delta=\delta^\star$} & \multicolumn{1}{p{.5in}}{$\delta=0$} & \multicolumn{1}{l}{Elastic} & \multicolumn{1}{l}{R-Learner} & \multicolumn{1}{l}{Weighting} & \multicolumn{1}{l|}{PS} &       & \multicolumn{1}{p{.5in}}{$\delta=\delta^\star$} & \multicolumn{1}{p{.5in}}{$\delta=0$} & \multicolumn{1}{l}{Elastic} & \multicolumn{1}{l}{R-Learner} & \multicolumn{1}{l}{Weighting} & \multicolumn{1}{l}{PS} \\
				\cmidrule{1-2}\cmidrule{4-9}\cmidrule{11-16}    0     & 1     &       & 0.953 & 0.949 & 0.945 & 0.975 & 1.000 & 0.978 &       & 0.945 & 0.933 & 0.905 & 0.953 & 0.998 & 0.943 \\
				0     & 1.2   &       & 0.990 & 0.953 & 0.868 & 0.932 & 0.998 & 0.975 &       & 0.983 & 0.946 & 0.848 & 0.935 & 0.998 & 0.938 \\
				0     & 1.5   &       & 0.995 & 0.945 & 0.750 & 0.902 & 0.990 & 0.978 &       & 0.992 & 0.950 & 0.842 & 0.901 & 0.998 & 0.938 \\
				0.2   & 1     &       & 0.972 & 0.925 & 0.887 & 0.955 & 0.998 & 0.960 &       & 0.988 & 0.950 & 0.838 & 0.900 & 9.998 & 0.948 \\
				0.2   & 1.2   &       & 0.996 & 0.913 & 0.738 & 0.815 & 0.995 & 0.958 &       & 1.000 & 0.929 & 0.571 & 0.788 & 0.998 & 0.950 \\
				0.2   & 1.5   &       & 0.998 & 0.900 & 0.630 & 0.805 & 0.993 & 0.955 &       & 0.996 & 0.942 & 0.750 & 0.792 & 0.998 & 0.948 \\
				0.5   & 1     &       & 0.977 & 0.792 & 0.568 & 0.792 & 1.000 & 0.963 &       & 0.975 & 0.879 & 0.600 & 0.850 & 1.000 & 0.948 \\
				0.5   & 1.2   &       & 0.997 & 0.800 & 0.435 & 0.507 & 1.000 & 0.963 &       & 0.996 & 0.840 & 0.450 & 0.631 & 1.000 & 0.950 \\
				0.5   & 1.5   &       & 1.000 & 0.797 & 0.405 & 0.552 & 0.993 & 0.965 &       & 1.000 & 0.904 & 0.537 & 0.619 & 1.000 & 0.950 \\
				\bottomrule
			\end{tabular}%
		\end{table}%
	\end{landscape}

	\subsubsection{Comparison in the favorable situation}
	
	To put the methods on equal footing, we consider the favorable setting with no unmeasured confounders and no generalizability bias. Regarding the covariate supports for the RCT and OS, we  either let the two supports be equal or let the RCT have a slightly larger support than the OS support. 
	
	More specifically, under the situation where the OS and RCT covariate supports are equal, we consider the following data-generating mechanism. There are five observed covariates $(X_1,\dots, X_5)$ independently distributed and each following the standard normal distribution. The potential outcome under control $Y(0)=10+4X_1-2X_2+3X_5+\epsilon$, where $\epsilon\sim N(0,1)$ and under treatment is $Y(1)=Y(0)$. The probability of selecting into the RCT is $expit(-1.5+0.1X_1+0.1X_2-0.3X_4)$ for any covariate value. Once selected into either the RCT or OS, the probability of being assigned to treatment is $1/2$ in the RCT and $expit(-2-0.3X_1+0.1X_3-0.2X_5)$ in the OS. Our simulation study creates several data-generating models by varying the total sample size $N$. 
	
	Alternatively, under the situation where the RCT support is bigger than the OS support, we consider the following data-generating mechanism. There are five observed covariates $(X_1,\dots, X_5, X_6)$ independently distributed, the first five following the standard normal distribution and $X_6$ is sampled as Bernoulli(0.10).  When $X_6$ is 1, the unit is placed into the RCT. Thus, this part of RCT, about 10\% of the total sample size, is outside of OS support. If $X_6$ is 0, the probability of selecting into the RCT is $expit(-1.5+0.1X_1+0.1X_2-0.3X_4)$ for any covariate value. Once selected into either the RCT or OS, the probability of being assigned to treatment is $1/2$ in the RCT and $expit(-2-0.3X_1+0.1X_3-0.2X_5)$ in the OS.
	The potential outcome under control $Y(0)=10+4X_1-2X_2+3X_5+2X_6+\epsilon$, where $\epsilon\sim N(0,1)$ and under treatment is $Y(1)=Y(0)$. Our simulation study creates several data-generating models by varying the total sample size $N$.

	The simulated MSE values are reported in Table \ref{tab:favorable}. We observe in this most simplified setting, where all methods are expected to perform well, that all methods indeed show decreasing MSE with increasing sample sizes. Still, the proposed method has the smallest MSE, while classical methods have notably large MSEs.

	\begin{table}[htbp]
		\centering
		\caption{Mean squared errors for estimating the ATOT comparing the proposed combined method with other competing state-of-the-art and classical methods in the favorable case of no unmeasured biases and no generalizability bias. The RCT support within the OS support shares 100\% of the OS's covariate support.}
		\begin{tabular}{r|ccccc}
			\toprule
			\multicolumn{6}{c}{RCT completely inside OS support} \\
			\midrule
			\multicolumn{1}{l|}{$N$} & \multicolumn{1}{l}{$\Gamma=\Gamma^\star=1, \delta=\delta^\star=0$} & \multicolumn{1}{l}{Elastic} & \multicolumn{1}{l}{R-Learner} & \multicolumn{1}{l}{Weighting} & \multicolumn{1}{l}{PS} \\
			\cmidrule{1-6}    2000  & 0.0023 & 0.025 & 0.008 & 0.5942 & 1.3791 \\
			3000  & 0.0014 & 0.016 & 0.004 & 0.3938 & 0.8158 \\
			5000  & 0.0008 & 0.008 & 0.002 & 0.1838 & 0.5570 \\
			\midrule
			\multicolumn{6}{c}{A fraction ($\sim$40\%) of RCT is outside OS support} \\
			\midrule
			\multicolumn{1}{l|}{$N$} & \multicolumn{1}{l}{$\Gamma=\Gamma^\star=1, \delta=\delta^\star=0$} & \multicolumn{1}{l}{Elastic} & \multicolumn{1}{l}{R-Learner} & \multicolumn{1}{l}{Weighting} & \multicolumn{1}{l}{PS} \\
			\cmidrule{1-6}    2000  & 0.0024 & 0.019 & 0.006 & 0.3291 & 1.1010 \\
			3000  & 0.0018 & 0.009 & 0.003 & 0.1724 & 0.7100 \\
			5000  & 0.0011 & 0.006 & 0.003 & 0.1196 & 0.4220 \\
			\bottomrule
		\end{tabular}%
		\label{tab:favorable}%
	\end{table}%
	\color{black}
	
	\section{Proofs of the technical results}

	\subsection{\bf Proof of Theorem 1.} 
	
	Let $\tau_{ij} = Y_{ij}^o(1) - Y_{ij}^o(0)$.

	\begin{assumption}\label{assm_thm1}
		
		* $\tau_{ij}\geq -M$ for some constant $M$. 
		
		* The strata are independent across.
		
		* $\frac{2}{I^2}\sum_i \sum_j Y_{ij}^o(0)^2 + \frac{2}{I^2}\sum_i \sum_j \tau_{ij}^2  \rightarrow 0$ almost surely.
		
		
		* $\frac{1}{I^2} \sum_i Var( \hat\tau_i^2)\rightarrow 0$, $\lim_{I\rightarrow\infty}\frac{1}{I} \sum_i E( \hat\tau_i) <\infty$, $\lim_{I\rightarrow\infty}\frac{1}{I}\sum_i Var(\hat{\tau}_i) \in (0,\infty)$, and $\lim_{I\rightarrow\infty}\frac{1}{I}\sum_i (\hat{\tau}_i-E\hat{\tau}_i)^2$ converges almost surely to a random variable with finite expectation.
		
	\end{assumption}
	
	{\bf Note:} $\lim_{I\rightarrow\infty}\frac{1}{I} \sum_i E( \hat\tau_i) <\infty$ is implied by an assumption $\lim_{I\rightarrow\infty}\frac{1}{I}\sum_i\sum_j E|Y_{ij}^o| <\infty$. 
	
	$\lim_{I\rightarrow\infty}\frac{1}{I}\sum_i Var(\hat{\tau}_i) \in (0,\infty)$ is implied by an assumption $\lim_{I\rightarrow\infty}\frac{1}{I}\sum_i J_i E( {Y_{ij}^o}^2) \in (0,\infty)$. 
	
	$\frac{1}{I^2} \sum_i Var( \hat\tau_i^2)\rightarrow 0$ is implied by an assumption $\lim_{I\rightarrow\infty} \frac{1}{I^2} \sum_i J_i^3\sum_j E {Y_{ij}^o}^4 = 0$.
	
	Using Kolmogorov's SLLN, if $\lim_{I\rightarrow\infty} \sum_i \frac{J_i^3}{i^2}\sum_j E{ Y_{ij}^o}^4 < \infty$ and $\lim_{I\rightarrow\infty}\frac{1}{I}\sum_i J_i E( {Y_{ij}^o}^2) <\infty$,  we have the final assumption that $\lim_{I\rightarrow\infty}\frac{1}{I}\sum_i (\hat{\tau}_i-E\hat{\tau}_i)^2$ converges almost surely to a random variable with finite expectation.
	
	---------------------------

	The $\tilde{\tau}_i$ statistic is unchanged if we redefine $Y_{ij}^o = Y_{ij}^o + MZ_{ij}^o$ and $\beta_0^\star=\beta_0^\star+M$. Thus, without loss of generality, assume $\tau_{ij}\geq 0$ and $\beta_0^\star\geq 0$.
	To simplify the notation write $\tilde\tau_i^{(\beta_0^\star)}$ as $\tilde\tau_i$.
	
	We first show that 
	$$\Pr\left\{ I\times se^2\left(\frac{1}{I} \sum_i \tilde{\tau}_i\right) \geq \frac{1}{I}Var\left( \sum_i \tilde{\tau}_i \right) \right\}\rightarrow 1.$$
	
	See that $I\times se^2\left(\frac{1}{I} \sum_i \tilde{\tau}_i\right) = \frac{1}{I-1}\sum_i (\tilde{\tau}_i - \bar{\tilde{\tau}})^2$. Also, $Var( \sum_i \tilde{\tau}_i ) = \sum_i \{E(\tilde{\tau}_i^2) - [ E \tilde{\tau}_i ]^2\}$. Thus,
	\begin{align*}
		& I\times se^2\left(\frac{1}{I} \sum_i \tilde{\tau}_i\right) -  \frac{1}{I}Var\left( \sum_i \tilde{\tau}_i \right)\\
		= & \frac{1}{I-1}\sum_i \tilde\tau_i^2 - \frac{I}{I-1} \bar{\tilde{\tau}}^2 - \frac{1}{I}\sum_i \{E(\tilde{\tau}_i^2) - [ E \tilde{\tau}_i ]^2\}\\
		\geq & \frac{1}{I-1}\sum_i (\tilde\tau_i^2 - E\tilde\tau_i^2) - \frac{I}{I-1}(\bar{\tilde{\tau}}^2 - (\frac{1}{I} \sum_i E \tilde\tau_i)^2) - \frac{I}{I-1}(\frac{1}{I} \sum_i E \tilde\tau_i)^2 + \frac{1}{I}\sum_i [ E \tilde{\tau}_i ]^2\\
		\geq & \frac{1}{I-1}\sum_i (\tilde\tau_i^2 - E\tilde\tau_i^2) - \frac{I}{I-1}(\bar{\tilde{\tau}}^2 - (\frac{1}{I} \sum_i E \tilde\tau_i)^2) + \left\{ 1 - \frac{I}{I-1}\right\}(\frac{1}{I} \sum_i E \tilde\tau_i)^2.
	\end{align*}
	It suffices to show that the three terms go to 0 in probability. The last term is obvious. For the first two terms, use Chebyshev's inequality. Since they have mean zero, it suffices to show that the variances of the terms go to zero as $I$ goes to infinity.
	
	For the first term,
	$$Var\left\{\frac{1}{I-1}\sum_i (\tilde\tau_i^2 - E\tilde\tau_i^2)\right\}
	\leq  \frac{1}{(I-1)^2} \sum_i Var(\tilde\tau_i^2).$$
	Similarly, for the second term,
	\begin{align*}
		& Var\left( \bar{\tilde{\tau}}^2 \right)\\
		= & \frac{1}{I^2} Var\left\{ \left(\sum_i \tilde\tau_i \right)^2 \right\}\\
		= & \frac{1}{I^2} \sum_i E \tilde\tau_i^2 + \frac{1}{I^2} \sum_{i\neq j} E \tilde\tau_i E \tilde\tau_j - \frac{1}{I^2}\left[E\sum_i \tilde\tau_i\right]^2\\
		= & \frac{1}{I^2} \sum_i E \tilde\tau_i^2 + \frac{1}{I^2}\left[ \sum_i E\tilde\tau_i \right]^2 - \frac{1}{I^2}\left(\sum_i E \tilde\tau_i\right)^2 - \frac{1}{I^2}\left[E\sum_i \tilde\tau_i\right]^2\\
		= & \frac{1}{I^2} \sum_i E \tilde\tau_i^2 - \frac{1}{I^2}\left(\sum_i E \tilde\tau_i\right)^2 = \frac{1}{I^2} \sum_i Var( \tilde\tau_i).
	\end{align*}
	
	Thus, we have proved as $I\rightarrow\infty$
	$$\Pr\left\{ I\times se^2\left(\frac{1}{I} \sum_i \tilde{\tau}_i\right) \geq \frac{1}{I}Var\left( \sum_i \tilde{\tau}_i \right) \right\}\rightarrow 1.$$
	
	Next, we show that $\tilde\tau_i^{(\beta_0^\star)}\leq 0$.
	We simplify the notation to write  $ \widetilde\eta_{ij}^{(\beta_0^\star)}$ as $\eta_{ij}$.
	
	Recall $\hat{\tau_i}$ 
	$$\hat{\tau_i} = \sum_j Z_{ij}^o(Y_{ij}^o-Z_{ij}^o\beta_0^\star) - \frac{1}{J_i-1}\sum_j (1-Z_{ij}^o)(Y_{ij}^o-Z_{ij}^o\beta_0^\star),$$
	and $$\overline{\hat{\tau}} = \frac{1}{I}\sum_i c_i\hat{\tau}_i.$$
	For our purpose $c_i=1$ throughout, but one may choose some other coefficients, e.g., $c_i=1/J_i$.
	
	Some calculations show $$\overline{\hat{\tau}} = \frac{1}{I}\sum_i c_i \sum_j Y_{ij}^o(0)Z_{ij}^o \frac{J_i}{J_i-1} - \frac{1}{I}\sum_i c_i
	\frac{1}{J_i-1}\sum_j Y_{ij}^o(0) - \beta_0^\star\bar{c} + \frac{1}{I}\sum_ic_i\sum_j Z_{ij}^o\tau_{ij}.$$
	
	Let $$E_\tau = \frac{1}{I}\sum_i c_i \sum_j Y_{ij}^o(0)\eta_{ij}\frac{J_i}{J_i-1} - \frac{1}{I}\sum_i c_i \frac{1}{J_i-1}\sum_j Y_{ij}^o(0) - \beta\bar{c} + \frac{1}{I}\sum_ic_i\sum_j \eta_{ij}\tau_{ij}.$$
	Subtracting the two,
	$$   \overline{\hat{\tau}} - E_\tau  =
	\frac{1}{I}\sum_i c_i \sum_j Y_{ij}^o(0)(Z_{ij}^o-\eta_{ij})\frac{J_i}{J_i-1} + \frac{1}{I}\sum_ic_i\sum_j (Z_{ij}^o - \eta_{ij})\tau_{ij}.$$
	
	Let $\eta'_{ij} = E(Z_{ij}^o\mid \mathcal{F})$.
	
	Let's use Chebyshev to say that we can approximate the above $\overline{\tilde{\tau}} - E_\tau$ by 
	\begin{equation*}\label{eq_approx}
		\frac{1}{I}\sum_i c_i \sum_j Y_{ij}^o(0)(\eta'_{ij}-\eta_{ij})\frac{J_i}{J_i-1} + \frac{1}{I}\sum_ic_i\sum_j (\eta'_{ij} - \eta_{ij})\tau_{ij}.    
	\end{equation*}
	The approximation follows since
	\begin{align*}
		&  Var\{\overline{\hat{\tau}} - E_\tau\mid \mathcal{F}\}\\
		& \leq \frac{1}{I^2}\sum_i \frac{c_i^2}{(J_i-1)^2} \sum_j (J_i Y_{ij}^o(0)+(J_i-1)\tau_{ij})^2 \eta'_{ij}(1-\eta'_{ij})\\
		& \leq \frac{1}{I^2}\sum_i \frac{c_i^2}{(J_i-1)^2}\sum_j 2(J_i^2Y_{ij}^o(0)^2 + (J_i-1)^2\tau_{ij}^2)\frac{1}{4}\\
		& \leq \frac{2}{I^2}\sum_i c_i^2  \sum_j Y_{ij}^o(0)^2 + \frac{2}{I^2}\sum_i c_i^2  \sum_j \tau_{ij}^2  \rightarrow 0.
	\end{align*}
	The limit follows from our assumptions.
	
	Since $\eta_{ij}$ maximizes $E(\bar{\hat{\tau}})$ for large enough $I$, \eqref{eq_approx} is less than or equal to 0. Thus, we have with probability going to 1, $\overline{\hat{\tau}} - E_\tau \leq 0$.

	However, $E_\tau$ cannot be calculated from the observed data. Consider instead,
	$$E_\tau' =  \frac{1}{I}\sum_i c_i\sum_j \eta_{ij} (Y_{ij}^o-Z_{ij}^o\beta) - \frac{1}{I}\sum_i c_i\frac{1}{J_i-1} \sum_j (1-\eta_{ij}) (Y_{ij}^o-Z_{ij}^o\beta_0^\star).$$
	
	\begin{align*}
		E_\tau' & =  \frac{1}{I}\sum_i c_i\sum_j \eta_{ij} (Y_{ij}^o-Z_{ij}^o\beta_0^\star) - \frac{1}{I}\sum_i c_i\frac{1}{J_i-1} \sum_j (1-\eta_{ij}) (Y_{ij}^o-Z_{ij}^o\beta_0^\star)\\
		& = \frac{1}{I}\sum_i c_i\left\{ \sum_j \eta_{ij}Z_{ij}^o(\tau_{ij}-\beta_0^\star) + \sum_j \eta_{ij}Y_{ij}^o(0) - \frac{1}{J_i-1}\sum_j(1-\eta_{ij})Z_{ij}^o(\tau_{ij}-\beta_0^\star)\right.\\
		& \ \ \ \left.-  \frac{1}{J_i-1}\sum_j(1-\eta_{ij})Y_{ij}^o(0)\right\}
	\end{align*}
	Thus,
	\begin{align*}
		E_\tau - E_\tau' & = -\beta_0^\star\bar{c} + \frac{1}{I}\sum_i c_i\sum_j \eta_{ij} \tau_{ij}- \frac{1}{I}\sum_i c_i\sum_j \eta_{ij} (\tau_{ij}-Z_{ij}^o\beta_0^\star) + \frac{1}{I}\sum_i c_i\frac{1}{J_i-1}\sum_j(1-\eta_{ij})Z_{ij}^o(\tau_{ij}-\beta_0^\star)\\ 
		& =  -\beta_0^\star\bar{c} - \frac{1}{I}\sum_i c_i\frac{1}{J_i-1}\sum_j \tau_{ij} (1-\eta_{ij})Z_{ij}^o + \beta_0^\star\frac{1}{I}\sum_i c_i\sum_j \{1-(1-\eta_{ij})/(J_i-1)\}Z_{ij}^o\\
		& = -\beta_0^\star\bar{c} + \beta_0^\star\frac{1}{I}\sum_i c_i \{1-1/(J_i-1)+\sum_j\eta_{ij}Z_{ij}^o/(J_i-1)\} - \frac{1}{I}\sum_i c_i\frac{1}{J_i-1}\sum_j \tau_{ij} (1-\eta_{ij})Z_{ij}^o\\
		& =  - \beta_0^\star\frac{1}{I}\sum_i c_i/(J_i-1)+\beta_0^\star\frac{1}{I}\sum_i c_i/(J_i-1)\sum_j\eta_{ij}Z_{ij}^o - \frac{1}{I}\sum_i c_i\frac{1}{J_i-1}\sum_j \tau_{ij} (1-\eta_{ij})Z_{ij}^o\\
		& \leq - \beta_0^\star\frac{1}{I}\sum_i c_i/(J_i-1)+\beta_0^\star\frac{1}{I}\sum_i c_i/(J_i-1) \\
		& \quad\quad \text{since, }\sum_jZ_{ij}^o\eta_{ij} \leq \sqrt{\sum_j(Z_{ij}^o)^2}\sqrt{\sum_j\eta_{ij}^2} \leq \sqrt{\sum_jZ_{ij}^o}\sqrt{\sum_j\eta_{ij}} = 1\\
		& = 0.
	\end{align*}
	In the above we have used that $\tau_{ij}\geq 0$ and $\beta_0^\star\geq 0$. 
	
	Thus, with probability going to 1,
	$$\overline{\hat\tau} - E_\tau' = \overline{\hat\tau} - E_\tau + E_\tau - E_\tau' \leq 0.$$
	Putting things together,   with probability going to 1,
	\begin{align*}
		& \overline{\hat\tau} - E_\tau'  \leq \overline{\hat\tau} - E(\overline{\hat\tau})\\
		\Rightarrow& \frac{\overline{\hat\tau} - E_\tau'}{\{Var(\frac{1}{I} \sum_i \tilde{\tau}_i)\}^{1/2}}  \leq \frac{\overline{\hat\tau} - E(\overline{\hat\tau})}{\{Var(\frac{1}{I} \sum_i \tilde{\tau}_i)\}^{1/2}}\\
		\Rightarrow& \frac{\overline{\hat\tau} - E_\tau'}{se\left(\frac{1}{I}\sum_i \tilde{\tau}_i\right)}\frac{se\left(\frac{1}{I}\sum_i \tilde{\tau}_i\right)}{\{Var(\frac{1}{I} \sum_i \tilde{\tau}_i)\}^{1/2}}  \leq \frac{\overline{\hat\tau} - E(\overline{\hat\tau})}{\{Var(\frac{1}{I} \sum_i \tilde{\tau}_i)\}^{1/2}}.
	\end{align*}
	Note, $\tilde{\tau}_i = \hat{\tau}_i - \left\{\sum_j \eta_{ij} (Y_{ij}^o-Z_{ij}^o\beta_0^\star) -\frac{1}{J_i-1} \sum_j (1-\eta_{ij}) (Y_{ij}^o-Z_{ij}^o\beta_0^\star)\right\}$.
	Recall, $c_i=1$ for all $i$; so, $\overline{\hat{\tau}} = \frac{1}{I}\sum_i \hat{\tau}_i$, and $E_\tau' = \frac{1}{I}\sum_i \left\{\sum_j \eta_{ij} (Y_{ij}^o-Z_{ij}^o\beta_0^\star) -\frac{1}{J_i-1} \sum_j (1-\eta_{ij}) (Y_{ij}^o-Z_{ij}^o\beta_0^\star)\right\}$. So, $\frac{1}{I}\sum_i \tilde{\tau}_i = \overline{\hat\tau} - E_\tau'$.
	
	Take $t\geq 0$.
	\begin{align*}
		&\Pr\left\{\frac{\overline{\hat\tau} - E_\tau'}{se\left(\frac{1}{I}\sum_i \tilde{\tau}_i\right)} \geq t\right\}\\
		=& \Pr\left\{\frac{\overline{\hat\tau} - E_\tau'}{se\left(\frac{1}{I}\sum_i \tilde{\tau}_i\right)}\frac{se\left(\frac{1}{I}\sum_i \tilde{\tau}_i\right)}{\{Var(\frac{1}{I} \sum_i \tilde{\tau}_i)\}^{1/2}} \geq t\frac{se\left(\frac{1}{I}\sum_i \tilde{\tau}_i\right)}{\{Var(\frac{1}{I} \sum_i \tilde{\tau}_i)\}^{1/2}}\right\}\\
		\leq & \Pr\left\{\frac{\overline{\hat\tau} - E(\overline{\hat\tau})}{\{Var(\frac{1}{I} \sum_i \tilde{\tau}_i)\}^{1/2}} \geq t\frac{se\left(\frac{1}{I}\sum_i \tilde{\tau}_i\right)}{\{Var(\frac{1}{I} \sum_i \tilde{\tau}_i)\}^{1/2}}\mid A_I^c\right\}\Pr(A_I^c) + \Pr(A_I)\\
		\leq & \Pr\left\{\frac{\overline{\hat\tau} - E(\overline{\hat\tau})}{\{Var(\frac{1}{I} \sum_i \tilde{\tau}_i)\}^{1/2}} \geq t\mid A_I^c, B_I^c\right\}\Pr(A_I^c)\Pr(B_I^c) + \Pr(A_I) + \Pr(B_I).
	\end{align*}
	Where $A_I=\{ \overline{\hat\tau} - E_\tau' > 0\}$ and $B_I=\{ {se\left(\frac{1}{I}\sum_i \tilde{\tau}_i\right)}< {\{Var(\frac{1}{I} \sum_i \tilde{\tau}_i)\}^{1/2}}\}$; $A_I^c$, $B_I^c$ are complements of these events. Note $\Pr(A_I)\rightarrow 0$ and $\Pr(B_I)\rightarrow 0$. Now,
	
	\begin{align*}
		& \Pr\left\{\frac{\overline{\hat\tau} - E(\overline{\hat\tau})}{\{Var(\frac{1}{I} \sum_i \tilde{\tau}_i)\}^{1/2}} \geq t\mid A_I^c, B_I^c\right\}\\
		= &  
		\Pr\left\{\frac{\overline{\hat\tau} - E(\overline{\hat\tau})}{\{Var(\frac{1}{I} \sum_i \tilde{\tau}_i)\}^{1/2}} \geq t\right\}\\
		& \quad - \Pr\left\{\frac{\overline{\hat\tau} - E(\overline{\hat\tau})}{\{Var(\frac{1}{I} \sum_i \tilde{\tau}_i)\}^{1/2}} \geq t\mid A_I \text{ or } B_I\right\}\Pr(A_I \text{ or } B_I)/\Pr(A_I^c, B_I^c).
	\end{align*}
	Since, $\Pr(A_I \text{ or } B_I) \leq \Pr(A_I) + \Pr( B_I) \rightarrow 0$ and $\Pr(A_I^c , B_I^c) \geq \Pr(A_I^c) + \Pr( B_I^c) - 1 \rightarrow 1$, we have,
	$$\lim\sup_{I\rightarrow \infty} \Pr\left\{\frac{\overline{\hat\tau} - E_\tau'}{se\left(\frac{1}{I}\sum_i \tilde{\tau}_i\right)} \geq t\right\} \leq \lim_{I\rightarrow \infty}\Pr\left\{\frac{\overline{\hat\tau} - E(\overline{\hat\tau})}{\{Var(\frac{1}{I} \sum_i \tilde{\tau}_i)\}^{1/2}} \geq t\right\}=1-\Phi(t).$$
	The last equality is by 
	using Lindeberg's CLT, as we establish below.
	
	Let $\sigma_i^2 = var(\hat{\tau}_i)$ and $s_I^2:=\sum_i \sigma_i^2$. By our assumption $\frac{1}{I}s_I^2$ converges almost surely to a positive number. Thus, $s_I\rightarrow\infty$ in probability. Thus, to establish Lindeberg's condition, it is enough to show that 
	$$\lim_{I\rightarrow\infty} \frac{1}{I}\sum_i E\left\{(\hat{\tau}_i - E(\hat{\tau}_i))^2\times I(| \hat{\tau}_i - E(\hat{\tau}_i)| >\epsilon s_I) \right\}=0.$$
	Write 
	$$\frac{1}{I}\sum_i E\left\{(\hat{\tau}_i - E(\hat{\tau}_i))^2\times I(| \hat{\tau}_i - E(\hat{\tau}_i)| >\epsilon s_I) \right\}
	= E\left\{\frac{1}{I}\sum_i (\hat{\tau}_i - E(\hat{\tau}_i))^2\times I(| \hat{\tau}_i - E(\hat{\tau}_i)| >\epsilon s_I) \right\}.$$
	We want to interchange the limit and expectation. We use a general version of the dominated convergence theorem. We check the conditions using our assumption. First, 
	$\frac{1}{I}\sum_i (\hat{\tau}_i - E(\hat{\tau}_i))^2\times I(| \hat{\tau}_i - E(\hat{\tau}_i)| >\epsilon s_I)\leq \frac{1}{I}\sum_i (\hat{\tau}_i - E(\hat{\tau}_i))^2$. Next, $E(\frac{1}{I}\sum_i (\hat{\tau}_i - E(\hat{\tau}_i))^2)= \frac{1}{I}\sum_i var(\hat{\tau}_i)$. And its limit is finite. Finally, $\frac{1}{I}\sum_i (\hat{\tau}_i - E(\hat{\tau}_i))^2$ converges almost surely to a random variable with finite expectation. Then, the general version of the dominated convergence theorem applies. 
	
	So, suffices to show $\lim_{I\rightarrow\infty}\frac{1}{I}\sum_i (\hat{\tau}_i - E(\hat{\tau}_i))^2\times I(| \hat{\tau}_i - E(\hat{\tau}_i)| >\epsilon s_I)$ goes to zero almost surely.
	
	To see this note that for any $M>0$, for large enough $I$, $\lim_{I\rightarrow\infty}\frac{1}{I}\sum_i (\hat{\tau}_i - E(\hat{\tau}_i))^2\times I(| \hat{\tau}_i - E(\hat{\tau}_i)| >\epsilon s_I) \leq \lim_{I\rightarrow\infty}\frac{1}{I}\sum_i (\hat{\tau}_i - E(\hat{\tau}_i))^2\times I(| \hat{\tau}_i - E(\hat{\tau}_i)| >M).$ Thus, $\lim_{I\rightarrow\infty}\frac{1}{I}\sum_i (\hat{\tau}_i - E(\hat{\tau}_i))^2\times I(| \hat{\tau}_i - E(\hat{\tau}_i)| >\epsilon s_I) \leq \lim_{M\rightarrow \infty}\lim_{I\rightarrow\infty}\frac{1}{I}\sum_i (\hat{\tau}_i - E(\hat{\tau}_i))^2\times I(| \hat{\tau}_i - E(\hat{\tau}_i)| >M).$ Since $\frac{1}{I}\sum_i (\hat{\tau}_i - E(\hat{\tau}_i))^2\times I(| \hat{\tau}_i - E(\hat{\tau}_i)| >M)$ is monotone in $M$, we can interchange the two limits. Thus, look at $\lim_{I\rightarrow\infty}\lim_{M\rightarrow\infty}\frac{1}{I}\sum_i (\hat{\tau}_i - E(\hat{\tau}_i))^2\times I(| \hat{\tau}_i - E(\hat{\tau}_i)| >M)$; which is zero. Thus, we have checked the Lindeberg condition for the CLT of the average of the $\hat{\tau}_i$s.
	
	Thus, in the theorem's original notation, for $t>0$,
	$$\lim\sup_{I\rightarrow \infty} 
	\Pr\left\{\frac{\frac{1}{I}\sum_i{\tilde\tau}_i^{(\beta_0^\star)}}{se\left(\frac{1}{I}\sum_i \tilde\tau_i^{(\beta_0^\star)}\right)} \geq t\right\} \leq 1-\Phi(t).$$
	Hence, we get an asymptotically valid confidence interval for the ATOT.\hfill Q.E.D.
	
	\medskip
	
	Throughout the proofs of Theorems 2 and 3, we refer to the result in the second paragraph of page 279, Section 19.4 of \cite{vandervaart1998asymptotic} as the GC class theorem and to Lemma 19.24 (See page 280, Section 19.2) of \cite{vandervaart1998asymptotic} as Donsker's theorem. These may be abuses of nomenclature, as there are other theorems with such names, but they simplify our presentation. 
	
	\subsection{\bf Proof of Theorem 2.}
	
	Write $C_m = \hat{\nu}(X_m)$, where $\hat{\nu}$ is estimated from the data. Notice that by construction $\hat{\nu}(x)=0$ for $x\not\in \mathcal{X}$.
	
	Let $\mathbf{X}=\{X_m^r:m=1,\ldots,n_r\}$.
	
	\begin{assumption}\label{assm_thm2}

		* $\hat{\nu}$ is in a Glivenko-Cantelli (GC) class, $\hat{\nu}(x)\rightarrow\nu(x)$ almost surely for all $x$ and the functions are bounded. Here and later,
		$$\nu(x) = \Pr(Z_l^o = 1 \mid X_l^o=x)\times \Pr(S_k=0\mid X_k=x)/\Pr(S_k=1\mid X_k=x).$$

		
		
		* $\max\{\tilde{\theta}_m, 1-\tilde{\theta}_m\}\leq \delta_{n_r}$ for all $m$. $\delta_{n_r}\rightarrow \delta \in (0,1)$.
		
		* $E(Y_m^r(z)^2)< \infty$ and $E[E(Y_m^r(z)\mid X_m)^2]< \infty$  for $z=0,1$.
		
		*  Let $\tilde{\theta}_{m,m'} = cov(Z_m^r, Z_{m'}^r\mid \mathbf{X})$. Assume $A_{n_r}\subseteq \{1,\ldots, n_r\}^2$ so that for $m,m'\in A_{n_r}$, $m\neq m'$ and ,  $\tilde{\theta}_{m,m'}\leq g(n_r)$, for some function $g$ with $g(n_r)\rightarrow 0$.
		
		*Also, $\frac{1}{n_r^2}\sum_{m\neq m'\not\in  A_{n_r}}E(|Y_m^r(1)|+|Y_m^r(0)|\mid \mathbf{X}) E(|Y_{m'}^r(1)|+|Y_{m'}^r(0)|\mid \mathbf{X}) \rightarrow 0$ almost surely.
		
		
		* Assume $S_l \perp (Y_l(1), Y_l(0)) \mid X_l$ and $(Y_l(1) - Y_l(0)) \perp Z_l^o \mid X_l, S_l=0$.

	\end{assumption}
	
	---------------------------

	Note that $\sum_m C_m/n_r$ converges almost surely to $E(\nu(X_m^r))$ by GC class theorem and our assumptions.  write,
	$$\frac{1}{{n_r}}\sum_m C_m \widehat{\beta^r_\chi}= \frac{1}{{n_r}}\sum_m C_m\left\{ \frac{Z^r_mY^r_m}{\tilde\theta_{m}} - \frac{(1-Z^r_m)Y^r_m}{1-\tilde\theta_{m}} \right\}.$$

	Given covariate data $X$, consider the conditional variance of the term. We show that it goes to zero. 
	
	Let $f_m:=f(Y_m(1), Y_m(0)) = \frac{Z^r_mY^r_m}{\tilde\theta_{m}} - \frac{(1-Z^r_m)Y^r_m}{1-\tilde\theta_{m}}$. Let $\mathbf{X}=\{X_m^r:m=1,\ldots,n_r\}$. Using the fact that $Z_m$s are independent of the potential outcomes given the covariates, the variance is
	\begin{align*}
		& \frac{1}{n_r^2}\sum_m C_m^2 \left\{\tilde{\theta}_m E(f_m^2\mid X) - \tilde{\theta}_m^2[E(f_m\mid \mathbf{X})]^2\right\}
		+ \frac{1}{n_r^2}\sum_{m\neq m'} C_mC_{m'}E(f_m f_{m'}\mid \mathbf{X})\tilde{\theta}_{m,m'}\\
		\leq & \frac{K^2}{n_r^2}\sum_m  \left\{ \frac{2}{\delta_{n_r}^2}E(Y_m^r(1)^2+Y_m^r(0)^2\mid \mathbf{X}) + \frac{2}{\delta_{n_r}^2}[E(|Y_m^r(1)|\mid X)^2+E(|Y_m^r(0)|\mid \mathbf{X})^2]\right\}\\
		& + \frac{K^2}{\delta_{n_r}^2n_r^2}\sum_{m\neq m'} E(|Y_m^r(1)|+|Y_m^r(0)|\mid \mathbf{X}) E(|Y_{m'}^r(1)|+|Y_{m'}^r(0)|\mid \mathbf{X})|\tilde{\theta}_{m,m'}|,
	\end{align*}
	where $K$ is the upper bound for the $C_m$s.
	
	The first term goes to zero by strong law of large number and our finite moment assumptions.
	
	For the second term
	\begin{align*}
		& \frac{K^2}{\delta_{n_r}^2n_r^2}\sum_{m\neq m'} E(|Y_m^r(1)|+|Y_m^r(0)|\mid \mathbf{X}) E(|Y_{m'}^r(1)|+|Y_{m'}^r(0)|\mid \mathbf{X})|\tilde{\theta}_{m,m'}|\\
		\leq & \frac{K^2g(n_r)}{\delta_{n_r}^2n_r^2}\sum_{m\neq m'\in  A_{n_r}} E(|Y_m^r(1)|+|Y_m^r(0)|\mid \mathbf{X}) E(|Y_{m'}^r(1)|+|Y_{m'}^r(0)|\mid \mathbf{X})\\
		& + \frac{K^2\times 1}{\delta_{n_r}^2n_r^2}\sum_{m\neq m'\not\in  A_{n_r}} E(|Y_m^r(1)|+|Y_m^r(0)|\mid \mathbf{X}) E(|Y_{m'}^r(1)|+|Y_{m'}^r(0)|\mid \mathbf{X})\\
		= & \frac{K^2g(n_r)}{\delta_{n_r}^2}\left(\frac{1}{n_r}\sum_{m} E(|Y_m^r(1)|+|Y_m^r(0)|\mid \mathbf{X}) \right)^2 - \frac{K^2g(n_r)}{\delta_{n_r}^2n_r^2}\sum_{m} E(|Y_m^r(1)|+|Y_m^r(0)|\mid \mathbf{X})^2\\
		& + \frac{K^2\times (1-g(n_r))}{\delta_{n_r}^2n_r^2}\sum_{m\neq m'\not\in  A_{n_r}} E(|Y_m^r(1)|+|Y_m^r(0)|\mid \mathbf{X}) E(|Y_{m'}^r(1)|+|Y_{m'}^r(0)|\mid \mathbf{X}).
	\end{align*}
	Use the strong law of large numbers for the averages in the first and the second terms. Then, by our assumptions, all three terms go to zero.
	
	So we can study the in-probability limit 
	$$\frac{1}{n_r}\sum_m C_m E\left[Y_m^r(1)-Y_m^r(0)\mid X_m^r\right].$$
	
	By the GC class theorem, the almost sure limit of this quantity is $E\left\{\nu(X_m^r)\left[ E(Y^r_m(1)\mid X_m^r) - E(Y^r_m(0)\mid X_m^r)\right]\right\}$.

	It remains to show:
	$$E\left\{\nu(X_m^r)\left[ E(Y^r_m(1)\mid X_m^r) - E(Y^r_m(0)\mid X_m^r)\right]\right\} = E\{\nu(X_m^r)\}E[Y_l^o(1) - Y_l^o(0) \mid X_l^o\in \mathcal{X}, Z_l^o=1].$$
	Start with the LHS
	\begin{align*}
		& E\left\{\nu(X_m^r)\left[ E(Y^r_m(1)\mid X_m^r) - E(Y^r_m(0)\mid X_m^r)\right]\right\}\\
		= & \int_\mathcal{X} \frac{\Pr(Z_l^o=1, S_l=0\mid X_l=x)}{\Pr(S_l=1\mid X_l=x)} E(Y_l(1) - Y_l(0)\mid X_l=x, S_l=1)f_{X_l\mid S_l=1}(x)\, dx\\
		= & \frac{\Pr(S_l=0, Z_l^o=1)}{\Pr(S_l=1)}\int_\mathcal{X}
		\frac{f_{X_l\mid S_l=0, Z_l^o=1}(x)}{
			f_{X_l\mid S_l=1}(x)} E(Y_l(1) -Y_l(0)\mid X_l=x, S_l=1)f_{X_l\mid S_l=1}(x)\, dx\\
		= & \frac{\Pr(S_l=0, Z_l^o=1)}{\Pr(S_l=1)}\int_\mathcal{X} E(Y_l(1) - Y_l(0)\mid X_l=x, S_l=1)f_{X_l\mid S_l=0, Z_l^o=1}(x)\, dx\\
		= & \frac{\Pr(S_l=0, Z_l^o=1)}{\Pr(S_l=1)}\int_\mathcal{X} E(Y_l(1) - Y_l(0)\mid X_l=x, S_l=0)f_{X_l\mid S_l=0, Z_l^o=1}(x)\, dx\\
		= & \frac{\Pr(S_l=0, Z_l^o=1)}{\Pr(S_l=1)}\int_\mathcal{X} E(Y_l(1) - Y_l(0)\mid X_l=x, S_l=0, Z_l^o=1)f_{X_l\mid S_l=0, Z_l^o=1}(x)\, dx.\\
	\end{align*}
	We have used the assumption $S_l \perp (Y_l(1), Y_l(0)) \mid X_l$ to go from line three to four and assumption $(Y_l(1) - Y_l(0)) \perp Z_l^o \mid X_l, S_l=0$ to get the final equality.
	
	We calculate,
	\begin{align*}
		E[\nu(X_m^r)]  = & \int_\chi \frac{\Pr(Z_l^o=1, S_l=0\mid X_l=x)}{\Pr(S_l=1\mid X_l=x)} \, dx\\
		= & \frac{\Pr(Z_l^o=1, S_l=0)}{\Pr(S_l=1)}\int_\chi \frac{f_{X_l\mid Z_l^o=1, S_l=0}(x)}{f_{X_l\mid S_l=1}(x)} f_{X_l\mid S_l=1}(x) \, dx\\
		= & \frac{\Pr(Z_l^o=1, S_l=0)}{\Pr(S_l=1)}\int_\chi f_{X_l\mid Z_l^o=1, S_l=0}(x) \, dx.
	\end{align*}
	
	Thus, we have proved the equality, $E\{\nu(X_m^r)[E(Y^r_m(1)\mid X_m^r) - E(Y^r_m(0)\mid X_m^r)]\} = E[\nu(X_m^r)] E(Y_l^o(1) - Y_l^o(0) \mid \X_l^o\in \mathcal{X}, Z_l^o=1)$.\hfill Q.E.D.
	
	
	\medskip
	
	\subsection{\bf Proof of Theorem 3 for completely randomized design.}
	
	Write $C_m = \hat{\nu}(X_m)$, where $\hat{\nu}$ is estimated from the data. Notice that by construction $\hat{\nu}(x)=0$ for $x\not\in \mathcal{X}$.

	\begin{assumption}\label{assm_thm3}

		* $E(Y_m^r(z)|X_m)^2$ for $z=0,1$ are subgaussian random variables.
		
		
		* A completely randomized design with $p_rn_r$ treated units and $(1-p_r)n_r$ control units. $p_r\rightarrow\bar{p}\in(0,1)$.
		
		* $\hat{\nu}$ is in a GC class, $\hat{\nu}(x)\rightarrow\nu(x)$ almost surely for all $x$ and the functions are bounded. 
		Here and later,
		$$\nu(x) = \Pr(Z_l^o = 1 \mid X_l^o=x)\times \Pr(S_k=0\mid X_k=x)/\Pr(S_k=1\mid X_k=x).$$
		
		* Assume that the class for the functions $\hat{\nu}$s is in a Donsker class. Also, assume $L_2$ convergence of $\hat{\nu}$ to $\nu$. 
		
		
		* $Var(Y_m^r(1)) < \infty$ for $z=0,1$. (Hence, $E(Var(Y_m^r(1)\mid X_m^r)) <\infty$ and $Var(E(Y_m^r(1)\mid X_m^r)) <\infty$.)

		* $E \left[\nu(X_m^r) \{ Var(Y_m^r(1)\mid X_m^r)/\tilde{\theta}_m + Var(Y_m^r(0)\mid X_m^r)/(1-\tilde{\theta}_m)\}\right]$ is positive.
		
		* Assume  $S_l \perp (Y_l(1), Y_l(0)) \mid X_l$ and $(Y_l(1) - Y_l(0)) \perp Z_l^o \mid X_l, S_l=0$.

	\end{assumption}
	
	---------------------------
	
	Note that $\sum_m C_m/n_r$ converges almost surely to $E(\nu(X_m^r))$ by the GC class theorem.
	
	We want to establish asymptotic normality of $\sqrt{n_r}\left\{\widehat{\beta^r_\X} - E[Y_l^o(1) - Y_l^o(0) \mid X_l^0\in \mathcal{X}, Z_l^o=1]\right\}$. 
	We instead study the asymptotic distribution of 
	$$\sqrt{n_r}\left\{\frac{\sum_m C_m}{n_r}\widehat{\beta^r_\X} - \frac{\sum_m C_m}{n_r}E[Y_l^o(1) - Y_l^o(0) \mid X_l^o\in \mathcal{X}, Z_l^o=1]\right\}$$. 
	
	Let $\mathbf{X}=\{X_m^o:m=1,\ldots, n_r\}$ and $\mathbf{Z} = \{Z_m^o:m=1,\ldots,n_r\}$.
	Write,
	\begin{align*}
		&\sqrt{n_r}\left\{\frac{\sum_m C_m}{n_r}\widehat{\beta^r_\X} - \frac{\sum_m C_m}{n_r}E[Y_l^o(1) - Y_l^o(0) \mid X_l^o\in \mathcal{X}, Z_l^o=1]\right\} \\
		= & \frac{1}{\sqrt{n_r}}\sum_m C_m\left\{ \frac{Z^r_mY^r_m}{\tilde\theta_{m}} - \frac{(1-Z^r_m)Y^r_m}{1-\tilde\theta_{m}} \right\} - \sqrt{n_r}\frac{\sum_m C_m}{n_r}E[Y_l^o(1) - Y_l^o(0) \mid X_l^o\in \mathcal{X}, Z_l^o=1]\\
		= & \underbrace{\frac{1}{\sqrt{n_r}}\sum_m C_m\left\{ \frac{Z^r_mY^r_m}{\tilde\theta_{m}} - \frac{(1-Z^r_m)Y^r_m}{1-\tilde\theta_{m}} - \frac{Z^r_m}{\tilde\theta_{m}}E(Y^r_m(1)\mid \mathbf{X}, \mathbf{Z}) + \frac{(1-Z^r_m)}{1-\tilde\theta_{m}}E(Y^r_m(0)\mid \mathbf{X}, \mathbf{Z}) \right\}}_{I_{n_r}} \\
		& + \underbrace{\frac{1}{\sqrt{n_r}}\sum_m C_m\left\{ \frac{Z^r_m}{\tilde\theta_{m}}E(Y^r_m(1)\mid \mathbf{X}, \mathbf{Z}) - \frac{(1-Z^r_m)}{1-\tilde\theta_{m}}E(Y^r_m(0)\mid \mathbf{X}, \mathbf{Z}) - E(Y^r_m(1)\mid \mathbf{X}, \mathbf{Z}) + E(Y^r_m(0)\mid \mathbf{X}, \mathbf{Z}) \right\}}_{II_{n_r}}\\
		& + \underbrace{\left\{ \frac{1}{\sqrt{n_r}}\sum_m C_m\left\{ E(Y^r_m(1)\mid \mathbf{X}, \mathbf{Z}) - E(Y^r_m(0)\mid \mathbf{X}, \mathbf{Z}) \right\} - \sqrt{n_r}\frac{\sum_m C_m}{n_r}E[Y_l^o(1) - Y_l^o(0) \mid X_l^o\in \mathcal{X}, Z_l^o=1]\right\}}_{III_{n_r}}\\
		= & I_{n_r} + II_{n_r} + III_{n_r}
	\end{align*}
	
	
	By the fact that the treatment is randomly assigned given $X$, and that we have $Y_m^r(z)$ independent of $X_{m'}^r$ for $m'\neq m$, we can replace $E(Y_m^r(z)\mid \mathbf{X}, \mathbf{Z})$ in the expressions of $I_{n_r}, II_{n_r}$ and $III_{n_r}$ with $E(Y_m^r(z)\mid X_m^r)$ for $z=0,1$. 
	
	For $I_{n_r}$, use Lindeberg's CLT for asymptotic of $I_{n_r}$ conditional on $\mathbf{X}$ and $\mathbf{Z}$. Given $\mathbf{X}$ and $\mathbf{Z}$, the only randomness is through the conditional distributions of $Y_m^r(z)$ given $X_m^r$. We verify Lindeberg's condition to find the asymptotic normality of this term. 
	
	Recall our assumption that the treatment assignment is completely randomized. Thus, notice that, after conditioning on $\mathbf{X}$ and $\mathbf{Z}$, $I_{n_r}$ is $1/\sqrt{n_r}$ times a sum of independent random variables,
	$$C_m\left\{ \frac{Z^r_mY^r_m}{\tilde\theta_{m}} - \frac{(1-Z^r_m)Y^r_m}{1-\tilde\theta_{m}} - \frac{Z^r_m}{\tilde\theta_{m}}E(Y^r_m(1)\mid \mathbf{X}, \mathbf{Z}) + \frac{(1-Z^r_m)}{1-\tilde\theta_{m}}E(Y^r_m(0)\mid \mathbf{X}, \mathbf{Z}) \right\}.$$
	Consider the variance of this term given $\mathbf{X}$ and $\mathbf{Z}$. It is 
	$$\sigma_m^2 := C_m^2\left\{\frac{Z_m^r}{\tilde\theta_m^2}Var(Y_m^r(1)\mid X_m^r) + \frac{1-Z_m^r}{(1-\tilde\theta_m)^2}Var(Y_m^r(0)\mid X_m^r)\right\}.$$
	The covariance term vanishes because $Z_m^r(1-Z_m^r)=0$. Then, with $s_{n_r} := \sum_m \sigma_m^2$, $s_{n_r}^2/n_r$ is 
	$$\frac{1}{n_r}\sum_m C_m^2\left\{\frac{Z_m^r}{\tilde\theta_m^2}Var(Y_m^r(1)\mid X_m^r) + \frac{1-Z_m^r}{(1-\tilde\theta_m)^2}Var(Y_m^r(0)\mid X_m^r)\right\}.$$
	
	Because $\hat{\nu}$'s are in a GC class, so are their squares. Hence, $(x,z)\mapsto \hat{g}(x,z):=\hat{\nu}(x)^2\{ z Var(Y_m^r(1)\mid X_m^r=x) + (1-z)Var(Y_m^r(0)\mid X_m^r = x) \}$ also belong to a GC class. Further, since $\hat{\nu}(x)\rightarrow \nu(x)$ almost surely, $\hat{g}(x, z)\rightarrow g(x, z):={\nu}(x)^2\{ z Var(Y_m^r(1)\mid X_m^r=x) + (1-z)Var(Y_m^r(0)\mid X_m^r = x) \}$ almost surely. Further, since $\hat{\nu}$ are bounded, $\hat{g}(x,z)$ is dominated by a constant times $\{ z Var(Y_m^r(1)\mid X_m^r=x) + (1-z)Var(Y_m^r(0)\mid X_m^r = x) \}$. Hence, we have, given
	$$s_{n_r}^2/n_r\rightarrow
	E \left[\nu(X_m^r) \{ Var(Y_m^r(1)\mid X_m^r)/\tilde{\theta}_m + Var(Y_m^r(0)\mid X_m^r)/(1-\tilde{\theta}_m)\}\right].
	$$
	almost surely. This limit is positive by our assumption.
	Thus, $s_{n_r}\rightarrow \infty$ in probability. 
	
	To check Lindeberg's condition, it is enough to show that, for all $\epsilon >0$,
	$$\lim_{K \rightarrow \infty} \frac{1}{n_r}
	\sum_m E(W_m^2\times I(|W_m|> \epsilon s_{n_r})) = 0,$$
	where $W_m = C_m\left\{ \frac{Z^r_mY^r_m}{\tilde\theta_{m}} - \frac{(1-Z^r_m)Y^r_m}{1-\tilde\theta_{m}} - \frac{Z^r_m}{\tilde\theta_{m}}E(Y^r_m(1)\mid \mathbf{X}, \mathbf{Z}) + \frac{(1-Z^r_m)}{1-\tilde\theta_{m}}E(Y^r_m(0)\mid \mathbf{X}, \mathbf{Z}) \right\}$.
	
	Write,
	$$\frac{1}{n_r}
	\sum_m E(W_m^2\times I(|W_m|> \epsilon s_{n_r})) = E( \frac{1}{n_r}
	\sum_mW_m^2\times I(|W_m|> \epsilon s_{n_r})).$$
	Note, $\frac{1}{n_r} \sum_mW_m^2\times I(|W_m|> \epsilon s_{n_r})\leq \frac{1}{n_r} \sum_mW_m^2$. Now, $E(\{ \frac{Z^r_mY^r_m}{\tilde\theta_{m}} - \frac{(1-Z^r_m)Y^r_m}{1-\tilde\theta_{m}} - \frac{Z^r_m}{\tilde\theta_{m}}E(Y^r_m(1)\mid \mathbf{X}, \mathbf{Z}) + \frac{(1-Z^r_m)}{1-\tilde\theta_{m}}E(Y^r_m(0)\mid \mathbf{X}, \mathbf{Z}) \}^2) \lesssim E(Var(  Y_m^r\mid X_m^r)) + E(Var(  Y_m^r\mid X_m^r)) < \infty$. Thus, using similar arguments as for the limit of $s_{n_r}^2$ the almost sure limit of $\frac{1}{n_r} \sum_mW_m^2$ is $E( \nu(X_m^r)^2\{ \frac{Z^r_mY^r_m}{\tilde\theta_{m}} - \frac{(1-Z^r_m)Y^r_m}{1-\tilde\theta_{m}} - \frac{Z^r_m}{\tilde\theta_{m}}E(Y^r_m(1)\mid \mathbf{X}, \mathbf{Z}) + \frac{(1-Z^r_m)}{1-\tilde\theta_{m}}E(Y^r_m(0)\mid \mathbf{X}, \mathbf{Z}) \}^2)$, which is finite. The Dominated Convergence Theorem gives that we can interchange the limit and the expectation. 
	
	Now,  $\frac{1}{n_r} \sum_m W_m^2\times I(|W_m|> \epsilon s_{n_r})$ converges almost surely to zero. Because, for any $M>0$, for large enough $n_r$, $\frac{1}{n_r} \sum_mW_m^2\times I(|W_m|> \epsilon s_{n_r})\leq \frac{1}{n_r} \sum_mW_m^2\times I(|W_m|> M)$. Thus, $\lim_{n_r\rightarrow\infty}\frac{1}{n_r} \sum_mW_m^2\times I(|W_m|> \epsilon s_{n_r})\leq \lim_{M\rightarrow\infty}\lim_{n_r\rightarrow\infty}\frac{1}{n_r} \sum_mW_m^2\times I(|W_m|> M)$; and since the terms are monotone in $M$, we can interchange the limits and have, $\lim_{n_r\rightarrow\infty}\lim_{M\rightarrow\infty}\frac{1}{n_r} \sum_mW_m^2\times I(|W_m|> M)=0$.
	
	
	Putting things together gives the proof of Lindeberg's condition. Hence, conditional on $X$ and $Z$, almost surely,
	$$I_{n_r} \rightarrow \text{Normal}(0, E \left[\nu(X_m^r) \{ Var(Y_m^r(1)\mid X_m^r)/\tilde{\theta}_m + Var(Y_m^r(0)\mid X_m^r)/(1-\tilde{\theta}_m)\}\right]).--------(*)$$

	For $II_{n_r}$, use results from sampling from a finite population to establish CLT given $\mathbf{X}$. Notice that $E(B_n\mid X) = 0$.  By Theorem 6 and Corollary 3 of Appendix 4 of \cite{lehmann2006nonparametrics}, for given covariate information,
	$$II_{n_r}/\sqrt{Var(II_{n_r}\mid \mathbf{X})} \;\Big{|}\; \mathbf{X} \rightarrow \text{ Normal(0,1)},$$
	when the following is satisfied:
	$$\frac{\max (D_m - \overline{D}_{n_r} )^2}{\sum_m (D_m - \bar{D}_{n_r} )^2/n_r} \text{ \ \ is finite \ \ as \ \ } n_r\rightarrow \infty,$$
	where $D_m = C_m\left\{ \frac{E(Y^r_m(1)\mid X_m^r)}{\tilde\theta_{m}} +  \frac{E(Y^r_m(0)\mid X_m^r)}{1-\tilde\theta_{m}}\right\}$ and $\overline{D}_{n_r} = \sum_m D_m/n_r$.
	
	By GC theorem, almost surely,
	$$\overline{D}_{n_r} = \sum_m D_m/n_r \ \ \text{ converges to } E_d:=E\left[ \nu(X_m^r)\left\{ \frac{E(Y^r_m(1)\mid X_m^r)}{\tilde\theta_{m}} +  \frac{E(Y^r_m(0)\mid X_m^r)}{1-\tilde\theta_{m}}\right\}\right].$$
	
	Next, for any $\lambda$, consider $\frac{1}{\lambda} \log ( \sum_m \exp(\lambda (D_m-\overline{D}_{n_r})^2) )$. Note that as $\lambda\rightarrow \infty$ this quantity goes to $\max_m (D_m-\overline{D}_{n_r})^2$.
	
	Now
	\begin{align*}
		& \frac{1}{\lambda} \log ( \sum_m \exp(\lambda (D_m-\overline{D}_{n_r})^2) )\\
		= & \frac{\log(n_r)}{\lambda} + \frac{1}{\lambda}\log ( \sum_m \exp(\lambda (D_m-\overline{D}_{n_r})^2)/n_r )\\
		= & \frac{\log(n_r)}{\lambda} + 
		2 (E_d-\overline{D}_{n_r})^2  + \frac{1}{\lambda}\log \left\{\frac{1}{n_r}\sum_m \exp(2\lambda (D_m-E_d)^2) \right\}\\ 
		\leq & \frac{\log(n_r)}{\lambda} + 
		2 (E_d-\overline{D}_{n_r})^2  + K\frac{1}{\lambda} \frac{1}{n_r}\sum_m 2\lambda (D_m-E_d)^2 \quad \text{for some constant } K, 
	\end{align*}
	almost surely as $n_r\rightarrow \infty$.
	
	$\Big[$For the final inequality in the above set of calculations, we use the following argument.
	
	By GC class theorem, for any $x$, 
	$$\frac{1}{n_r}\sum_m 1\left(D_m^2 > x\right)\rightarrow \Pr\left\{|\nu(X_m^r)\{ E(Y_m^r(1)\mid X_m^r)/\tilde\theta_m + E(Y_m^r(0)\mid X_m^r)/(1-\tilde\theta_m)\} |^2 > x\right\}.$$
	Now, we have assumed $\nu(X_m^r)$s are bounded. 
	the fact that $|\nu(X_m^r)\{ E(Y_m^r(1)\mid X_m^r)/\tilde\theta_m + E(Y_m^r(0)\mid X_m^r)/(1-\tilde\theta_m)\} |^2$ is subgaussian since $E(Y_m^r(z)\mid X_m^r)^2$ are subgaussian for $z=0,1$.Thus the above limit is bounded by $\leq K'\exp(-x^2/2a^2)$, for some constants $a$ and $K'$.
	
	Denote by $E_{n_r}$ the expectation with respect to the empirical distribution of the data. Then, the above display implies that, for some constant $K''$
	$$E_{n_r} \{ \exp( t (D_m- E_d)^2) \} \leq K''\exp(t^2 a^2/2)\exp( E_{n_r}(D_m- E_d)^2).$$
	Thus we have the inequality with $t=1$ and $K=K''\exp(a^2/2)$.$\qquad\quad\Big]$
	
	Letting $\lambda = n_r$ and letting $n_r\rightarrow\infty$,
	$$\lim_{n_r\rightarrow\infty} \frac{1}{\lambda} \log ( \sum_m \exp(\lambda (D_m-\overline{D}_{n_r})^2 ) ) \leq \lim_{n_r\rightarrow\infty} 2K \frac{1}{n_r}\sum_m  (D_m-E_d)^2.$$
	
	Thus, almost surely in $\mathbf{X}$
	$$II_{n_r}/\sqrt{Var(II_{n_r} \mid \mathbf{X})} \rightarrow \text{Normal(0,1)}.$$
	Next, note that as we have a completely randomized treatment assignment where $p_rn_r$ units are treated and the rest are in control (by using Example 4 of the Appendix of Lehmann Nonparametrics)
	$var(B_n\mid \mathbf{X}) = p_r(1-p_r) \frac{1}{n_r-1}\sum_m (D_m - \overline{D}_{n_r})^2$.
	
	By GC theorem, $\frac{1}{n_r-1}\sum_m (D_m - \overline{D}_{n_r})^2$ converges almost surely to $Var(\nu(X_m^r)\{E(Y_m^r(1)\mid X_m^r)/\tilde\theta_m + E(Y_m^r(0)\mid X_m^r)/(1-\tilde\theta_m)\})$
	
	Thus, by Slutkey's theorem, conditionally on $\mathbf{X}$, almost surely
	$$II_{n_r}\rightarrow Normal\left\{0, \bar{p}(1-\bar{p}))Var(\nu(X_m^r)\{E(Y_m^r(1)\mid X_m^r)/\tilde\theta_m + E(Y_m^r(0)\mid X_m^r)/(1-\tilde\theta_m)\})\right\}.-----(**)$$ 
	
	For $III_{n_r}$, use the fact that $C_m$s are from a Donsker's class and converge in $L_2$. Write, $III_{n_r}$ as
	$$\frac{1}{\sqrt{n_r}}\sum_m C_m\left\{ E(Y^r_m(1)\mid X_m^r) - E(Y^r_m(0)\mid X_m^r) -E[Y_l^o(1) - Y_l^o(0) \mid X_l^0\in \mathcal{X}, Z_l^o=1]\right\}.$$
	
	First, 
	\begin{align*}
		\sqrt{n_r}\bigg[&
		\frac{1}{n_r}\sum_m C_m\left\{ E(Y^r_m(1)\mid X_m^r) - E(Y^r_m(0)\mid X_m^r) -E[Y_l^o(1) - Y_l^o(0) \mid X_l^o\in \mathcal{X}, Z_l^o=1]\right\} \\
		&- E\left\{\nu(X_m^r)\left[ E(Y^r_m(1)\mid X_m^r) - E(Y^r_m(0)\mid X_m^r)\right]\right\} +E\{\nu(X_m^r)\}E[Y_l^o(1) - Y_l^o(0) \mid X_l^o\in \mathcal{X}, Z_l^o=1]\bigg] 
	\end{align*}
	converges to a normal distribution with mean zero and variance 
	$Var( \nu(X_m^r)[E(Y^r_m(1)\mid X_m^r) - E(Y^r_m(0)\mid X_m^r)]) + Var(\nu(X_m^r))\{E[Y_l^o(1) - Y_l^o(0) \mid X_l^o\in \mathcal{X}, Z_l^o=1]\}^2.$
	
	It remains to show:
	$$E\left\{\nu(X_m^r)\left[ E(Y^r_m(1)\mid X_m^r) - E(Y^r_m(0)\mid X_m^r)\right]\right\} = E\{\nu(X_m^r)\}E[Y_l^o(1) - Y_l^o(0) \mid X_l^o\in \mathcal{X}, Z_l^o=1].$$
	Start with the LHS
	\begin{align*}
		& E\left\{\nu(X_m^r)\left[ E(Y^r_m(1)\mid X_m^r) - E(Y^r_m(0)\mid X_m^r)\right]\right\}\\
		= & \int_\mathcal{X} \frac{\Pr(Z_l^o=1, S_l=0\mid X_l=x)}{\Pr(S_l=1\mid X_l=x)} E(Y_l(1) - Y_l(0)\mid X_l=x, S_l=1)f_{X_l\mid S_l=1}(x)\, dx\\
		= & \frac{\Pr(S_l=0, Z_l^o=1)}{\Pr(S_l=1)}\int_\mathcal{X}
		\frac{f_{X_l\mid S_l=0, Z_l^o=1}(x)}{
			f_{X_l\mid S_l=1}(x)} E(Y_l(1) -Y_l(0)\mid X_l=x, S_l=1)f_{X_l\mid S_l=1}(x)\, dx\\
		= & \frac{\Pr(S_l=0, Z_l^o=1)}{\Pr(S_l=1)}\int_\mathcal{X} E(Y_l(1) - Y_l(0)\mid X_l=x, S_l=1)f_{X_l\mid S_l=0, Z_l^o=1}(x)\, dx\\
		= & \frac{\Pr(S_l=0, Z_l^o=1)}{\Pr(S_l=1)}\int_\mathcal{X} E(Y_l(1) - Y_l(0)\mid X_l=x, S_l=0)f_{X_l\mid S_l=0, Z_l^o=1}(x)\, dx\\
		= & \frac{\Pr(S_l=0, Z_l^o=1)}{\Pr(S_l=1)}\int_\mathcal{X} E(Y_l(1) - Y_l(0)\mid X_l=x, S_l=0, Z_l^o=1)f_{X_l\mid S_l=0, Z_l^o=1}(x)\, dx.
	\end{align*}
	We have used the assumption $S_l \perp (Y_l(1), Y_l(0)) \mid X_l$ to go from line three to four and assumption $(Y_l(1) - Y_l(0)) \perp Z_l^o \mid X_l, S_l=0$ to get the final equality.
	
	We calculate,
	\begin{align*}
		E[\nu(X_m^r)]  = & \int_\chi \frac{\Pr(Z_l^o=1, S_l=0\mid X_l=x)}{\Pr(S_l=1\mid X_l=x)} \, dx\\
		= & \frac{\Pr(Z_l^o=1, S_l=0)}{\Pr(S_l=1)}\int_\chi \frac{f_{X_l\mid Z_l^o=1, S_l=0}(x)}{f_{X_l\mid S_l=1}(x)} f_{X_l\mid S_l=1}(x) \, dx\\
		= & \frac{\Pr(Z_l^o=1, S_l=0)}{\Pr(S_l=1)}\int_\chi f_{X_l\mid Z_l^o=1, S_l=0}(x) \, dx.
	\end{align*}
	
	Thus, we have proved the equality, $E\{\nu(X_m^r)[E(Y^r_m(1)\mid X_m^r) - E(Y^r_m(0)\mid X_m^r)]\} = E[\nu(X_m^r)] E(Y_l^o(1) - Y_l^o(0) \mid \X_m^r\in \mathcal{X}, Z_l^o=1)$.
	
	Hence,
	\begin{align*}
		III_{n_r}\rightarrow Normal&\left\{0, Var( \nu(X_m^r)[E(Y^r_m(1)\mid X_m^r) - E(Y^r_m(0)\mid X_m^r)])\right.\\
		&\qquad \left.+ Var(\nu(X_m^r))\{E[Y_l^o(1) - Y_l^o(0) \mid X_l^o\in \mathcal{X}, Z_l^o=1]\}^2\right\}.-----(***)
	\end{align*}

	Using Proposition \ref{prop_sumCLT} we combine  (*),(**), and (***) to show that $I_{n_r}+II_{n_r}+III_{n_r}$ converges in distribution to centered normal with variance $V_I + V_{II} + V_{III}$.
	
	Thus, by $\sum_m C_m/n_r$ almost surely converging to $E(\nu(X_m^r))$, we have
	$$\sqrt{n_r}\left\{\widehat{\beta^r_\X} - E[Y_l^o(1) - Y_l^o(0) \mid X_l^o\in \mathcal{X}, Z_l^o=1]\right\}$$ 
	converges to a normal distribution with mean zero and variance
	$E(\nu(X_m^r))^{-2}\times ( V_I + V_{II} + V_{III})$
	where 
	$$V_I = E \left[\nu(X_m^r) \{ Var(Y_m^r(1)\mid X_m^r)/\tilde{\theta}_m + Var(Y_m^r(0)\mid X_m^r)/(1-\tilde{\theta}_m)\}\right]$$
	$$V_{II} = \bar{p}(1-\bar{p}))Var(\nu(X_m^r)\{E(Y_m^r(1)\mid X_m^r)/\tilde\theta_m + E(Y_m^r(0)\mid X_m^r)/(1-\tilde\theta_m)\})$$
	and 
	$$V_{III} = Var( \nu(X_m^r)[E(Y^r_m(1)-Y^r_m(0)\mid X_m^r)]) + Var(\nu(X_m^r))\{E[Y_l^o(1) - Y_l^o(0) \mid X_l^0\in \mathcal{X}, Z_l^o=1]\}^2.$$\hfill Q.E.D.

	\medskip
	
	\subsection{\bf Proof of Theorem 3 for stratified designs.}
	
	\begin{assumption}\label{assm_thm3_stratified}
		
		* Assume a stratified design where the number of strata $S$ increases to infinity in the asymptotic. Assume fixed stratum size $J$ and fixed number of treated units $t$ in each stratum. Thus, we use the indexing $sj$ for the $j$th unit in stratum $s$.

		* Assume $(Y_{sj}^r(1), Y_{sj}^r(0), X_{sj}, Z_{sj}^r : j=1,\ldots, J)$ are sampled i.i.d.~across $s$. Also, assume the covariates are the same in each stratum, i.e., $X_{s1}=\cdots=X_{sJ}$ for all $s$. 

		* $C_{sj}$s are bounded, belong to a GC class and $C_{sj}\rightarrow\nu(X_{sj})$ almost surely.
		
		* $C_{sj}$s belong to a Donsker class and converges in $L_2$ to $\nu(X_{sj})$.
		
		* Assume finite second moments of the potential outcomes.

		* $E\left[Var\left(\sum_j \nu(X_{sj})\left\{\frac{Z_{sj}^r Y_{sj}^r}{\tilde\theta_{sj}} - \frac{(1-Z_{sj}^r) Y_{sj}^r}{1-\tilde\theta_{sj}} \right\}\Big{|} X_s\right)\right]$ is positive.

		* Assume $S_{sj} \perp (Y_{sj}(1), Y_{sj}(0)) \mid X_{sj}$ and $(Y_{sj}(1) - Y_{sj}(0)) \perp Z_{sj}^o \mid X_{sj}, S_{sj}=0$.

	\end{assumption}
	
	-------------------------------
	
	Recall, 
	$$\widehat{\beta_\mathcal{X}^r} = \frac{1}{\sum_s\sum_j C_{sj}}\sum_s\sum_j C_{sj}\left\{\frac{Z_{sj}^r Y_{sj}^r}{\tilde\theta_{sj}} - \frac{(1-Z_{sj}^r) Y_{sj}^r}{1-\tilde\theta_{sj}}\right\}.$$ 
	
	We want to establish the asymptotic normality of $\sqrt{S}\left\{\widehat{\beta^r_\X} - E[Y_l^o(1) - Y_l^o(0) \mid X_l^0\in \mathcal{X}, Z_l^o=1]\right\}$. We instead study the asymptotic distribution of 
	$$\sqrt{S}\left\{\frac{\sum_s\sum_j C_{sj}}{S}\widehat{\beta^r_\X} - \frac{\sum_s\sum_j C_{sj}}{S}E[Y_l^o(1) - Y_l^o(0) \mid X_l^0\in \mathcal{X}, Z_l^o=1]\right\}.$$
	
	Write,
	\begin{align*}
		& \sqrt{S}\left\{\frac{\sum_s\sum_j C_{sj}}{S}\widehat{\beta^r_\X} - \frac{\sum_s\sum_j C_{sj}}{S}E[Y_l^o(1) - Y_l^o(0) \mid X_l^o\in \mathcal{X}, Z_l^o=1]\right\}\\
		= & \frac{1}{\sqrt{S}}\sum_s\sum_j C_{sj}\left\{\frac{Z_{sj}^r Y_{sj}^r}{\tilde\theta_{sj}} - \frac{(1-Z_{sj}^r) Y_{sj}^r}{1-\tilde\theta_{sj}}\right\} - \sqrt{S}\frac{\sum_s\sum_j C_{sj}}{S}E[Y_l^o(1) - Y_l^o(0) \mid X_l^o\in \mathcal{X}, Z_l^o=1]\\
		= & \underbrace{\frac{1}{\sqrt{S}}\sum_s\left[\sum_j C_{sj}\left\{\frac{Z_{sj}^r Y_{sj}^r}{\tilde\theta_{sj}} - \frac{(1-Z_{sj}^r) Y_{sj}^r}{1-\tilde\theta_{sj}} - E[Y_{sj}^r(1)\mid \mathbf{X}_s] + E[Y_{sj}^r(0)\mid \mathbf{X}_s]\right\}\right]}_{I_{n_r}}\\
		& + \underbrace{\frac{1}{\sqrt{S}}\sum_s\left[\sum_j 
			C_{sj}\left\{E[Y_{sj}^r(1)\mid \mathbf{X}_s] - E[Y_{sj}^r(0)\mid \mathbf{X}_s]  - E[Y_l^o(1) - Y_l^o(0) \mid X_l^o\in \mathcal{X}, Z_l^o=1]\right\}
			\right]}_{II_{n_r}}\\
		= & I_{S} + II_{S}
	\end{align*}
	
	Here $\mathbf{X}_s$ is $(X_{sj}: j=1,\ldots, J)$, which are all equal by our assumption. Let $\mathbf{X}=\{X_{sj}:s=1,\ldots,S, j=1,\ldots,J\}$.
	To establish the asymptotic normality, we first show the asymptotic normality of $I_{S}$ conditional on $X$. This uses Lindeberg's CLT. Then, the asymptotic normality of $II_S$ will follow from Donsker's theorem. 
	
	Consider $I_S$. Notice that $C_{sj}$ is only a function of $X$ and that the conditional expectation of $\frac{Z_{sj}^r Y_{sj}^r}{\tilde\theta_{sj}} - \frac{(1-Z_{sj}^r) Y_{sj}^r}{1-\tilde\theta_{sj}}$ given $\mathbf{X}$ is $E[Y_{sj}^r(1)\mid \mathbf{X}_s] - E[Y_{sj}^r(0)\mid \mathbf{X}_s]$. Let, 
	$$\sigma_{s}^2 = Var\left(\sum_j C_{sj}\left\{\frac{Z_{sj}^r Y_{sj}^r}{\tilde\theta_{sj}} - \frac{(1-Z_{sj}^r) Y_{sj}^r}{1-\tilde\theta_{sj}} \right\} \, \bigg{|}\, \mathbf{X}\right)$$
	and $\Omega_S^2 = \sum_s \sigma_{s}^2$. Notice that, by GC theorem, almost surely, (justify the required assumptions term by term by expanding the variance of the sum)
	$$\frac{1}{S}\Omega_S^2 \rightarrow E\left[Var\left(\sum_j \nu(X_{sj})\left\{\frac{Z_{sj}^r Y_{sj}^r}{\tilde\theta_{sj}} - \frac{(1-Z_{sj}^r) Y_{sj}^r}{1-\tilde\theta_{sj}} \right\}\Big{|} \mathbf{X}_s\right)\right].$$
	By our assumption, this limit is positive. Thus, $\Omega_S^2$ converges to infinity in probability and almost surely. 
	
	To check Lindeberg's condition, it is enough to show that, for all $\epsilon > 0$,
	$$\lim_{S\rightarrow\infty} \frac{1}{S}\sum_sE(W_s^2\times I(|W_s|>\epsilon \Omega_S)\mid x) = 0,$$
	where, $W_{st} = \sum_j C_{sj}\left\{\frac{Z_{sj}^r Y_{sj}^r}{\tilde\theta_{sj}} - \frac{(1-Z_{sj}^r) Y_{sj}^r}{1-\tilde\theta_{sj}} - E[Y_{sj}^r(1)\mid X_s] + E[Y_{sj}^r(0)\mid X_s]\right\}$. By the i.i.d.~assumption on the strata, it suffices to show $\lim_{S\rightarrow\infty} \sum_sE(W_s^2\times I(|W_s|>M))=0$ as $M, S\rightarrow \infty$. To see this, use the dominated convergence theorem with the following facts (i) $W_s^2\times I(|W_s|>M) \leq W_s^2$, (ii) $E(W_s^2 \mid X) = Var(\sum_j C_{sj}\left\{\frac{Z_{sj}^r Y_{sj}^r}{\tilde\theta_{sj}} - \frac{(1-Z_{sj}^r) Y_{sj}^r}{1-\tilde\theta_{sj}}\right\} \mid X)\leq K \sum_j \{Var(Y_{sj}^r(1)mid X_s)+Var(Y_{sj}^r(0)\mid X_s) \}<\infty$ (for some constant $K$; by the finite second moment of the potential outcomes)  and (iii) $\lim_{M\rightarrow \infty} W_s^2\times I(|W_s|>M) = 0$ pointwise.
	
	Thus, by Lindeberg's CLT, conditional on $X$, almost surely 
	$$I_S\rightarrow Normal\left\{0, Var\left(\sum_j \nu(X_{sj})\left\{\frac{Z_{sj}^r Y_{sj}^r}{\tilde\theta_{sj}} - \frac{(1-Z_{sj}^r) Y_{sj}^r}{1-\tilde\theta_{sj}} \right\}\right)\right\}.------(*)$$
	
	Next, consider $II_S$. Recall, by our assumption, $C_{sj}$s belong to a Donsker class. Hence, so do the functions 
	$$\sum_j 
	C_{sj}\left\{E[Y_{sj}^r(1)\mid \mathbf{X}_s] - E[Y_{sj}^r(0)\mid \mathbf{X}_s]  - E[Y_l^o(1) - Y_l^o(0) \mid X_l^o\in \mathcal{X}, Z_l^o=1]\right\},$$
	as a function of $(X_{sj} : j=1,\ldots,J)$. Thus, by Donsker's theorem, we have the asymptotic normality of $II_{n_r}$ using our assumption that we have $L_2$ convergence of $C_{sj}$s.
	
	The variance of the limiting distribution is 
	$$Var\left(\sum_j 
	\nu(X_{sj})\left\{E[Y_{sj}^r(1)\mid \mathbf{X}_s] - E[Y_{sj}^r(0)\mid \mathbf{X}_s]  - E[Y_l^o(1) - Y_l^o(0) \mid X_l^o\in \mathcal{X}, Z_l^o=1]\right\} \right).$$
	
	It remains to check that 
	$$E\left(\sum_j 
	\nu(X_{sj})\left\{E[Y_{sj}^r(1) - Y_{sj}^r(0)\mid \mathbf{X}_s]  - E[Y_l^o(1) - Y_l^o(0) \mid X_l^0\in \mathcal{X}, Z_l^o=1]\right\} \right) = 0.$$
	Equivalently,
	$$\frac{1}{\sum_j \nu(X_{sj})}E\left(\sum_j 
	\nu(X_{sj})E[Y_{sj}^r(1) - Y_{sj}^r(0)\mid X_s]\right) = 
	E[Y_l^o(1) - Y_l^o(0) \mid X_l^o\in \mathcal{X}, Z_l^o=1].$$
	
	Since we assume that the units are stratified perfectly on the covariates, using calculations in the previous proof, we have the equality. 
	
	By Proposition \ref{prop_sumCLT}, we combine (*) and the asymptotic normality of $II_{n_r}$ to get that $I_{n_r}+II_{n_r}$ converges in distribution to centered normal.
	
	Finally, since the almost sure limit (by the GC class theorem) of $\frac{1}{S}\sum_{s,j}C_{sj}$ is $E(\sum_j \nu(X_{sj}))$, we have completed the asymptotic normality.
	
	The limiting variance is 
	\begin{align*}
		\{E(\sum_j & \nu(X_{sj}))\}^{-2}\Big[E\left[Var\left(\sum_j \nu(X_{sj})\left\{\frac{Z_{sj}^r Y_{sj}^r}{\tilde\theta_{sj}} - \frac{(1-Z_{sj}^r) Y_{sj}^r}{1-\tilde\theta_{sj}} \right\}\Big{|} \mathbf{X}_s\right)\right]\\
		& + 
		Var\left(\sum_j 
		\nu(X_{sj})\left\{E[Y_{sj}^r(1) -Y_{sj}^r(0)\mid \mathbf{X}_s]  - E[Y_l^o(1) - Y_l^o(0) \mid X_l^o\in \mathcal{X}, Z_l^o=1]\right\} \right)\Big].
	\end{align*}
	\hfill Q.E.D.
	
	\begin{proposition}\label{prop_sumCLT}
		Consider the sequence of $l$ random vectors $\mathbf{X}_{1,n},\ldots,\mathbf{X}_{L,n}$. Consider the sequence of random vectors $Y_1=f_1(\mathbf{X}_{1,n},\ldots,\mathbf{X}_{L,n})$, $Y_2=f_2(\mathbf{X}_{2,n},\ldots,\mathbf{X}_{L,n})$, $\ldots$, $Y_L=f_L( \mathbf{X}_{L,n})$.
		
		Suppose $Y_{l,n}$ given $\mathbf{X}_{l+1,n},\ldots,\mathbf{X}_{L,n}$ converges in law to the distribution $\mathcal{L}_l$, for $l=1,\ldots,L$. Here $\mathbf{X}_{L+1,n}=\emptyset$.
		
		Then, $\sum_l Y_l$ converges in law to the distribution that is a convolution of the distributions $\mathcal{L}_1,\ldots,\mathcal{L}_l$.
	\end{proposition}
	
	\noindent{\bf Proof.} \ We first prove it for $L=2$. Let $\Psi_l(t)$ be the characteristic function of $\mathcal{L}_l$ for $l=1,...,L$. \vspace{-15pt}
	\begin{align*}
		&\Big|E( e^{\iota t(Y_{1,n}+Y_{2,n})}) - \Psi_1(t)\Psi_2(t)\Big|\\
		= & \Big| E\left\{e^{\iota tY_{2,n}} E(e^{\iota tY_{1,n}}\mid \mathbf{X}_{2,n})\right\} - \Psi_1(t)Ee^{\iota tY_{2,n}}\Big| + \Big|\Psi_1(t)Ee^{\iota tY_{2,n}} - \Psi_1(t)\Psi_2(t) \Big|  \\
		\leq  & E\left\{|e^{\iota tY_{2,n}}| \Big|  E(e^{\iota tY_{1,n}}\mid \mathbf{X}_{2,n})- \Psi_1(t)\Big|\right\} + |\Psi_1(t)|\Big|Ee^{\iota tY_{2,n}} - \Psi_2(t) \Big|\\
		\leq & E\Big|  E(e^{\iota tY_{1,n}}\mid \mathbf{X}_{2,n})- \Psi_1(t)\Big| + \Big|Ee^{\iota tY_{2,n}} - \Psi_2(t) \Big|\vspace{-15pt}
	\end{align*}
	By the convergence in distribution of $Y_{2,n}$, the second term converges to 0. For the first term, by the convergence in distribution of $Y_{1,n}$ given $\mathbf{X}_{2,n}$ to $\mathcal{L}_1$ almost surely, $E(e^{\iota tY_{1,n}}\mid \mathbf{X}_{2,n})$ converges to $ \Psi_1(t)$ almost surely and their difference is bounded by 2. Thus, by the dominated convergence theorem $\Big|E( e^{\iota t(Y_{1,n}+Y_{2,n})}) - \Psi_1(t)\Psi_2(t)\Big|$ goes to zero. Hence, the proof for $L=2$.

	Now consider proof by induction. Suppose we have the result for $L-1$ variables $Y_{2,n},\ldots, Y_{L,n}$
	and we want to show it for $Y_{1,n},\ldots, Y_{L,n}$.
	Thus, we know that $Y_{2,n}+\cdots+ Y_{L,n}$ converges in law to distribution that is a convolution of $\mathcal{L}_2,\ldots, \mathcal{L}_L$. Thus,\vspace{8pt}
	\begingroup
	\setlength{\abovedisplayskip}{-2pt} 
	\setlength{\belowdisplayskip}{-2pt}
	\[
	E\, e^{\iota t (Y_{2,n}+\cdots+ Y_{L,n})} \rightarrow \Psi_2(t)\times \cdots \times \Psi_L(t). \tag{*}
	\]
	Further,
	\[
	E\Big\{ e^{\iota t Y_{1,n}}\,\big|\, \mathbf{X}_{2,n},\ldots, X_{L,n} \Big\} \rightarrow \Psi_1(t). \tag{**}
	\]
	\endgroup
	almost surely. Thus,\vspace{-19pt}
	\begin{align*}
		&\Big| E e^{\iota t (Y_{1,n}+\cdots+Y_{L,n})} - \Psi_1(t)\times \cdots\times \Psi_L(t)\Big|\\
		= & \Big| E \left\{ e^{\iota t(Y_{2,n}+\cdots+Y_{L,n})} E\left( e^{\iota t Y_{1,n}} \mid \mathbf{X}_{2,n},\ldots, \mathbf{X}_{L,n}\right)\right\} - \Psi_1(t)\times \cdots\times \Psi_L(t)\Big|\\
		\leq &  \Big| E \left\{ e^{\iota t(Y_{2,n}+\cdots+Y_{L,n})} E\left( e^{\iota t Y_{1,n}} \mid \mathbf{X}_{2,n},\ldots, \mathbf{X}_{L,n}\right)\right\} -  \Psi_1(t) E \left\{ e^{\iota t(Y_{2,n}+\cdots+Y_{L,n})}\right\}\Big|\\
		& \ \ \ + 
		\Big| \Psi_1(t) E \left\{ e^{\iota t(Y_{2,n}+\cdots+Y_{L,n})}\right\} - 
		\Psi_1(t)\times \cdots\times \Psi_L(t)\Big|\\
		\leq &   E\left[\Big|\left\{ e^{\iota t(Y_{2,n}+\cdots+Y_{L,n})} \right\}\Big| \Big|E\left( e^{\iota t Y_{1,n}} \mid \mathbf{X}_{2,n},\ldots, \mathbf{X}_{L,n}\right)-\Psi_1(t)\Big|\right]\\
		& \ \ \ + 
		\Big| E \left\{ e^{\iota t(Y_{2,n}+\cdots+Y_{L,n})}\right\} - 
		\Psi_2(t)\times\cdots\times\Psi_L(t)\Big|\Big|\Psi_1(t)\Big|\\
		\leq & E \Big|E\left( e^{\iota t Y_{1,n}} \mid \mathbf{X}_{2,n},\ldots, \mathbf{X}_{L,n}\right)-\Psi_1(t)\Big| + \Big| E \left\{ e^{\iota t(Y_{2,n}+\cdots+Y_{L,n})}\right\} - 
		\Psi_2(t)\times\cdots\times\Psi_L(t)\Big|\\
		& \rightarrow 0.
	\end{align*}
	Where the first term goes to zero by the dominated convergence theorem using the bound 2 and (**). The second term goes to zero by (*). Hence, the proof by induction is complete.\hfill Q.E.D.
	
	\medskip
	
	\subsection{\bf Proofs of Theorems 4 and 5.} 
	
	Define the sensitivity analysis $p$-value for testing 
	$$H_0: \beta^\star=\beta^\star_0 \;\; vs\;\;
	H_1: \beta^\star>\beta^\star_0,$$
	as in the main text for the OS study $OS_s$ and RCT $RCT_s$. Denote them as $p_{\beta_0^\star}^{OS_s}$ and $p_{\beta_0^\star}^{RCT_s}$ respectively.
	
	Construct lower sided $(1-\alpha)$ confidence regions as 
	$$(-\infty, \widehat\beta_{U,OS_s,\alpha}^\star] \;\;\;\text{ where }\;\;\;\widehat\beta_{U,OS_s,\alpha}^\star = \sup\{\beta^\star : p_{\beta_0^\star}^{OS_s} \geq \alpha\},$$
	$$(-\infty, \widehat\beta_{U,RCT_s,\alpha}^\star] \;\;\;\text{ where }\;\;\;\widehat\beta_{U,RCT_s,\alpha}^\star = \sup\{\beta^\star : p_{\beta_0^\star}^{RCT_s} \geq \alpha\}.$$
	
	Note now the calculation of the combined $(1-\alpha)$ confidence interval as
	$$(-\infty, \widehat\beta_{U,\alpha}^\star]\;\;\;\text{ where } \widehat\beta_{U,\alpha}^\star = \sup\{\beta^\star : \widehat\beta_{U,RCT_s,\alpha}^\star\times \widehat\beta_{U,OS_s,\alpha}^\star \geq \kappa_\alpha \},$$
	where $\kappa_\alpha = \exp(-1/2\chi^2_{4,1-\alpha})$, with $\chi^2_{4,1-\alpha}$ denoting the $(1-\alpha)$th quantile of the $chi^2$ distribution of $4$ degrees of freedom.
	
	\medskip
	
	\subsubsection{\bf Proof of Theorem 4.}
	The proof is straightforward from the fact that (i) the sensitivity analysis $p$-values are valid and hence are stochastically larger than uniform random variables, (ii) they are independent, and (iii) the $\kappa_\alpha$ is the $(1-\alpha)$th quantile of the product of two independent uniform$(0,1)$ random variables.\hfill Q.E.D.
	\medskip
	
	In Theorem 5, we aim to show that the combined C.I. is ``better" than the individual confidence intervals of the same confidence level. The theoretical result considers an asymptotic situation where the $OS_s$ and $RCT_s$ both increase in size, perhaps are different rates as $s\rightarrow \infty$.
	
	Let $\alpha_s\rightarrow 0$ as $s\rightarrow \infty$ be a sequence that gives an increasing sequence of confidence levels $(1-\alpha_s)\times 100\% \rightarrow 100\%$. In this asymptotic situation we compare $\widehat\beta_{U,OS_s,\alpha_s}^\star$ and $\widehat\beta_{U,RCT_s,\alpha_s}^\star$ to $\widehat\beta_{U,\alpha_s}^\star$.
	
	Recall our assumptions:
	
	\normalfont
	\textbf{Assumption 4}    
	\begin{enumerate}
		\item[4.1] The two sequences of $p$-values $p^{os_s}_{\beta_0^\star}$ and $p^{rct_s}_{\beta_0^\star}$ are monotone in $\beta_0^\star$.
		\item[4.2] $p^{os_s}_{\beta_0^\star}$ and $p^{rct_s}_{\beta_0^\star}$ are continuous in $\beta_0^\star$.
		\item[4.3] $\lim_{s\rightarrow \infty} [\widehat\beta^\star_{U, OS_s, \alpha_s}-\widehat\beta^\star_{U, RCT_s, \alpha_s} ] = 0$.
		Thus, $p^{os_s}_{\beta_0^\star}\rightarrow 0$ and $p^{rct_s}_{\beta_0^\star}\rightarrow 0$ for any $\beta_0^\star>\lim_{s\rightarrow \infty}\widehat\beta^\star_{U, OS_s, \alpha_s}$.
	\end{enumerate}
	
	\subsubsection{\bf Proof of Theorems 5.}
	
	We show that for large enough $s$ 
	$$\widehat\beta^\star_{U,\alpha_s} < \widehat\beta^\star_{U,OS_s,\alpha_s}. \cdots\cdots(*)$$
	The proof for the RCT confidence interval upper limit is similar.
	
	It is enough to show the following to establish $(*)$. 
	$$p_{\beta^\star_0, OS_s}\times p_{\beta^\star_0, RCT_s} < \kappa_{\alpha_s}\;\;\text{ for } \beta^\star_0 = \widehat\beta^\star_{U,OS_s,\alpha_s}.$$
	
	By assumption 
	4.2,
	$p_{\widehat\beta^\star_{U,OS_s,\alpha_s}} = \alpha_s$. Hence, we want to show 
	$$\alpha_s\times p_{\widehat\beta^\star_{U,OS_s,\alpha_s}, RCT_s} > \kappa_{\alpha_s}.$$
	We use the following result from probability theory
	$$\Pr(\chi^2_d \geq d+(2+a)x) \leq \exp(-x)
	\;\;\;\text{ for any } x\geq \frac{4d}{a^2},$$
	where $\chi^2_d$ denotes a $\chi^2$ random variable with degrees of freedom $d$. Thus,
	$$\log\Pr(\chi^2_4\geq y) \leq -\left(\frac{y-4}{2+a}\right) \;\;\;\text{ for } \frac{y-4}{2+a} \geq \frac{4d}{a^2}.$$
	Take $y = -2\log\left(\alpha_s\times p_{\widehat\beta^\star_{U,OS_s,\alpha_s}, RCT_s}\right)$. Then, the upper bound of the probability is 
	$$\frac{4}{2+a} + \frac{2\log\left(\alpha_s\times p_{\widehat\beta^\star_{U,OS_s,\alpha_s}, RCT_s}\right)}{2+a}.$$
	It is enough to show that this number is strictly less than $\log \alpha_s$. Or enough to show
	$$ \frac{a}{2+a}\log \alpha_s - \frac{2\log\left(\alpha_s\times p_{\widehat\beta^\star_{U,OS_s,\alpha_s}, RCT_s}\right)}{2+a} - \frac{4}{2+a} > 0\cdots\cdots(I)$$
	where $a$ is such that 
	$$-\frac{2}{2+a}\log \alpha_s - \frac{2}{2+a}\log p_{\widehat\beta^\star_{U,OS_s,\alpha_s}, RCT_s} \geq \frac{4}{2+a} + \frac{16}{a^2}.\cdots\cdots(II)$$
	We show that we can choose $a$ so that for large enough $s$ $(I)$ and $(II)$ are simultaneously satisfied.
	
	Notice that, $(II)$ is satisfied if, ($a$ is $\geq 1$)
	$$-\log \alpha_s - 2\log p_{\widehat\beta^\star_{U,OS_s,\alpha_s}, RCT_s} \geq 4 + \frac{16(2+a)}{a},$$
	or 
	$$a\geq \frac{32}{-2\log \alpha_s - 2\log p_{\widehat\beta^\star_{U,OS_s,\alpha_s}, RCT_s} - 20}\cdots\cdots(III).$$
	By assumption 
	4.3, choose $s$ large enough so that 
	$$\widehat\beta^\star_{U,OS_s,\alpha_s} > \beta^\star_0 \;\;\;\text{ and } p_{\beta_0^\star, RCT_s} < \exp(-10) \cdots\cdots(IV)$$
	Then we can choose
	$a = \frac{32}{-2\log(\alpha_s)}+2$  to satisfy  (III)  and hence (II). 
	Now in $(I)$ the left hand side multiplied by $(2+a)$ is
	\begin{align*}
		&(a-2)\log \alpha_s - 2\log p_{\widehat\beta^\star_{U,OS_s,\alpha_s}, RCT_s} - 4\\
		& = -16 -2\log p_{\widehat\beta^\star_{U,OS_s,\alpha_s}, RCT_s} - 4\\
		\text{by }(IV) & > -20 -2\log e^{-10} = 0.
	\end{align*}
	Hence the proof.\hfill Q.E.D.

{
\small
\begin{singlespace}
  \bibliography{bib}    
\end{singlespace}
}

\end{document}